%% file: hole.tex
\documentstyle[pre,aps,eqsecnum]{revtex}

\begin{document}
\title{Rate-and-State Theory of Plastic Deformation Near a 
  Circular Hole}
\author{J.~S. Langer}
\address{Department of Physics, University of California, Santa
  Barbara, CA 93106}
\author{Alexander E.~Lobkovsky}
\address{Institute for Theoretical Physics, University of California,
  Santa Barbara, CA 93106}
\draft
\date{\today}

\maketitle

\begin{abstract}
  We show that a simple rate-and-state theory accounts for
  most features of both time-independent and time-dependent
  plasticity in a spatially inhomogeneous situation,
  specifically, a circular hole in a large stressed plate.
  Those features include linear viscoelastic flow at small
  applied stresses, strain hardening at larger stresses, and
  a dynamic transition to viscoplasticity at a yield stress.
  In the static limit, this theory predicts the existence of
  a plastic zone near the hole for some but not all ranges
  of parameters.  The rate-and-state theory also predicts
  dynamic failure modes that we believe may be relevant to
  fracture mechanics.
\end{abstract}

\section{Introduction}
\label{sec:intro}

Since the work of Hart \cite{hart} in the 1960's, scientists have
understood that a satisfactory theory of plastic deformation in solids
must include dynamic variables that describe the internal states of
materials.  The deformation field itself cannot be sufficient.  It
cannot, in any natural way, describe the irreversible changes that
lead to hysteretic stress-strain curves, or to the transition from
nonlinear viscoelastic to viscoplastic behavior with increasing
applied stress.  Conventional theories of plasticity cope with these
limitations by specifying phenomenological rules to suit various
situations and histories of deformation.  For example, strain
hardening curves, viscoplastic laws, or the distinctions between
loading and unloading behaviors are determined from experiment and
used as needed in computations.  In most treatments there is also a
sharp distinction between time-independent and time-dependent
behaviors, with little or no indication of how these properties may be
related to one another. We believe that a deeper, more nearly
fundamental level of phenomenology is required for modern
applications, for example, for computing deformations near moving
crack tips.  Plasticity is an intrinsically time dependent phenomenon;
time-independent descriptions should emerge as static limits of fully
dynamic theories.

In a recent paper \cite{fl}, M. Falk and one of the present authors
(JSL) proposed a theory of plastic deformation in amorphous solids in
which they introduced an internal state variable to describe the
orientations of what they called ``shear-transformation zones.'' (We
refer to this as the ``STZ'' theory.) The resulting ``rate-and-state''
equations (a concept that is widely used in the seismological
literature \cite{ruina,dieterich} and in recent theories of friction
\cite{cb}) successfully describe the full range of viscoelastic and
viscoplastic phenomena, including hysteretic effects.  The basic
structure of the STZ theory appears to be broadly applicable.  The new
internal state variable might equally well describe, for example,
anisotropy in the way dislocations pile up near defects in crystalline
materials.

Reference \cite{fl} discusses plasticity only in spatially uniform
situations.  Our purpose here is to apply a simple version of the STZ
theory to a spatially inhomogeneous situation and to make contact with
conventional plasticity theory.  We especially want to learn whether
the conventional picture of a time-independent plastic zone appears in
the static limit.  Looking ahead to fully dynamic situations such as
fracture, we also want to understand the dynamics of plastic flow in
regions of concentrated stress.  We shall show that the conventional
time-independent concepts --- the ``yield-surface'' hypotheses --- do
emerge from dynamic theories in an approximate way in many normal
situations.  As we shall argue, however, the rate-and-state theory is
simpler, richer and more general than the conventional approaches.

The problem of describing spatially inhomogeneous plastic deformation
is best approached by looking at a simple example where questions of
time dependence and compatibility are not obscured by mathematical
details.  A growing circular hole in a very large stressed plate
satisfies the criterion of simplicity.  It shares important features
with the case of plastic deformation near a crack tip; the tractions
are applied at a distance and the stresses are concentrated near the
hole.  Several researchers have addressed the problem of dynamic hole
growth in the context of ductile fracture \cite{mcclintock} and
spallation \cite{johnson}.  Carroll and Holt and later Johnson
included inertial effects but neglected rate dependent plasticity
\cite{carroll,johnson}.  Bodner and Partom used a version of a
conventional rate dependent plasticity theory but did not study a
stress controlled situation \cite{bodner}.

The scheme of this paper is as follows.  In Section \ref{sec:dynplast},
we derive equations of motion for the radius of the circular hole and
the surrounding stress field, assuming a general form for the
constitutive relation that governs the rate of plastic deformation.
Section \ref{sec:conventional} contains a brief summary of several
results of conventional plasticity theory which will serve as points
of comparison for the rate-and-state analysis. A simplified version of
the STZ model is introduced in Section \ref{sec:STZBasic} where, for
completeness, we outline some important properties of this model that
were reported previously in \cite{fl} and \cite{LL}.  We describe the
dynamic behavior of the STZ model for the hole problem in Section
\ref{sec:STZCircle}.  The paper concludes with remarks about the
implications of these results in Section \ref{sec:discussion}.

\section{Dynamic Plasticity in a Circular Geometry}
\label{sec:dynplast}

Throughout this analysis, we consider only a two-dimensional solid in
a state of plane strain, and assume that inertial effects are
negligible.  Suppose that a circular hole in this system has radius
$R(t)$ at time $t$. Outward tractions at the distant edges of the
plate cause the pressure $p$ far from the hole to be
$p\to-\sigma_{\infty}$.  We introduce polar coordinates $r$ and
$\theta$ that define an Eulerian reference frame, so that $u(r,t)$ is
the radial displacement, measured from some initial reference state,
of the material currently at position $r$.  The function $u(r,t)$ is
the only degree of freedom in the problem.

The total rate-of-deformation tensor (including both elastic 
and plastic parts) is diagonal with components:
\begin{equation}
  \label{Dtot}
  {\cal D}_{rr}^{tot}={\partial v\over \partial r};~~~~~{\cal
    D}_{\theta\theta}^{tot}={v\over r},
\end{equation}
where $v$ is the material velocity:
\begin{equation}
  v={\partial u/\partial t\over 1 - \partial u/\partial r}.
\end{equation}
For small strains, the tensor ${\cal D}^{tot}$ is approximately equal
to the total strain-rate tensor $\dot\varepsilon^{tot}$.
Eq.(\ref{Dtot}) implies that the components of ${\cal D}^{tot}$
satisfy the compatibility condition:
\begin{equation}
  \label{compatibility}
  {\partial \over \partial r}\left(r{\cal D}_{\theta \theta}^{tot}\right)= 
{\cal D}_{rr}^{tot}.
\end{equation}
The stress tensor has the form:
\begin{equation}
\sigma_{rr}=-p-s,~~~~~\sigma_{\theta\theta}=-p+s.
\end{equation}
Here, $s=(\sigma_{\theta\theta}-\sigma_{rr})/2$ is the 
deviatoric stress. The elastic stress-strain relations are:
\begin{equation}
  \label{elast}
  2\mu\,\varepsilon_{rr}^{el}=-(1-2\nu)p-s,~~~~  
  2\mu\,\varepsilon_{\theta\theta}^{el}=-(1-2\nu)p + s,
\end{equation}
where $\mu$ is the shear modulus and $\nu$ is Poisson's ratio. In the
absence of inertial effects, balance of forces implies
\begin{equation}
  \label{force}
  {\partial p \over \partial r} = -{1 \over r^2}\, {\partial \over
    \partial r}\,(r^2s).
\end{equation}
The boundary condition at $r = R$ is
\begin{equation}
\label{BC}
 -\sigma_{rr}(R,t)=p(R,t)+s(R,t)=0.
\end{equation}
For simplicity, we neglect surface tension. Also at $r=R$, we have
\begin{equation}
\label{DtotR}
{\dot R\over R}={\cal D}_{\theta\theta}^{tot}[s(R), ...].
\end{equation}
Here and elsewhere, dots above symbols denote derivatives 
with respect to time $t$. 

For incompressible plasticity, the constitutive relation has the form:
\begin{equation}
  \label{plast}
  {\cal D}_{\theta\theta}^{pl}=-{\cal D}_{rr}^{pl}\equiv{\cal 
    D}(s,...),
\end{equation}
where ${\cal D}$ is some function of $s$ and possibly other  
variables as indicated by the ellipsis.  Combining 
(\ref{plast}) with the compatibility condition 
(\ref{compatibility}), the elasticity equations 
(\ref{elast}), and force balance (\ref{force}), we find
\begin{equation}
  \label{sdoteqn}
  {\cal D}(s,...)+{(1 - \nu) \over \mu}\,\dot s = {C(t) \over r^2},
\end{equation}
where $C(t)$ is an $r$-independent constant of integration.  
Further analysis using the boundary conditions (\ref{BC}) 
and (\ref{DtotR}) at $r=R$ yields 
\begin{equation}
\label{Ceqn}
C(t)=\left[1+\left({1-2\nu\over 
\mu}\right)\,s(R,t)\right]\,R\dot R; 
\end{equation}
and
\begin{equation}
  \label{Rdoteqn}
  \left[1 - {1 \over \mu}\, s(R, t) \right]\, {\dot R \over R} = 
  2\int_R^{\infty}{dr \over r}\, {\cal D}[s(r, t), ...].
\end{equation}
Eqs. (\ref{sdoteqn}) through (\ref{Rdoteqn}), supplemented 
by equations of motion for other arguments of ${\cal D}(s, 
...)$, constitute a coupled set of first-order differential 
equations suitable for computing the time evolution of 
$s(r,t)$ and $R(t)$.  

If $R(t)$ is the only length scale in the problem, we can look for
self-similar solutions in which $\dot R/R = \omega = {\rm constant}$
so that the hole radius is growing exponentially.  All functions
depend only on $\xi = r/R(t)$.  Combining the expression for $C(t)$ in
(\ref{Ceqn}) with (\ref{sdoteqn}), and transforming to the scaling
variable $\xi$, we find
\begin{equation}
\label{sdoteqn2}
{\cal D}(\tilde s, ...) - {(1 - \nu)\over \mu}\,\omega\xi\, 
{d\tilde s\over d\xi}= {\omega\over 
\xi^2}\,\left[1+\left({1-2\nu\over \mu}\right)\,\tilde 
s(1)\right].
\end{equation}
Force balance (\ref{force}) plus the boundary condition 
(\ref{BC}) imply
\begin{equation}
\label{BC2}
\sigma_{\infty}=2\int_1^{\infty} {d\xi\over \xi}\,\tilde 
s(\xi).
\end{equation}
If these self-similar solutions exist for $\omega > 0$, they 
describe unbounded plastic failure of the material.

\section{Conventional Theories}
\label{sec:conventional}

In a typical time-independent approach to this problem
\cite{lubliner}, one assumes that there exists a maximum value of $s$,
say $s_y$, and that, in any sufficiently slow deformation, the
material adjusts its state so that $s\le s_y$ everywhere.
Technically, the condition $s=s_y$ is a special case of a Tresca yield
surface in the space of stress components.  For present purposes, we
take this assumption to mean that the hole is surrounded by a plastic
zone, $R<r<R_1$, within which the force-balance equation (\ref{force})
remains valid but the condition $s=s_y$ replaces Hookean elasticity
(\ref{elast}). Outside this zone, $r > R_1$, $p = -\sigma_\infty$ and
$s = s_y R_1^2/r^2$.  Continuity of stress at $R_1$ means that, within
the zone, $p = -\sigma_{\infty}-2 s_y \ln(r/R_1)$.  Then the boundary
condition (\ref{BC}) at $r=R$ implies that
\begin{equation}
  \label{plastzone}
  \ln{R_1\over R}={1\over 2}\left({\sigma_{\infty}\over s_y}-
    1\right).
\end{equation}
Thus, these assumptions predict that a stationary state with a
non-vanishing plastic zone exists for $\sigma_{\infty}>s_y$.  A
calculation of the displacements similar to that described by Hill
\cite{HILL} shows that the hole radius $R$ diverges as
$\sigma_{\infty}$ approaches an upper threshold stress
$\sigma_\infty^{th}$ which (for the case $s_y/\mu \ll 1$) is given by:
\begin{equation}
  \label{sigmathresh}
  {\sigma_{\infty}^{th}\over s_y} = 1 + \ln \left({\mu \over 2 s_y(1 -
      \nu)}\right).
\end{equation}
To see what happens above this stress, we must consider a 
time-dependent theory.

The simplest conventional time-dependent hypothesis is the 
Bingham law which, for $s\ge 0$, we can write in the form:
\begin{equation}
  \label{bingham}
  {\cal D}(s) = \cases{\alpha(s-s_y)&for $s\ge s_y$\cr 
0&otherwise,}
\end{equation}
where $\alpha$ is a response coefficient.  Essentially by definition,
the combination of (\ref{bingham}) with
(\ref{sdoteqn})--(\ref{Rdoteqn}) describes an elastic
perfectly-plastic material whose stationary states, for $s_y <
\sigma_{\infty} < \sigma_{\infty}^{th}$, are the same as those
described in the preceding paragraph.  We have checked that these
states are stable attractors by using (\ref{bingham}) to integrate
(\ref{sdoteqn})--(\ref{Rdoteqn}), using the initial condition $s(r,0)=
\sigma_{\infty}R^2(0)/r^2$ (for $r>R(0)$).  We found no surprises; a
plastic zone consistent with (\ref{plastzone}) forms around the
growing hole.

We also can compute the self-similar solutions of (\ref{sdoteqn2}) for
the Bingham model.  For ease of analysis, we write these for the case
of incompressible elasticity, $\nu = 1/2$. In the outer, elastic
region, $\xi > \xi_1 = (\mu/s_y)^{1/2}$:
\begin{equation}
  \tilde s(\xi)=s_y\,\left({\xi_1\over\xi}\right)^2;
\end{equation}
and, in the plastic region, $s\ge s_y$, $1 < \xi <\xi_1$:
\begin{equation}
  \label{stilde}
  \tilde s(\xi)= s_y + \left({\mu\omega\over 
      \mu\alpha+\omega}\right)\,{1\over\xi^2}\,\left[1-
    \left({\xi\over \xi_1}\right)^{\beta+2}\right],
\end{equation}
where $\beta(\omega)=2\mu\alpha/\omega$. Note that the 
exponent $\beta$ becomes indefinitely large in the limit of 
small $\omega$; thus the second term in the square brackets 
in (\ref{stilde}) produces a function $\tilde s(\xi)$ that 
is sharply bent but continuous at $\xi=\xi_1$.  To complete 
the calculation, we use (\ref{stilde}) to evaluate the 
right-hand side of (\ref{BC2}).  After rearranging and 
taking the limit of small $\omega$, we find:
\begin{equation}
\omega\approx {\alpha\mu\over s_y}\,(\sigma_{\infty}-
\sigma_{\infty}^{th}),
\end{equation}
where  $\sigma_{\infty}^{th}$ is the upper threshold defined 
in (\ref{sigmathresh}).  Thus, as expected, the dynamic 
failure modes start where the time-independent theory breaks 
down.

\section{The STZ Model}
\subsection{Basic properties}
\label{sec:STZBasic}

For present purposes, it will be sufficient to use the simplified
version of the STZ model that we introduced in an earlier
one-dimensional analysis \cite{LL}.  As in the Bingham case, the model
is specified by the function ${\cal D}(s, ...)$ defined in
(\ref{plast}).  For two-dimensions, with circular symmetry, we write:
\begin{equation}
  \label{STZ1}
  {\cal D}(s, \Delta) = {1 \over \tau}\,(\lambda s -\Delta),
\end{equation}
and supplement this by an equation of motion for the state 
variable $\Delta$:
\begin{equation}
  \label{STZ2}
  \dot \Delta = {\cal D}(s, \Delta)\,(1 - \gamma s\,\Delta).
\end{equation}
$\Delta(r,t)$ is the single independent element of a 
diagonal, traceless tensor which describes the local 
anisotropy of the shear-transformation zones.  We are 
omitting the other state variable in \cite{fl} that 
describes the density of STZ's on the assumption that this 
quantity quickly reaches its equilibrium value.  More 
importantly, this simplified version of the STZ model omits 
the strongly $s$-dependent rate factor that governs memory 
effects.  This version is qualitatively sensible if we load 
the system only once in one direction, but it does not 
behave properly if the loading is cycled in any way.

The inverse stress $\gamma$ can be eliminated in favor of a group of
parameters that plays the role of a dynamic yield stress,
specifically, $s_y=1/\sqrt{\lambda\gamma}$.  Note what is happening
here for spatially uniform situations.  For $s<s_y$, $\Delta(t)$ has
its stable fixed points on the line $\Delta=\lambda s$, where ${\cal
  D} \cong \dot\varepsilon_{\theta\theta}^{pl}=0$.  In this region,
the material is nonlinearly viscoelastic.  For $s\ll s_y$ or,
equivalently, $\gamma\to 0$, it obeys a conventional creep-compliance
law \cite{LL}:
\begin{equation}
\varepsilon^{tot}(t)=(1+\lambda)\,s(t)-\lambda\,\int_{-
\infty}^t dt'\,\exp\left[-{1\over\tau}\,(t-t')\right]\,\dot 
s(t').
\end{equation}

We obtain a particularly important result by supposing that the system
is initially in a state with $\varepsilon_{\theta \theta}^{pl} =
\Delta = 0$ and that a stress $s < s_y$ is applied suddenly at time $t
= 0$.  A simple calculation then yields:
\begin{equation}
  \label{hardening}
  \varepsilon_{\rm final}^{pl}\equiv\varepsilon_{\theta\theta}^{pl} (t
  \to \infty) =- {\lambda s_y^2\over s}\,\ln\left(1-{s^2\over 
      s_y^2}\right).
\end{equation}
$\varepsilon^{pl}(t)$ approaches $\varepsilon_{\rm final}^{pl}$
exponentially in time with a relaxation time $\tau_{\rm relax}$ that
diverges as $s\to s_y$:
\begin{equation}
  \label{relax}
  \tau_{\rm relax}={\tau \over 1 - (s/s_y)^2}.
\end{equation}
Eq.~(\ref{hardening}) is a strain-hardening curve; that is,
$\varepsilon_{\rm final}^{pl}$ is the non-recoverable plastic strain
produced after an infinitely long time by the deviatoric stress $s$.
In the limit $\lambda \to 0$, with $s_y$ held constant,
$\varepsilon_{\rm final}^{pl}$ vanishes for $s < s_y$ but can have any
positive value for $s \cong s_y$.  Thus the parameter $\lambda$ is a
measure of the deviation from perfect plasticity.  The diverging
relaxation time near $s = s_y$, however, has no simple analog in
conventional descriptions of strain hardening.

This one-parameter fit (\ref{hardening}) to the shape of the
strain-hardening curve is much too simple even for the fully
nonlinear STZ model, where the behaviors at small stresses and at
stresses near $s_y$ are determined by different groups of parameters.
Moreover, the small-$s$ behavior of this $t \to \infty$ curve is not
what is measured experimentally.  In real materials and in the full
STZ theory, the plastic deformation rates at small stresses are too
small to be observed, and the material behaves as if it were purely
elastic.  For an illustration of this behavior, see Fig.~5 in
\cite{fl}.  

For $s > s_y$, $\Delta$ goes to $1/\gamma s$ and 
\begin{equation}
  \label{syeqn}
  \dot\varepsilon_{\theta\theta}^{pl}\to {\lambda\over \tau  
    s}\,\left(s^2-s_y^2\right) \cong {2\lambda\over \tau}\,(s-
  s_y).
\end{equation}
In dynamic situations at large stress, therefore, the STZ 
model looks like a Bingham plastic.  In short, even this 
highly simplified version of the STZ model describes much of 
both static and dynamic plasticity. 

\subsection{Dynamically growing hole}
\label{sec:STZCircle}

We return now to the circle problem.  As a first investigation, we
have numerically integrated (\ref{sdoteqn})--(\ref{Rdoteqn}),
supplemented by (\ref{STZ2}), to find the time evolution of $s(r,t)$,
$\Delta(r,t)$, and $R(t)$.  In Figs.~\ref{fig:lowlambda} and
\ref{fig:delta_lowlam}, we show what happens if we suddenly apply the
stress $s(r,0) = \sigma_{\infty} R^2(0)/r^2$.  We set $\lambda s_y =
0.005$, $s_y = 0.1\mu$, and $\sigma_{\infty}= 0.2\mu$ and assume a
previously undeformed system, $\Delta(r, 0) = 0$.  For these values of
the parameters, the hole grows for a while and then stops, and a
plastic zone with $s \cong s_y$, $\Delta \cong \lambda s_y$ forms
around it.  Apart from the fact that the outer boundary of the plastic
zone is smooth rather than sharply defined, this static limit of the
time-dependent deformation is qualitatively consistent with
conventional, time-independent plasticity theory.

In Fig.~\ref{fig:highlambda}, we show an analogous set of curves for a
substantially larger value of $\lambda$, specifically, $\lambda s_y =
0.5$. According to (\ref{hardening}), this system deviates appreciably
from perfect plasticity.  A plastic zone does form around the hole,
but it has no sharp outer boundary. If we estimate the position of
this boundary, say, by finding the point of inflection in the final
curve $s(r)$, we find that the relation (\ref{plastzone}) is strongly
violated.

As in the Bingham model discussed in Section \ref{sec:conventional},
the STZ model predicts the existence of unbounded failure modes for
sufficiently large $\sigma_{\infty}$.  The stationary states of the
kind shown in Figs.~\ref{fig:lowlambda} and \ref{fig:highlambda} cease
to exist beyond $\sigma_{\infty}$.  Indeed, as we show in
Fig.~\ref{fig:Rfinal}) the radius of the hole diverges at
$\sigma_{\infty}^{th}$.  We find these dynamically growing states via
the scaling analysis of Eqs.~(\ref{sdoteqn2}) and (\ref{BC2}).  It is
useful to make the following changes of variables:
\begin{equation}
  {R^2(t)\over r^2} = {1 \over \xi^2} = \zeta;~~~~
  s(r,t) = s_y \psi(\zeta);~~~~\Delta(r,t) = \lambda 
  s_y \varphi(\zeta).
\end{equation}
We find:
\begin{equation}
  \label{sdoteqn3}
  \psi - \varphi + {2(1 - \nu) \omega \tau \over \mu \lambda}\,
  \zeta\, {d\psi \over d\zeta} = {\omega \tau \over \lambda s_y}\,
  \zeta\, \left[1 + (1 - 2\nu)\, {s_y \over \mu}\psi(1)\right].
\end{equation}
The equation of motion for $\Delta$, i.e. (\ref{STZ2}), 
becomes:
\begin{equation}
  \label{STZ3}
  2\omega \tau \zeta{d\varphi \over d\zeta} = (\psi - \varphi)\, (1 -
  \varphi \psi).
\end{equation}
Finally, (\ref{BC2}) becomes:
\begin{equation}
  \label{BC3}
  \sigma_{\infty} = s_y \int_0^1 {d\zeta \over \zeta}\, \psi(\zeta).
\end{equation}

As a first step in interpreting these equations, we compute the
threshold $\sigma_{\infty}^{th}$ by taking the limit $\omega\to 0$.
To avoid unnecessary complication, we again set $\nu = 1/2$.  In this
case, (\ref{sdoteqn2}) and (\ref{STZ3}) reduce to $\psi \approx
\varphi$ and, after a simple integration:
\begin{equation}
  \label{eq:zeroth_order}
  \lambda s_y \ln \left({1 + \psi \over 1 -\psi} \right) + {s_y \over
    \mu}\, \psi = \zeta.
\end{equation}
Note that, if $s_y/\mu \ll 1$ (which is generally true for realistic
situations), then
\begin{equation}
\psi(\zeta)\cong \tanh\left({\zeta\over 2\lambda 
s_y}\right).
\end{equation}
This solution exhibits a plastic zone with a smooth elastic-plastic
boundary only for $\lambda s_y \ll 1$, in which case there is a region
between $\zeta = 2 \lambda s_y$ and $\zeta = 1$ in which $\psi \approx
1$.
 
It is easy to compute $\psi(\zeta)$ without making the latter
approximation and, via (\ref{BC3}), to obtain the threshold stress
$\sigma_{\infty}^{th}$. If $(s_y/\mu)(1 + 2\lambda \mu) \ll 1$, then
\begin{equation}
  {\sigma_{\infty}^{th} \over s_y} \approx 1 + \ln\left({\mu \over 
      s_y}\right) - \ln(1 + 2\lambda \mu),
\end{equation}
which agrees with (\ref{sigmathresh}) when $\lambda = 0$ and $\nu =
1/2$, i.e. in the limit of perfect plasticity.  Note that we have
chosen the parameters in Fig.~\ref{fig:lowlambda} to lie within this
range.  If, on the other hand, $\lambda s_y \gg 1$, then
$\sigma_{\infty}^{th} \approx 1/ 2\lambda$. Here the threshold lies
below $s_y$; the material is highly deformable and conventional
plasticity theory has no range of validity.  We illustrate this point
in Fig.~\ref{fig:lam_thresh}.

Expanding the solutions of (\ref{sdoteqn3}) and (\ref{STZ3}) to first
order in $\omega$, we find the following behavior near threshold: For
$s_y/\mu \ll 2\lambda s_y \ll 1$,
\begin{equation}
  \omega\approx {2 \lambda \over \tau}\, [1 + 2 \lambda s_y\,
  \ln(\lambda s_y)]\,(\sigma_{\infty} - \sigma_{\infty}^{th});
\end{equation}
and, for $\lambda s_y\gg 1$,
\begin{equation}
  \omega \approx {\lambda \over \tau}\,(\sigma_{\infty} -
  \sigma_{\infty}^{th}).
\end{equation}
In each of the last two equations, the quantity $\sigma_{\infty}^{th}$
has the value computed in the corresponding limit in the previous
paragraph.

We have studied the behavior of the similarity solutions
(\ref{sdoteqn3})--(\ref{BC3}) numerically.  Just as in the Bingham
plastic, the deviatoric stress at the surface of the hole in the STZ
material grows with the hole expansion rate $\omega$ as shown in
Fig.~\ref{fig:sim_omega}.  While intuitively obvious, this result is
relevant to understanding stress transmission to brittle crack tips.
In Fig.~\ref{fig:sim_lambda} we show the $\lambda$-dependence of the
stress for a slowly expanding hole.  This solution is essentially the
$\omega \to 0$ limit obtained in Eq.~(\ref{eq:zeroth_order}).  In a
``softer'' material with larger $\lambda$, the plastic zone shrinks
and disappears completely for a large enough $\lambda.$ This softening
of the material for larger $\lambda$ leads to the decrease in the
threshold stress $\sigma_\infty^{th}$ above which dynamic failure
modes exist.

\section{Discussion}
\label{sec:discussion}

Our principal point is that the STZ model, a simple example of a
rate-and-state theory, provides an extremely compact and physically
motivated description of essentially all of plasticity theory, both
static and time dependent.  In just two constitutive relations,
(\ref{STZ1}) and (\ref{STZ2}), containing just two dimensionless
groups of parameters, $s_y/\mu$ and $\lambda s_y$, plus one time
constant $\tau$, we capture linear viscoelasticity at small stress,
strain hardening at larger stress, and a dynamic transition to
viscoplasticity at a yield stress.  All of these properties have been
described previously in \cite{fl} for homogeneous situations and in
\cite{LL} for an inhomogeneous one dimensional case.

In this paper, our principal interest has been to make contact with
conventional theories of plasticity by looking at deformation near a
circular hole.  As a rule, we recover conventional results when both
$s_y \ll \mu$ and $\lambda s_y \ll 1$.  The first of these conditions,
that the yield stress must be much less than the shear modulus, is
generally true for ordinary materials.  The second condition,
according to (\ref{hardening}), says that the plastic strain at which
large-scale deformations start to occur must be much smaller than
unity.  This too is ordinarily satisfied by rigid solids.  However, it
remains to be seen whether this $\lambda$-dependent condition might be
modified in more realistic parameterizations of rate-and-state
theories.

One place where the rate-and-state theory differs qualitatively from
conventional plasticity is in the interpretation of the strain
hardening curve (\ref{hardening}).  Here, this curve must be
interpreted as the $t \to \infty$ limit of the dynamic inelastic
response to an externally applied stress, not as an instantaneous
response that is mathematically equivalent to nonlinear elasticity.
Moreover, the parameter $\lambda$ that determines the deviation from
perfect plasticity in (\ref{hardening}) is the same parameter that
appears in the viscoplastic law (\ref{syeqn}), and is also the same
parameter that determines both the shape of the plastic zone and the
threshold $\sigma_{\infty}^{th}$ at which static solutions give way to
time-dependent, unbounded failure modes.

Perhaps the single most important advantage of the STZ theory is that
the material in the plastic zone is characterized not just by the
stress and corresponding displacement fields but also by the state
variable $\Delta$.  This variable tells us in a natural way that the
material inside the zone will respond differently to subsequent
changes in stress than will the undeformed material outside it. To
take advantage of this feature, however, we must go beyond the
truncated version of the STZ model that we have used here and include
the strongly nonlinear, $s$-dependent rate factor derived in
\cite{fl}.

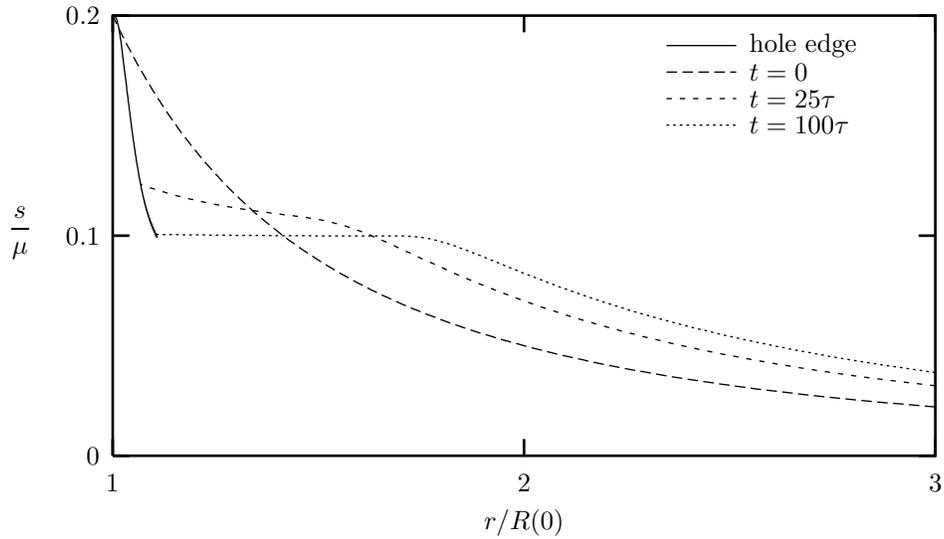
\begin{figure}[htbp]
  \begin{center}
    \input{lowlambda.tex}
    \caption{Deviatoric stress $s/\mu$ as a function of the distance
      from the center of the hole shown at several times after the
      application of the stress $\sigma_\infty = 0.2\mu$.  The yield
      stress $s_y = 0.1\mu$, Poisson's ratio $\nu = 0.3$, $\lambda s_y
      = 0.005$.}
    \label{fig:lowlambda}
  \end{center}
\end{figure}

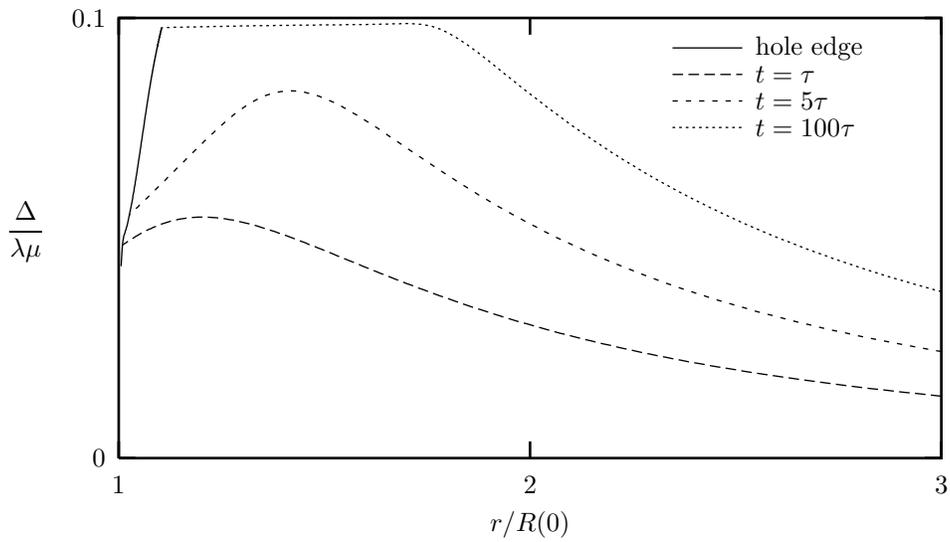
\begin{figure}[htbp]
  \begin{center}
    \input{delta_lowlam.tex}
    \caption{Time evolution of $\Delta$ for the same values of the
      parameters as in Fig.~\ref{fig:lowlambda}.} 
    \label{fig:delta_lowlam}
  \end{center}
\end{figure}

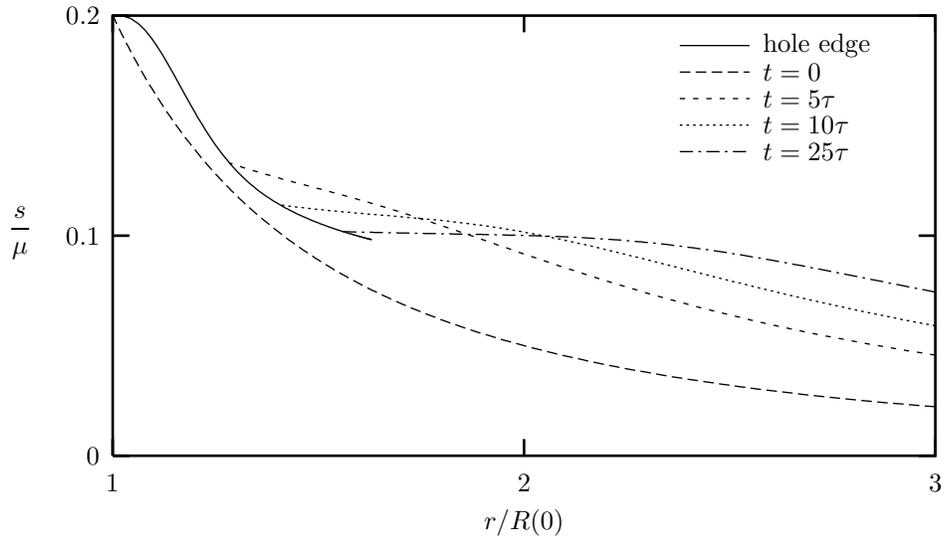
\begin{figure}[htbp]
  \begin{center}
    \input{highlambda.tex}
    \caption{Same as in Fig.~\ref{fig:lowlambda} but wuth $\lambda s_y
      = 0.05$.  Note that the approach to a stationary state is faster
      and that the plastic zone is much less pronounced.}
    \label{fig:highlambda}
  \end{center}
\end{figure}

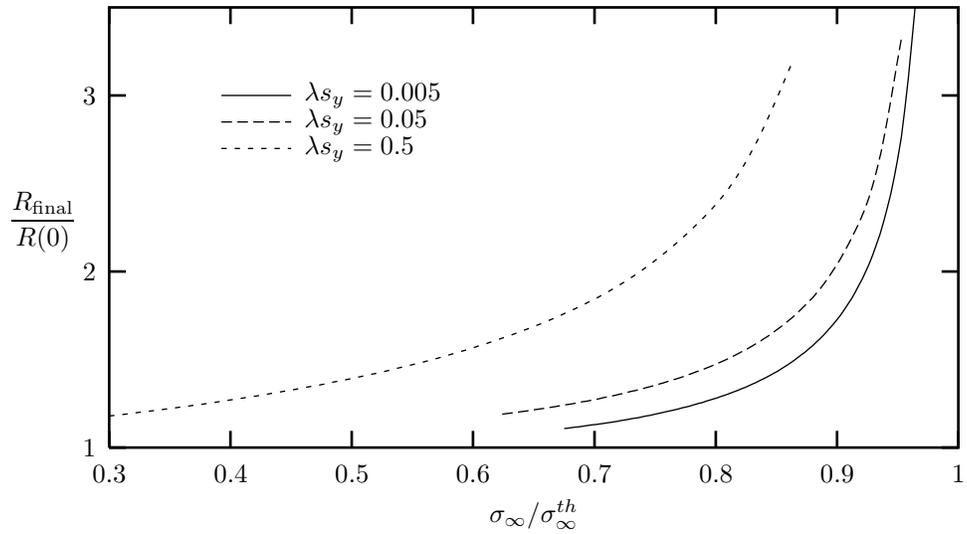
\begin{figure}[htbp]
  \begin{center}
    \input{Rfinal.tex}
    \caption{The final radius of the hole versus the applied stress in
      units of the corresponding threshold stress for three values of
      $\lambda s_y$.  $s_y = 0.1\mu$.  The solid line is the prediction
      for the Tresca plastic.}
    \label{fig:Rfinal}
  \end{center}
\end{figure}

\begin{figure}[htbp]
  \begin{center}
    \input{lam_thresh.tex}
    \caption{Threshold stress in units of the yield stress as a
      function of $\lambda$.  Note that for soft materials, the
      thresold for unbounded failure is smaller than the yield
      stress.}
    \label{fig:lam_thresh}
  \end{center}
\end{figure}
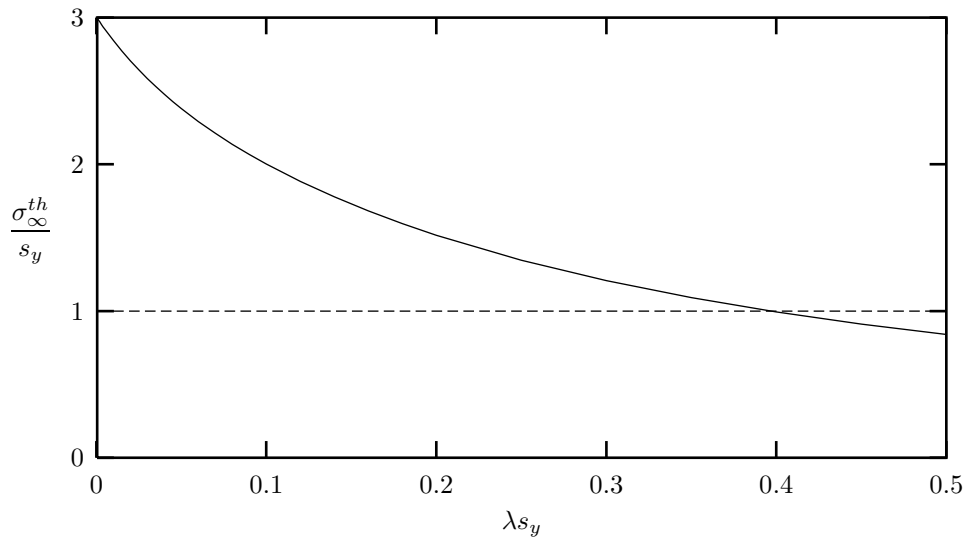

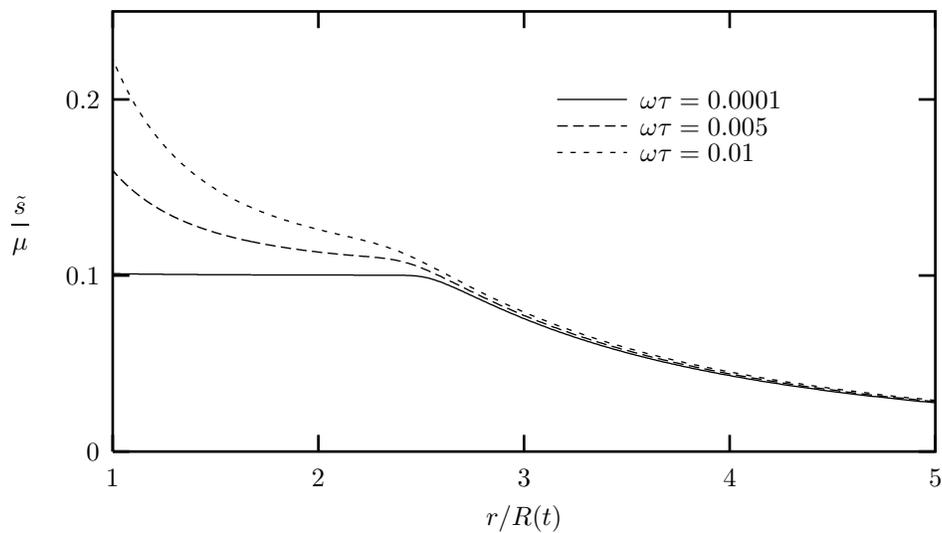
\begin{figure}[htbp]
  \begin{center}
    \input{sim_omega.tex}
    \caption{Similarity solutions $\tilde s(r/R(t))$ for three
      different values of the hole growth rate $\omega$.  $\lambda s_y
      = 0.005,$ $s_y = 0.1\mu$.}
    \label{fig:sim_omega}
  \end{center}
\end{figure}

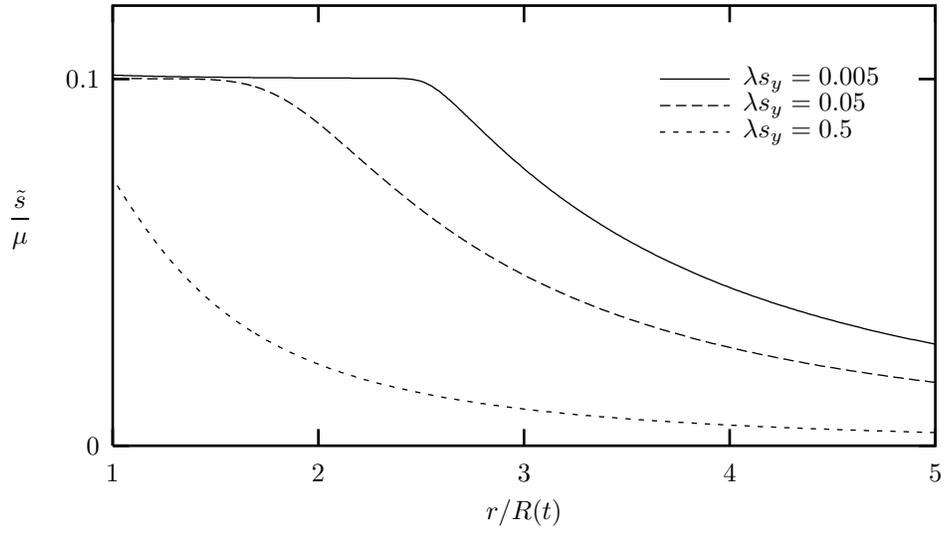
\begin{figure}[htbp]
  \begin{center}
    \input{sim_lambda.tex}
    \caption{Similarity solutions $\tilde s(r/R(t))$ for three
      different values of $\lambda s_y$ in the case of a small growth
      rate $\omega\tau = 0.0001$.} 
    \label{fig:sim_lambda}
  \end{center}
\end{figure}

Without doing any further calculations, we can see how the nonlinear
rate factor produces memory effects.  Suppose that we start in a
stressed configuration such as the one shown in
Figs.~\ref{fig:lowlambda} and \ref{fig:delta_lowlam} in which a
plastic zone has formed around the hole; and suppose that we then
unload the system by quickly reducing the external stress to zero.  At
$s = 0$, the rate factor becomes extremely small; in the absence of a
stress that can induce transitions from one orientation to another,
the zones remain as they were in the previously stressed state.  That
is, $\Delta(r)$ remains unchanged on unloading.  The system relaxes
elastically, but the plastic degrees of freedom do not revert to their
earlier values.  Accordingly, there must be residual stresses in the
region near the hole.  The material will ``remember'' the magnitude
and direction of its previous loading because it ``knows'' the
function $\Delta(r)$, which will determine via the nonlinear
generalizations of the constitutive equations (\ref{STZ1}) and
(\ref{STZ2}) what happens when new stresses are applied.

The other feature that we have emphasized in this analysis is the
dynamic failure that occurs at $\sigma_{\infty}^{th}$, a feature that
also occurs in the conventional time-dependent theories.  We believe
that the dynamic growth modes may provide a clue for solving a
long-standing puzzle in the theory of fracture, specifically, the
question of how breaking stresses can be transmitted through plastic
zones to crack tips.  This puzzle pertains to quasi-static crack
advance, which would seem to be impossible if, as in conventional
plasticity theory, the stresses near a crack tip are constrained to be
less than or equal to the yield stress.  The new modes, especially
those that occur when there is appreciable deviation from perfect
plasticity, raise the possibility that, like the growing hole, crack
growth could occur via plastic flow near the tip.  We hope to explore
this possibility in a subsequent report.

\acknowledgements

We thank Zhigang Suo for suggesting that we study the hole problem as
a way to understand the issues raised in this paper. This research has
been supported primarily by U.S.  DOE Grants DE-FG03-84ER45108 and
DE-FG03-99ER45762, and in part by the MRSEC Program of the NSF under
award number DMR96-32716.
  
\bibliographystyle{prsty}
\bibliography{circle}











\end{document}

%% file: lowlambda.tex
\setlength{\unitlength}{0.1bp}
\special{!
/gnudict 120 dict def
gnudict begin
/Color false def
/Solid false def
/gnulinewidth 5.000 def
/userlinewidth gnulinewidth def
/vshift -33 def
/dl {10 mul} def
/hpt_ 31.5 def
/vpt_ 31.5 def
/hpt hpt_ def
/vpt vpt_ def
/M {moveto} bind def
/L {lineto} bind def
/R {rmoveto} bind def
/V {rlineto} bind def
/vpt2 vpt 2 mul def
/hpt2 hpt 2 mul def
/Lshow { currentpoint stroke M
  0 vshift R show } def
/Rshow { currentpoint stroke M
  dup stringwidth pop neg vshift R show } def
/Cshow { currentpoint stroke M
  dup stringwidth pop -2 div vshift R show } def
/UP { dup vpt_ mul /vpt exch def hpt_ mul /hpt exch def
  /hpt2 hpt 2 mul def /vpt2 vpt 2 mul def } def
/DL { Color {setrgbcolor Solid {pop []} if 0 setdash }
 {pop pop pop Solid {pop []} if 0 setdash} ifelse } def
/BL { stroke gnulinewidth 2 mul setlinewidth } def
/AL { stroke gnulinewidth 2 div setlinewidth } def
/UL { gnulinewidth mul /userlinewidth exch def } def
/PL { stroke userlinewidth setlinewidth } def
/LTb { BL [] 0 0 0 DL } def
/LTa { AL [1 dl 2 dl] 0 setdash 0 0 0 setrgbcolor } def
/LT0 { PL [] 0 1 0 DL } def
/LT1 { PL [4 dl 2 dl] 0 0 1 DL } def
/LT2 { PL [2 dl 3 dl] 1 0 0 DL } def
/LT3 { PL [1 dl 1.5 dl] 1 0 1 DL } def
/LT4 { PL [5 dl 2 dl 1 dl 2 dl] 0 1 1 DL } def
/LT5 { PL [4 dl 3 dl 1 dl 3 dl] 1 1 0 DL } def
/LT6 { PL [2 dl 2 dl 2 dl 4 dl] 0 0 0 DL } def
/LT7 { PL [2 dl 2 dl 2 dl 2 dl 2 dl 4 dl] 1 0.3 0 DL } def
/LT8 { PL [2 dl 2 dl 2 dl 2 dl 2 dl 2 dl 2 dl 4 dl] 0.5 0.5 0.5 DL } def
/Pnt { stroke [] 0 setdash
   gsave 1 setlinecap M 0 0 V stroke grestore } def
/Dia { stroke [] 0 setdash 2 copy vpt add M
  hpt neg vpt neg V hpt vpt neg V
  hpt vpt V hpt neg vpt V closepath stroke
  Pnt } def
/Pls { stroke [] 0 setdash vpt sub M 0 vpt2 V
  currentpoint stroke M
  hpt neg vpt neg R hpt2 0 V stroke
  } def
/Box { stroke [] 0 setdash 2 copy exch hpt sub exch vpt add M
  0 vpt2 neg V hpt2 0 V 0 vpt2 V
  hpt2 neg 0 V closepath stroke
  Pnt } def
/Crs { stroke [] 0 setdash exch hpt sub exch vpt add M
  hpt2 vpt2 neg V currentpoint stroke M
  hpt2 neg 0 R hpt2 vpt2 V stroke } def
/TriU { stroke [] 0 setdash 2 copy vpt 1.12 mul add M
  hpt neg vpt -1.62 mul V
  hpt 2 mul 0 V
  hpt neg vpt 1.62 mul V closepath stroke
  Pnt  } def
/Star { 2 copy Pls Crs } def
/BoxF { stroke [] 0 setdash exch hpt sub exch vpt add M
  0 vpt2 neg V  hpt2 0 V  0 vpt2 V
  hpt2 neg 0 V  closepath fill } def
/TriUF { stroke [] 0 setdash vpt 1.12 mul add M
  hpt neg vpt -1.62 mul V
  hpt 2 mul 0 V
  hpt neg vpt 1.62 mul V closepath fill } def
/TriD { stroke [] 0 setdash 2 copy vpt 1.12 mul sub M
  hpt neg vpt 1.62 mul V
  hpt 2 mul 0 V
  hpt neg vpt -1.62 mul V closepath stroke
  Pnt  } def
/TriDF { stroke [] 0 setdash vpt 1.12 mul sub M
  hpt neg vpt 1.62 mul V
  hpt 2 mul 0 V
  hpt neg vpt -1.62 mul V closepath fill} def
/DiaF { stroke [] 0 setdash vpt add M
  hpt neg vpt neg V hpt vpt neg V
  hpt vpt V hpt neg vpt V closepath fill } def
/Pent { stroke [] 0 setdash 2 copy gsave
  translate 0 hpt M 4 {72 rotate 0 hpt L} repeat
  closepath stroke grestore Pnt } def
/PentF { stroke [] 0 setdash gsave
  translate 0 hpt M 4 {72 rotate 0 hpt L} repeat
  closepath fill grestore } def
/Circle { stroke [] 0 setdash 2 copy
  hpt 0 360 arc stroke Pnt } def
/CircleF { stroke [] 0 setdash hpt 0 360 arc fill } def
/C0 { BL [] 0 setdash 2 copy moveto vpt 90 450  arc } bind def
/C1 { BL [] 0 setdash 2 copy        moveto
       2 copy  vpt 0 90 arc closepath fill
               vpt 0 360 arc closepath } bind def
/C2 { BL [] 0 setdash 2 copy moveto
       2 copy  vpt 90 180 arc closepath fill
               vpt 0 360 arc closepath } bind def
/C3 { BL [] 0 setdash 2 copy moveto
       2 copy  vpt 0 180 arc closepath fill
               vpt 0 360 arc closepath } bind def
/C4 { BL [] 0 setdash 2 copy moveto
       2 copy  vpt 180 270 arc closepath fill
               vpt 0 360 arc closepath } bind def
/C5 { BL [] 0 setdash 2 copy moveto
       2 copy  vpt 0 90 arc
       2 copy moveto
       2 copy  vpt 180 270 arc closepath fill
               vpt 0 360 arc } bind def
/C6 { BL [] 0 setdash 2 copy moveto
      2 copy  vpt 90 270 arc closepath fill
              vpt 0 360 arc closepath } bind def
/C7 { BL [] 0 setdash 2 copy moveto
      2 copy  vpt 0 270 arc closepath fill
              vpt 0 360 arc closepath } bind def
/C8 { BL [] 0 setdash 2 copy moveto
      2 copy vpt 270 360 arc closepath fill
              vpt 0 360 arc closepath } bind def
/C9 { BL [] 0 setdash 2 copy moveto
      2 copy  vpt 270 450 arc closepath fill
              vpt 0 360 arc closepath } bind def
/C10 { BL [] 0 setdash 2 copy 2 copy moveto vpt 270 360 arc closepath fill
       2 copy moveto
       2 copy vpt 90 180 arc closepath fill
               vpt 0 360 arc closepath } bind def
/C11 { BL [] 0 setdash 2 copy moveto
       2 copy  vpt 0 180 arc closepath fill
       2 copy moveto
       2 copy  vpt 270 360 arc closepath fill
               vpt 0 360 arc closepath } bind def
/C12 { BL [] 0 setdash 2 copy moveto
       2 copy  vpt 180 360 arc closepath fill
               vpt 0 360 arc closepath } bind def
/C13 { BL [] 0 setdash  2 copy moveto
       2 copy  vpt 0 90 arc closepath fill
       2 copy moveto
       2 copy  vpt 180 360 arc closepath fill
               vpt 0 360 arc closepath } bind def
/C14 { BL [] 0 setdash 2 copy moveto
       2 copy  vpt 90 360 arc closepath fill
               vpt 0 360 arc } bind def
/C15 { BL [] 0 setdash 2 copy vpt 0 360 arc closepath fill
               vpt 0 360 arc closepath } bind def
/Rec   { newpath 4 2 roll moveto 1 index 0 rlineto 0 exch rlineto
       neg 0 rlineto closepath } bind def
/Square { dup Rec } bind def
/Bsquare { vpt sub exch vpt sub exch vpt2 Square } bind def
/S0 { BL [] 0 setdash 2 copy moveto 0 vpt rlineto BL Bsquare } bind def
/S1 { BL [] 0 setdash 2 copy vpt Square fill Bsquare } bind def
/S2 { BL [] 0 setdash 2 copy exch vpt sub exch vpt Square fill Bsquare } bind def
/S3 { BL [] 0 setdash 2 copy exch vpt sub exch vpt2 vpt Rec fill Bsquare } bind def
/S4 { BL [] 0 setdash 2 copy exch vpt sub exch vpt sub vpt Square fill Bsquare } bind def
/S5 { BL [] 0 setdash 2 copy 2 copy vpt Square fill
       exch vpt sub exch vpt sub vpt Square fill Bsquare } bind def
/S6 { BL [] 0 setdash 2 copy exch vpt sub exch vpt sub vpt vpt2 Rec fill Bsquare } bind def
/S7 { BL [] 0 setdash 2 copy exch vpt sub exch vpt sub vpt vpt2 Rec fill
       2 copy vpt Square fill
       Bsquare } bind def
/S8 { BL [] 0 setdash 2 copy vpt sub vpt Square fill Bsquare } bind def
/S9 { BL [] 0 setdash 2 copy vpt sub vpt vpt2 Rec fill Bsquare } bind def
/S10 { BL [] 0 setdash 2 copy vpt sub vpt Square fill 2 copy exch vpt sub exch vpt Square fill
       Bsquare } bind def
/S11 { BL [] 0 setdash 2 copy vpt sub vpt Square fill 2 copy exch vpt sub exch vpt2 vpt Rec fill
       Bsquare } bind def
/S12 { BL [] 0 setdash 2 copy exch vpt sub exch vpt sub vpt2 vpt Rec fill Bsquare } bind def
/S13 { BL [] 0 setdash 2 copy exch vpt sub exch vpt sub vpt2 vpt Rec fill
       2 copy vpt Square fill Bsquare } bind def
/S14 { BL [] 0 setdash 2 copy exch vpt sub exch vpt sub vpt2 vpt Rec fill
       2 copy exch vpt sub exch vpt Square fill Bsquare } bind def
/S15 { BL [] 0 setdash 2 copy Bsquare fill Bsquare } bind def
/D0 { gsave translate 45 rotate 0 0 S0 stroke grestore } bind def
/D1 { gsave translate 45 rotate 0 0 S1 stroke grestore } bind def
/D2 { gsave translate 45 rotate 0 0 S2 stroke grestore } bind def
/D3 { gsave translate 45 rotate 0 0 S3 stroke grestore } bind def
/D4 { gsave translate 45 rotate 0 0 S4 stroke grestore } bind def
/D5 { gsave translate 45 rotate 0 0 S5 stroke grestore } bind def
/D6 { gsave translate 45 rotate 0 0 S6 stroke grestore } bind def
/D7 { gsave translate 45 rotate 0 0 S7 stroke grestore } bind def
/D8 { gsave translate 45 rotate 0 0 S8 stroke grestore } bind def
/D9 { gsave translate 45 rotate 0 0 S9 stroke grestore } bind def
/D10 { gsave translate 45 rotate 0 0 S10 stroke grestore } bind def
/D11 { gsave translate 45 rotate 0 0 S11 stroke grestore } bind def
/D12 { gsave translate 45 rotate 0 0 S12 stroke grestore } bind def
/D13 { gsave translate 45 rotate 0 0 S13 stroke grestore } bind def
/D14 { gsave translate 45 rotate 0 0 S14 stroke grestore } bind def
/D15 { gsave translate 45 rotate 0 0 S15 stroke grestore } bind def
/DiaE { stroke [] 0 setdash vpt add M
  hpt neg vpt neg V hpt vpt neg V
  hpt vpt V hpt neg vpt V closepath stroke } def
/BoxE { stroke [] 0 setdash exch hpt sub exch vpt add M
  0 vpt2 neg V hpt2 0 V 0 vpt2 V
  hpt2 neg 0 V closepath stroke } def
/TriUE { stroke [] 0 setdash vpt 1.12 mul add M
  hpt neg vpt -1.62 mul V
  hpt 2 mul 0 V
  hpt neg vpt 1.62 mul V closepath stroke } def
/TriDE { stroke [] 0 setdash vpt 1.12 mul sub M
  hpt neg vpt 1.62 mul V
  hpt 2 mul 0 V
  hpt neg vpt -1.62 mul V closepath stroke } def
/PentE { stroke [] 0 setdash gsave
  translate 0 hpt M 4 {72 rotate 0 hpt L} repeat
  closepath stroke grestore } def
/CircE { stroke [] 0 setdash 
  hpt 0 360 arc stroke } def
/BoxFill { gsave Rec 1 setgray fill grestore } def
end
}
\begin{picture}(3600,2160)(0,0)
\special{"
gnudict begin
gsave
0 0 translate
0.100 0.100 scale
0 setgray
newpath
LTb
450 400 M
63 0 V
3037 0 R
-63 0 V
450 1230 M
63 0 V
3037 0 R
-63 0 V
450 2060 M
63 0 V
3037 0 R
-63 0 V
450 400 M
0 63 V
0 1597 R
0 -63 V
2000 400 M
0 63 V
0 1597 R
0 -63 V
3550 400 M
0 63 V
0 1597 R
0 -63 V
LTb
450 400 M
3100 0 V
0 1660 V
-3100 0 V
450 400 L
1.000 UL
LT0
2537 1947 M
263 0 V
450 2060 M
4 0 V
3 -1 V
2 -3 V
3 -4 V
2 -4 V
2 -6 V
2 -6 V
1 -7 V
2 -7 V
2 -8 V
1 -8 V
1 -8 V
2 -8 V
1 -8 V
1 -9 V
2 -8 V
1 -9 V
1 -8 V
1 -9 V
1 -8 V
1 -9 V
1 -8 V
1 -9 V
1 -8 V
1 -8 V
1 -9 V
1 -8 V
1 -8 V
1 -8 V
1 -8 V
1 -7 V
1 -8 V
1 -8 V
1 -7 V
1 -8 V
1 -7 V
1 -7 V
1 -8 V
0 -7 V
1 -7 V
1 -7 V
1 -7 V
1 -6 V
1 -7 V
0 -7 V
1 -6 V
1 -6 V
1 -7 V
1 -6 V
0 -6 V
1 -6 V
1 -6 V
1 -6 V
0 -6 V
1 -6 V
1 -6 V
1 -5 V
0 -6 V
1 -5 V
1 -6 V
0 -5 V
1 -5 V
1 -5 V
0 -5 V
1 -5 V
1 -5 V
0 -5 V
1 -5 V
1 -5 V
0 -4 V
1 -5 V
1 -5 V
0 -4 V
1 -5 V
1 -4 V
0 -4 V
1 -5 V
0 -4 V
1 -4 V
1 -4 V
0 -4 V
1 -4 V
0 -4 V
1 -4 V
0 -3 V
1 -4 V
1 -4 V
0 -4 V
1 -3 V
0 -4 V
1 -3 V
0 -4 V
1 -3 V
0 -3 V
1 -4 V
0 -3 V
1 -3 V
1 -3 V
0 -3 V
1 -3 V
0 -3 V
1 -3 V
0 -3 V
1 -3 V
0 -3 V
1 -3 V
0 -3 V
0 -2 V
1 -3 V
0 -3 V
1 -2 V
0 -3 V
1 -3 V
0 -2 V
1 -3 V
0 -2 V
1 -3 V
0 -2 V
1 -2 V
0 -3 V
0 -2 V
1 -2 V
0 -2 V
1 -3 V
0 -2 V
1 -2 V
0 -2 V
0 -2 V
1 -2 V
0 -2 V
1 -2 V
0 -2 V
0 -2 V
1 -2 V
0 -2 V
1 -2 V
0 -2 V
0 -1 V
1 -2 V
0 -2 V
1 -2 V
0 -1 V
0 -2 V
1 -2 V
0 -1 V
0 -2 V
1 -2 V
0 -1 V
0 -2 V
1 -1 V
0 -2 V
0 -1 V
1 -2 V
0 -1 V
1 -2 V
0 -1 V
0 -2 V
1 -1 V
0 -1 V
0 -2 V
1 -1 V
0 -1 V
0 -2 V
0 -1 V
1 -1 V
0 -2 V
0 -1 V
1 -1 V
0 -1 V
0 -2 V
1 -1 V
0 -1 V
0 -1 V
1 -1 V
0 -1 V
0 -1 V
0 -2 V
1 -1 V
0 -1 V
0 -1 V
1 -1 V
0 -1 V
0 -1 V
0 -1 V
1 -1 V
0 -1 V
0 -1 V
1 -1 V
0 -1 V
0 -1 V
0 -1 V
1 -1 V
0 -1 V
0 -1 V
1 -1 V
0 -1 V
0 -1 V
0 -1 V
1 -1 V
0 -1 V
0 -1 V
1 -1 V
0 -1 V
0 -1 V
1 -1 V
0 -1 V
0 -1 V
1 -1 V
0 -1 V
0 -1 V
0 -1 V
1 -1 V
0 -1 V
0 -1 V
1 0 V
0 -1 V
0 -1 V
0 -1 V
1 -1 V
0 -1 V
0 -1 V
1 -1 V
0 -1 V
0 -1 V
1 -1 V
0 -1 V
0 -1 V
1 0 V
0 -1 V
0 -1 V
0 -1 V
1 0 V
0 -1 V
0 -1 V
0 -1 V
1 0 V
0 -1 V
0 -1 V
0 -1 V
1 0 V
0 -1 V
0 -1 V
1 -1 V
0 -1 V
0 -1 V
1 0 V
0 -1 V
0 -1 V
1 -1 V
0 -1 V
0 -1 V
1 0 V
0 -1 V
0 -1 V
0 -1 V
1 0 V
0 -1 V
0 -1 V
1 -1 V
0 -1 V
0 -1 V
1 0 V
0 -1 V
0 -1 V
1 0 V
0 -1 V
0 -1 V
0 -1 V
1 0 V
0 -1 V
0 -1 V
1 0 V
0 -1 V
0 -1 V
1 -1 V
0 -1 V
0 -1 V
1 0 V
0 -1 V
0 -1 V
1 0 V
0 -1 V
0 -1 V
0 -1 V
1 0 V
0 -1 V
0 -1 V
1 0 V
0 -1 V
0 -1 V
1 0 V
0 -1 V
0 -1 V
1 0 V
0 -1 V
0 -1 V
0 -1 V
1 0 V
0 -1 V
0 -1 V
1 0 V
0 -1 V
0 -1 V
1 0 V
0 -1 V
0 -1 V
1 0 V
0 -1 V
0 -1 V
1 0 V
0 -1 V
0 -1 V
0 -1 V
1 0 V
0 -1 V
0 -1 V
1 0 V
0 -1 V
0 -1 V
1 0 V
0 -1 V
0 -1 V
1.000 UL
LT1
2537 1847 M
263 0 V
3550 584 M
-46 4 V
-45 4 V
-44 4 V
-44 3 V
-42 4 V
-42 4 V
-41 4 V
-40 4 V
-40 4 V
-39 4 V
-38 4 V
-37 4 V
-37 5 V
-36 4 V
-35 4 V
-35 4 V
-35 4 V
-33 5 V
-33 4 V
-33 5 V
-32 4 V
-31 5 V
-31 4 V
-31 5 V
-30 4 V
-29 5 V
-29 4 V
-29 5 V
-28 5 V
-28 5 V
-27 5 V
-27 4 V
-27 5 V
-26 5 V
-25 5 V
-26 5 V
-25 5 V
-24 5 V
-25 5 V
-23 6 V
-24 5 V
-23 5 V
-23 5 V
-23 5 V
-22 6 V
-22 5 V
-22 6 V
-21 5 V
-21 5 V
-21 6 V
-20 6 V
-21 5 V
-20 6 V
-19 5 V
-20 6 V
-19 6 V
-19 6 V
-19 5 V
-18 6 V
-19 6 V
-18 6 V
-18 6 V
-17 6 V
-18 6 V
-17 6 V
-17 6 V
-16 6 V
-17 7 V
-16 6 V
-17 6 V
-16 6 V
-15 7 V
-16 6 V
-16 6 V
-15 7 V
-15 6 V
-15 7 V
-15 6 V
-14 7 V
-15 7 V
-14 6 V
-14 7 V
-14 7 V
-14 6 V
-13 7 V
-14 7 V
-13 7 V
-14 7 V
-13 7 V
-13 7 V
-12 7 V
-13 7 V
-13 7 V
-12 7 V
-12 7 V
-12 8 V
-13 7 V
-11 7 V
-12 7 V
-12 8 V
-11 7 V
-12 8 V
-11 7 V
-12 8 V
-11 7 V
-11 8 V
-11 7 V
-10 8 V
-11 8 V
-11 7 V
-10 8 V
-11 8 V
-10 8 V
-10 8 V
-10 8 V
-10 8 V
-10 8 V
-10 8 V
-10 8 V
-9 8 V
-10 8 V
-9 8 V
-10 8 V
-9 8 V
-9 9 V
-9 8 V
-9 8 V
-9 9 V
-9 8 V
-9 9 V
-9 8 V
-9 9 V
-8 8 V
-9 9 V
-8 9 V
-9 8 V
-8 9 V
-8 9 V
-8 9 V
-8 8 V
-8 9 V
-8 9 V
-8 9 V
-8 9 V
-8 9 V
-8 9 V
-7 9 V
-8 9 V
-7 10 V
-8 9 V
-7 9 V
-8 9 V
-7 10 V
-7 9 V
-7 9 V
-8 10 V
-7 9 V
-7 10 V
-7 9 V
-6 10 V
-7 9 V
-7 10 V
-7 10 V
-7 10 V
-6 9 V
-7 10 V
-6 10 V
-7 10 V
-6 10 V
-7 10 V
-6 10 V
-6 10 V
-7 10 V
-6 10 V
-6 10 V
-6 10 V
-6 10 V
-6 10 V
-6 11 V
-6 10 V
-6 10 V
-6 11 V
-6 10 V
-6 11 V
-5 10 V
-6 11 V
-6 10 V
-5 11 V
-6 11 V
-6 10 V
-5 11 V
-5 11 V
-6 10 V
-5 11 V
-6 11 V
-5 11 V
-5 11 V
-6 11 V
-5 11 V
-5 11 V
1.000 UL
LT2
2537 1747 M
263 0 V
3550 664 M
-9 1 V
-43 5 V
-43 5 V
-41 5 V
-41 5 V
-40 5 V
-39 5 V
-38 5 V
-38 5 V
-37 6 V
-37 5 V
-36 5 V
-35 6 V
-35 5 V
-34 6 V
-34 5 V
-33 6 V
-33 6 V
-32 5 V
-31 6 V
-31 6 V
-31 6 V
-30 5 V
-29 6 V
-30 6 V
-28 6 V
-29 6 V
-27 6 V
-28 6 V
-27 6 V
-27 7 V
-26 6 V
-26 6 V
-25 6 V
-25 7 V
-25 6 V
-25 6 V
-24 7 V
-24 6 V
-23 7 V
-23 6 V
-23 7 V
-23 6 V
-22 7 V
-22 7 V
-21 6 V
-22 7 V
-21 7 V
-21 7 V
-20 6 V
-20 7 V
-20 7 V
-20 7 V
-20 7 V
-19 7 V
-19 7 V
-19 7 V
-18 7 V
-19 7 V
-18 7 V
-18 7 V
-17 7 V
-18 7 V
-17 7 V
-17 7 V
-17 7 V
-17 7 V
-16 7 V
-17 7 V
-16 8 V
-16 7 V
-15 7 V
-16 7 V
-15 7 V
-16 7 V
-15 7 V
-15 6 V
-14 7 V
-15 7 V
-14 7 V
-15 6 V
-14 7 V
-14 6 V
-14 7 V
-13 6 V
-14 6 V
-13 6 V
-14 6 V
-13 6 V
-13 5 V
-12 6 V
-13 5 V
-13 5 V
-12 5 V
-13 4 V
-12 5 V
-12 4 V
-12 4 V
-12 4 V
-12 3 V
-11 3 V
-12 3 V
-11 3 V
-12 3 V
-11 2 V
-11 3 V
-11 2 V
-11 2 V
-11 2 V
-10 1 V
-11 2 V
-10 2 V
-11 1 V
-10 2 V
-10 1 V
-10 1 V
-10 2 V
-10 1 V
-10 1 V
-10 1 V
-10 1 V
-9 2 V
-10 1 V
-9 1 V
-9 1 V
-10 1 V
-9 2 V
-9 1 V
-9 1 V
-9 1 V
-9 1 V
-9 2 V
-8 1 V
-9 1 V
-9 1 V
-8 2 V
-9 1 V
-8 1 V
-8 1 V
-9 2 V
-8 1 V
-8 1 V
-8 2 V
-8 1 V
-8 1 V
-8 2 V
-7 1 V
-8 2 V
-8 1 V
-7 1 V
-8 2 V
-7 1 V
-8 2 V
-7 1 V
-8 2 V
-7 1 V
-7 2 V
-7 1 V
-7 2 V
-7 1 V
-7 2 V
-7 2 V
-7 1 V
-7 2 V
-7 1 V
-6 2 V
-7 2 V
-7 1 V
-6 2 V
-7 2 V
-6 1 V
-7 2 V
-6 2 V
-7 2 V
-6 1 V
-6 2 V
-6 2 V
-7 2 V
-6 2 V
-6 1 V
-6 2 V
-6 2 V
-6 2 V
-6 2 V
-6 2 V
-6 2 V
-5 1 V
-6 2 V
-6 2 V
-5 2 V
-6 2 V
-6 2 V
-5 2 V
-6 2 V
1.000 UL
LT3
2537 1647 M
263 0 V
3550 714 M
-23 3 V
-41 6 V
-41 6 V
-40 6 V
-39 5 V
-38 6 V
-38 6 V
-37 6 V
-37 6 V
-36 6 V
-35 6 V
-35 7 V
-34 6 V
-34 6 V
-33 6 V
-33 7 V
-32 6 V
-31 7 V
-31 6 V
-31 7 V
-30 6 V
-30 7 V
-29 6 V
-29 7 V
-28 7 V
-28 7 V
-28 6 V
-27 7 V
-27 7 V
-26 7 V
-26 7 V
-26 7 V
-25 7 V
-25 7 V
-25 7 V
-24 7 V
-24 7 V
-24 8 V
-23 7 V
-23 7 V
-23 7 V
-22 8 V
-22 7 V
-22 7 V
-21 8 V
-22 7 V
-21 8 V
-20 7 V
-21 7 V
-20 8 V
-20 7 V
-19 8 V
-20 7 V
-19 8 V
-19 7 V
-19 7 V
-18 8 V
-19 7 V
-18 7 V
-18 8 V
-17 7 V
-18 7 V
-17 7 V
-17 7 V
-17 7 V
-17 6 V
-16 7 V
-16 6 V
-17 6 V
-16 6 V
-15 6 V
-16 5 V
-15 5 V
-16 4 V
-15 5 V
-15 3 V
-14 4 V
-15 2 V
-14 3 V
-15 2 V
-14 1 V
-14 1 V
-14 1 V
-13 1 V
-14 0 V
-13 0 V
-14 0 V
-13 0 V
-13 0 V
-13 0 V
-13 0 V
-12 0 V
-13 0 V
-12 0 V
-13 0 V
-12 0 V
-12 0 V
-12 0 V
-11 0 V
-12 0 V
-12 0 V
-11 0 V
-12 0 V
-11 0 V
-11 0 V
-11 0 V
-11 0 V
-11 0 V
-11 0 V
-10 0 V
-11 0 V
-10 0 V
-11 0 V
-10 1 V
-10 0 V
-10 0 V
-10 0 V
-10 0 V
-10 0 V
-10 0 V
-10 0 V
-9 0 V
-10 0 V
-9 0 V
-9 0 V
-10 0 V
-9 0 V
-9 0 V
-9 0 V
-9 1 V
-9 0 V
-9 0 V
-8 0 V
-9 0 V
-9 0 V
-8 0 V
-9 0 V
-8 0 V
-8 0 V
-9 0 V
-8 0 V
-8 0 V
-8 0 V
-8 0 V
-8 0 V
-8 1 V
-7 0 V
-8 0 V
-8 0 V
-8 0 V
-7 0 V
-8 0 V
-7 0 V
-7 0 V
-8 0 V
-7 0 V
-7 0 V
-7 0 V
-8 0 V
-7 0 V
-7 1 V
-7 0 V
-7 0 V
-6 0 V
-7 0 V
-7 0 V
-7 0 V
-6 0 V
-7 0 V
-7 0 V
-6 0 V
-7 0 V
-6 0 V
-6 0 V
-7 1 V
-6 0 V
-6 0 V
-7 0 V
-6 0 V
-6 0 V
-6 0 V
-6 0 V
-6 0 V
-6 0 V
-6 0 V
-6 0 V
-6 0 V
-5 1 V
-6 0 V
-6 0 V
stroke
grestore
end
showpage
}
\put(2850,1647){\makebox(0,0)[l]{$t = 100\tau$}}
\put(2850,1747){\makebox(0,0)[l]{$t =25\tau$}}
\put(2850,1847){\makebox(0,0)[l]{$t = 0$}}
\put(2850,1947){\makebox(0,0)[l]{hole edge}}
\put(2000,150){\makebox(0,0){$r/R(0)$}}
\put(100,1230){\makebox(0,0)[b]{\shortstack{$\displaystyle{s \over \mu}$}}}
\put(3550,300){\makebox(0,0){3}}
\put(2000,300){\makebox(0,0){2}}
\put(450,300){\makebox(0,0){1}}
\put(400,2060){\makebox(0,0)[r]{0.2}}
\put(400,1230){\makebox(0,0)[r]{0.1}}
\put(400,400){\makebox(0,0)[r]{0}}
\end{picture}

%% file: delta_lowlam.tex
\setlength{\unitlength}{0.1bp}
\special{!
/gnudict 120 dict def
gnudict begin
/Color false def
/Solid false def
/gnulinewidth 5.000 def
/userlinewidth gnulinewidth def
/vshift -33 def
/dl {10 mul} def
/hpt_ 31.5 def
/vpt_ 31.5 def
/hpt hpt_ def
/vpt vpt_ def
/M {moveto} bind def
/L {lineto} bind def
/R {rmoveto} bind def
/V {rlineto} bind def
/vpt2 vpt 2 mul def
/hpt2 hpt 2 mul def
/Lshow { currentpoint stroke M
  0 vshift R show } def
/Rshow { currentpoint stroke M
  dup stringwidth pop neg vshift R show } def
/Cshow { currentpoint stroke M
  dup stringwidth pop -2 div vshift R show } def
/UP { dup vpt_ mul /vpt exch def hpt_ mul /hpt exch def
  /hpt2 hpt 2 mul def /vpt2 vpt 2 mul def } def
/DL { Color {setrgbcolor Solid {pop []} if 0 setdash }
 {pop pop pop Solid {pop []} if 0 setdash} ifelse } def
/BL { stroke gnulinewidth 2 mul setlinewidth } def
/AL { stroke gnulinewidth 2 div setlinewidth } def
/UL { gnulinewidth mul /userlinewidth exch def } def
/PL { stroke userlinewidth setlinewidth } def
/LTb { BL [] 0 0 0 DL } def
/LTa { AL [1 dl 2 dl] 0 setdash 0 0 0 setrgbcolor } def
/LT0 { PL [] 0 1 0 DL } def
/LT1 { PL [4 dl 2 dl] 0 0 1 DL } def
/LT2 { PL [2 dl 3 dl] 1 0 0 DL } def
/LT3 { PL [1 dl 1.5 dl] 1 0 1 DL } def
/LT4 { PL [5 dl 2 dl 1 dl 2 dl] 0 1 1 DL } def
/LT5 { PL [4 dl 3 dl 1 dl 3 dl] 1 1 0 DL } def
/LT6 { PL [2 dl 2 dl 2 dl 4 dl] 0 0 0 DL } def
/LT7 { PL [2 dl 2 dl 2 dl 2 dl 2 dl 4 dl] 1 0.3 0 DL } def
/LT8 { PL [2 dl 2 dl 2 dl 2 dl 2 dl 2 dl 2 dl 4 dl] 0.5 0.5 0.5 DL } def
/Pnt { stroke [] 0 setdash
   gsave 1 setlinecap M 0 0 V stroke grestore } def
/Dia { stroke [] 0 setdash 2 copy vpt add M
  hpt neg vpt neg V hpt vpt neg V
  hpt vpt V hpt neg vpt V closepath stroke
  Pnt } def
/Pls { stroke [] 0 setdash vpt sub M 0 vpt2 V
  currentpoint stroke M
  hpt neg vpt neg R hpt2 0 V stroke
  } def
/Box { stroke [] 0 setdash 2 copy exch hpt sub exch vpt add M
  0 vpt2 neg V hpt2 0 V 0 vpt2 V
  hpt2 neg 0 V closepath stroke
  Pnt } def
/Crs { stroke [] 0 setdash exch hpt sub exch vpt add M
  hpt2 vpt2 neg V currentpoint stroke M
  hpt2 neg 0 R hpt2 vpt2 V stroke } def
/TriU { stroke [] 0 setdash 2 copy vpt 1.12 mul add M
  hpt neg vpt -1.62 mul V
  hpt 2 mul 0 V
  hpt neg vpt 1.62 mul V closepath stroke
  Pnt  } def
/Star { 2 copy Pls Crs } def
/BoxF { stroke [] 0 setdash exch hpt sub exch vpt add M
  0 vpt2 neg V  hpt2 0 V  0 vpt2 V
  hpt2 neg 0 V  closepath fill } def
/TriUF { stroke [] 0 setdash vpt 1.12 mul add M
  hpt neg vpt -1.62 mul V
  hpt 2 mul 0 V
  hpt neg vpt 1.62 mul V closepath fill } def
/TriD { stroke [] 0 setdash 2 copy vpt 1.12 mul sub M
  hpt neg vpt 1.62 mul V
  hpt 2 mul 0 V
  hpt neg vpt -1.62 mul V closepath stroke
  Pnt  } def
/TriDF { stroke [] 0 setdash vpt 1.12 mul sub M
  hpt neg vpt 1.62 mul V
  hpt 2 mul 0 V
  hpt neg vpt -1.62 mul V closepath fill} def
/DiaF { stroke [] 0 setdash vpt add M
  hpt neg vpt neg V hpt vpt neg V
  hpt vpt V hpt neg vpt V closepath fill } def
/Pent { stroke [] 0 setdash 2 copy gsave
  translate 0 hpt M 4 {72 rotate 0 hpt L} repeat
  closepath stroke grestore Pnt } def
/PentF { stroke [] 0 setdash gsave
  translate 0 hpt M 4 {72 rotate 0 hpt L} repeat
  closepath fill grestore } def
/Circle { stroke [] 0 setdash 2 copy
  hpt 0 360 arc stroke Pnt } def
/CircleF { stroke [] 0 setdash hpt 0 360 arc fill } def
/C0 { BL [] 0 setdash 2 copy moveto vpt 90 450  arc } bind def
/C1 { BL [] 0 setdash 2 copy        moveto
       2 copy  vpt 0 90 arc closepath fill
               vpt 0 360 arc closepath } bind def
/C2 { BL [] 0 setdash 2 copy moveto
       2 copy  vpt 90 180 arc closepath fill
               vpt 0 360 arc closepath } bind def
/C3 { BL [] 0 setdash 2 copy moveto
       2 copy  vpt 0 180 arc closepath fill
               vpt 0 360 arc closepath } bind def
/C4 { BL [] 0 setdash 2 copy moveto
       2 copy  vpt 180 270 arc closepath fill
               vpt 0 360 arc closepath } bind def
/C5 { BL [] 0 setdash 2 copy moveto
       2 copy  vpt 0 90 arc
       2 copy moveto
       2 copy  vpt 180 270 arc closepath fill
               vpt 0 360 arc } bind def
/C6 { BL [] 0 setdash 2 copy moveto
      2 copy  vpt 90 270 arc closepath fill
              vpt 0 360 arc closepath } bind def
/C7 { BL [] 0 setdash 2 copy moveto
      2 copy  vpt 0 270 arc closepath fill
              vpt 0 360 arc closepath } bind def
/C8 { BL [] 0 setdash 2 copy moveto
      2 copy vpt 270 360 arc closepath fill
              vpt 0 360 arc closepath } bind def
/C9 { BL [] 0 setdash 2 copy moveto
      2 copy  vpt 270 450 arc closepath fill
              vpt 0 360 arc closepath } bind def
/C10 { BL [] 0 setdash 2 copy 2 copy moveto vpt 270 360 arc closepath fill
       2 copy moveto
       2 copy vpt 90 180 arc closepath fill
               vpt 0 360 arc closepath } bind def
/C11 { BL [] 0 setdash 2 copy moveto
       2 copy  vpt 0 180 arc closepath fill
       2 copy moveto
       2 copy  vpt 270 360 arc closepath fill
               vpt 0 360 arc closepath } bind def
/C12 { BL [] 0 setdash 2 copy moveto
       2 copy  vpt 180 360 arc closepath fill
               vpt 0 360 arc closepath } bind def
/C13 { BL [] 0 setdash  2 copy moveto
       2 copy  vpt 0 90 arc closepath fill
       2 copy moveto
       2 copy  vpt 180 360 arc closepath fill
               vpt 0 360 arc closepath } bind def
/C14 { BL [] 0 setdash 2 copy moveto
       2 copy  vpt 90 360 arc closepath fill
               vpt 0 360 arc } bind def
/C15 { BL [] 0 setdash 2 copy vpt 0 360 arc closepath fill
               vpt 0 360 arc closepath } bind def
/Rec   { newpath 4 2 roll moveto 1 index 0 rlineto 0 exch rlineto
       neg 0 rlineto closepath } bind def
/Square { dup Rec } bind def
/Bsquare { vpt sub exch vpt sub exch vpt2 Square } bind def
/S0 { BL [] 0 setdash 2 copy moveto 0 vpt rlineto BL Bsquare } bind def
/S1 { BL [] 0 setdash 2 copy vpt Square fill Bsquare } bind def
/S2 { BL [] 0 setdash 2 copy exch vpt sub exch vpt Square fill Bsquare } bind def
/S3 { BL [] 0 setdash 2 copy exch vpt sub exch vpt2 vpt Rec fill Bsquare } bind def
/S4 { BL [] 0 setdash 2 copy exch vpt sub exch vpt sub vpt Square fill Bsquare } bind def
/S5 { BL [] 0 setdash 2 copy 2 copy vpt Square fill
       exch vpt sub exch vpt sub vpt Square fill Bsquare } bind def
/S6 { BL [] 0 setdash 2 copy exch vpt sub exch vpt sub vpt vpt2 Rec fill Bsquare } bind def
/S7 { BL [] 0 setdash 2 copy exch vpt sub exch vpt sub vpt vpt2 Rec fill
       2 copy vpt Square fill
       Bsquare } bind def
/S8 { BL [] 0 setdash 2 copy vpt sub vpt Square fill Bsquare } bind def
/S9 { BL [] 0 setdash 2 copy vpt sub vpt vpt2 Rec fill Bsquare } bind def
/S10 { BL [] 0 setdash 2 copy vpt sub vpt Square fill 2 copy exch vpt sub exch vpt Square fill
       Bsquare } bind def
/S11 { BL [] 0 setdash 2 copy vpt sub vpt Square fill 2 copy exch vpt sub exch vpt2 vpt Rec fill
       Bsquare } bind def
/S12 { BL [] 0 setdash 2 copy exch vpt sub exch vpt sub vpt2 vpt Rec fill Bsquare } bind def
/S13 { BL [] 0 setdash 2 copy exch vpt sub exch vpt sub vpt2 vpt Rec fill
       2 copy vpt Square fill Bsquare } bind def
/S14 { BL [] 0 setdash 2 copy exch vpt sub exch vpt sub vpt2 vpt Rec fill
       2 copy exch vpt sub exch vpt Square fill Bsquare } bind def
/S15 { BL [] 0 setdash 2 copy Bsquare fill Bsquare } bind def
/D0 { gsave translate 45 rotate 0 0 S0 stroke grestore } bind def
/D1 { gsave translate 45 rotate 0 0 S1 stroke grestore } bind def
/D2 { gsave translate 45 rotate 0 0 S2 stroke grestore } bind def
/D3 { gsave translate 45 rotate 0 0 S3 stroke grestore } bind def
/D4 { gsave translate 45 rotate 0 0 S4 stroke grestore } bind def
/D5 { gsave translate 45 rotate 0 0 S5 stroke grestore } bind def
/D6 { gsave translate 45 rotate 0 0 S6 stroke grestore } bind def
/D7 { gsave translate 45 rotate 0 0 S7 stroke grestore } bind def
/D8 { gsave translate 45 rotate 0 0 S8 stroke grestore } bind def
/D9 { gsave translate 45 rotate 0 0 S9 stroke grestore } bind def
/D10 { gsave translate 45 rotate 0 0 S10 stroke grestore } bind def
/D11 { gsave translate 45 rotate 0 0 S11 stroke grestore } bind def
/D12 { gsave translate 45 rotate 0 0 S12 stroke grestore } bind def
/D13 { gsave translate 45 rotate 0 0 S13 stroke grestore } bind def
/D14 { gsave translate 45 rotate 0 0 S14 stroke grestore } bind def
/D15 { gsave translate 45 rotate 0 0 S15 stroke grestore } bind def
/DiaE { stroke [] 0 setdash vpt add M
  hpt neg vpt neg V hpt vpt neg V
  hpt vpt V hpt neg vpt V closepath stroke } def
/BoxE { stroke [] 0 setdash exch hpt sub exch vpt add M
  0 vpt2 neg V hpt2 0 V 0 vpt2 V
  hpt2 neg 0 V closepath stroke } def
/TriUE { stroke [] 0 setdash vpt 1.12 mul add M
  hpt neg vpt -1.62 mul V
  hpt 2 mul 0 V
  hpt neg vpt 1.62 mul V closepath stroke } def
/TriDE { stroke [] 0 setdash vpt 1.12 mul sub M
  hpt neg vpt 1.62 mul V
  hpt 2 mul 0 V
  hpt neg vpt -1.62 mul V closepath stroke } def
/PentE { stroke [] 0 setdash gsave
  translate 0 hpt M 4 {72 rotate 0 hpt L} repeat
  closepath stroke grestore } def
/CircE { stroke [] 0 setdash 
  hpt 0 360 arc stroke } def
/BoxFill { gsave Rec 1 setgray fill grestore } def
end
}
\begin{picture}(3600,2160)(0,0)
\special{"
gnudict begin
gsave
0 0 translate
0.100 0.100 scale
0 setgray
newpath
LTb
450 400 M
63 0 V
3037 0 R
-63 0 V
450 2060 M
63 0 V
3037 0 R
-63 0 V
450 400 M
0 63 V
0 1597 R
0 -63 V
2000 400 M
0 63 V
0 1597 R
0 -63 V
3550 400 M
0 63 V
0 1597 R
0 -63 V
LTb
450 400 M
3100 0 V
0 1660 V
-3100 0 V
450 400 L
1.000 UL
LT0
2537 1947 M
263 0 V
459 1124 M
3 51 V
2 28 V
2 17 V
2 10 V
1 8 V
2 5 V
2 5 V
1 4 V
1 5 V
2 4 V
1 4 V
1 5 V
2 4 V
1 5 V
1 4 V
1 5 V
1 5 V
1 4 V
1 5 V
1 5 V
1 5 V
1 5 V
1 5 V
1 5 V
1 4 V
1 5 V
1 5 V
1 5 V
1 5 V
1 5 V
1 5 V
1 5 V
1 5 V
1 5 V
1 5 V
0 5 V
1 5 V
1 5 V
1 5 V
1 4 V
1 5 V
0 5 V
1 5 V
1 5 V
1 5 V
1 4 V
0 5 V
1 5 V
1 5 V
1 4 V
0 5 V
1 5 V
1 4 V
1 5 V
0 4 V
1 5 V
1 4 V
0 5 V
1 5 V
1 4 V
0 4 V
1 5 V
1 4 V
0 5 V
1 4 V
1 4 V
0 5 V
1 4 V
1 4 V
0 4 V
1 5 V
1 4 V
0 4 V
1 4 V
0 4 V
1 4 V
1 4 V
0 4 V
1 4 V
0 4 V
1 4 V
0 4 V
1 4 V
1 3 V
0 4 V
1 4 V
0 4 V
1 3 V
0 4 V
1 4 V
0 3 V
1 4 V
0 3 V
1 4 V
1 4 V
0 3 V
1 3 V
0 4 V
1 3 V
0 4 V
1 3 V
0 3 V
1 4 V
0 3 V
0 3 V
1 3 V
0 3 V
1 4 V
0 3 V
1 3 V
0 3 V
1 3 V
0 3 V
1 3 V
0 3 V
1 3 V
0 3 V
0 2 V
1 3 V
0 3 V
1 3 V
0 3 V
1 2 V
0 3 V
0 3 V
1 2 V
0 3 V
1 3 V
0 2 V
0 3 V
1 2 V
0 3 V
1 2 V
0 3 V
0 2 V
1 3 V
0 2 V
1 3 V
0 2 V
0 2 V
1 3 V
0 2 V
0 2 V
1 2 V
0 3 V
0 2 V
1 2 V
0 2 V
0 2 V
1 3 V
0 2 V
1 2 V
0 2 V
0 2 V
1 2 V
0 2 V
0 2 V
1 2 V
0 2 V
0 2 V
0 2 V
1 1 V
0 2 V
0 2 V
1 2 V
0 2 V
0 2 V
1 1 V
0 2 V
0 2 V
1 2 V
0 1 V
0 2 V
0 2 V
1 2 V
0 1 V
0 2 V
1 1 V
0 2 V
0 2 V
0 1 V
1 2 V
0 1 V
0 2 V
1 1 V
0 2 V
0 1 V
0 2 V
1 1 V
0 2 V
0 1 V
0 2 V
1 1 V
0 2 V
0 1 V
0 1 V
1 2 V
0 1 V
0 1 V
0 2 V
1 1 V
0 1 V
0 2 V
0 1 V
1 1 V
0 1 V
0 2 V
0 1 V
1 1 V
0 1 V
0 2 V
0 1 V
0 1 V
1 1 V
0 1 V
0 1 V
0 2 V
1 1 V
0 1 V
0 1 V
0 1 V
0 1 V
1 1 V
0 1 V
0 1 V
0 1 V
0 1 V
1 1 V
0 1 V
0 1 V
0 1 V
0 1 V
1 1 V
0 1 V
0 1 V
0 1 V
0 1 V
1 1 V
0 1 V
0 1 V
0 1 V
0 1 V
1 1 V
0 1 V
0 1 V
0 1 V
0 1 V
1 1 V
0 1 V
0 1 V
0 1 V
0 1 V
1 1 V
0 1 V
0 1 V
0 1 V
0 1 V
1 1 V
0 1 V
0 1 V
0 1 V
0 1 V
1 1 V
0 1 V
0 1 V
0 1 V
1 1 V
0 1 V
0 1 V
0 1 V
0 1 V
1 1 V
0 1 V
0 1 V
0 1 V
0 1 V
1 0 V
0 1 V
0 1 V
0 1 V
0 1 V
0 1 V
1 0 V
0 1 V
0 1 V
0 1 V
0 1 V
1 1 V
0 1 V
0 1 V
0 1 V
0 1 V
1 0 V
0 1 V
0 1 V
0 1 V
0 1 V
1 1 V
0 1 V
0 1 V
0 1 V
0 1 V
1 0 V
0 1 V
0 1 V
0 1 V
0 1 V
1 1 V
0 1 V
0 1 V
0 1 V
0 1 V
1 0 V
0 1 V
0 1 V
0 1 V
0 1 V
1 0 V
0 1 V
0 1 V
0 1 V
0 1 V
0 1 V
1 0 V
0 1 V
0 1 V
0 1 V
0 1 V
1 0 V
0 1 V
0 1 V
0 1 V
0 1 V
1 1 V
0 1 V
0 1 V
0 1 V
0 1 V
1 0 V
0 1 V
0 1 V
0 1 V
0 1 V
1 0 V
0 1 V
0 1 V
0 1 V
0 1 V
1 0 V
0 1 V
0 1 V
0 1 V
0 1 V
0 1 V
1 0 V
0 1 V
0 1 V
1.000 UL
LT1
2537 1847 M
263 0 V
3550 634 M
-5 0 V
-45 5 V
-45 5 V
-44 4 V
-43 5 V
-42 5 V
-41 4 V
-41 5 V
-40 5 V
-39 5 V
-38 5 V
-38 5 V
-37 5 V
-37 5 V
-35 5 V
-36 5 V
-34 5 V
-34 5 V
-34 5 V
-32 5 V
-33 6 V
-31 5 V
-32 5 V
-31 5 V
-30 6 V
-30 5 V
-29 5 V
-29 6 V
-28 5 V
-28 6 V
-28 5 V
-27 6 V
-27 5 V
-26 6 V
-26 6 V
-25 5 V
-26 6 V
-24 6 V
-25 5 V
-24 6 V
-24 6 V
-23 5 V
-23 6 V
-23 6 V
-22 6 V
-23 5 V
-21 6 V
-22 6 V
-21 6 V
-21 6 V
-21 6 V
-20 6 V
-21 5 V
-20 6 V
-19 6 V
-20 6 V
-19 6 V
-19 6 V
-18 6 V
-19 6 V
-18 6 V
-18 5 V
-18 6 V
-17 6 V
-18 6 V
-17 6 V
-17 6 V
-16 6 V
-17 6 V
-16 5 V
-16 6 V
-16 6 V
-16 6 V
-16 6 V
-15 5 V
-15 6 V
-15 6 V
-15 5 V
-15 6 V
-15 6 V
-14 5 V
-14 6 V
-14 5 V
-14 6 V
-14 5 V
-14 6 V
-13 5 V
-14 6 V
-13 5 V
-13 5 V
-13 6 V
-12 5 V
-13 5 V
-13 5 V
-12 5 V
-12 5 V
-13 5 V
-12 5 V
-11 5 V
-12 5 V
-12 4 V
-11 5 V
-12 5 V
-11 4 V
-11 5 V
-12 4 V
-11 4 V
-10 5 V
-11 4 V
-11 4 V
-11 4 V
-10 4 V
-10 4 V
-11 4 V
-10 4 V
-10 3 V
-10 4 V
-10 3 V
-10 4 V
-9 3 V
-10 3 V
-10 3 V
-9 3 V
-10 3 V
-9 3 V
-9 3 V
-9 3 V
-9 2 V
-9 3 V
-9 2 V
-9 2 V
-9 2 V
-8 2 V
-9 2 V
-9 2 V
-8 2 V
-8 2 V
-9 1 V
-8 1 V
-8 2 V
-8 1 V
-8 1 V
-8 1 V
-8 1 V
-8 1 V
-8 0 V
-8 1 V
-7 0 V
-8 1 V
-7 0 V
-8 0 V
-7 0 V
-8 0 V
-7 0 V
-7 -1 V
-7 0 V
-8 -1 V
-7 0 V
-7 -1 V
-7 -1 V
-6 -1 V
-7 -1 V
-7 -1 V
-7 -1 V
-7 -2 V
-6 -1 V
-7 -2 V
-6 -1 V
-7 -2 V
-6 -2 V
-7 -2 V
-6 -2 V
-6 -2 V
-7 -2 V
-6 -2 V
-6 -3 V
-6 -2 V
-6 -3 V
-6 -2 V
-6 -3 V
-6 -3 V
-6 -2 V
-6 -3 V
-6 -3 V
-6 -3 V
-6 -3 V
-5 -3 V
-6 -3 V
-6 -4 V
-5 -3 V
-6 -3 V
-5 -4 V
-6 -3 V
-5 -3 V
-6 -4 V
-5 -3 V
-5 -4 V
-6 -4 V
-5 -3 V
-5 -4 V
1.000 UL
LT2
2537 1747 M
263 0 V
3550 802 M
-18 3 V
-44 8 V
-44 8 V
-43 8 V
-42 8 V
-41 8 V
-41 8 V
-39 8 V
-39 9 V
-39 8 V
-38 8 V
-37 9 V
-36 9 V
-36 8 V
-35 9 V
-34 9 V
-34 9 V
-34 8 V
-33 9 V
-32 10 V
-32 9 V
-31 9 V
-31 9 V
-30 9 V
-30 10 V
-29 9 V
-29 10 V
-29 9 V
-28 10 V
-27 9 V
-27 10 V
-27 10 V
-27 10 V
-26 10 V
-25 10 V
-25 10 V
-25 10 V
-25 10 V
-24 10 V
-24 10 V
-23 10 V
-24 10 V
-22 11 V
-23 10 V
-22 11 V
-22 10 V
-22 10 V
-21 11 V
-21 10 V
-21 11 V
-20 11 V
-21 10 V
-20 11 V
-19 11 V
-20 10 V
-19 11 V
-19 11 V
-19 10 V
-18 11 V
-19 11 V
-18 11 V
-18 10 V
-17 11 V
-18 11 V
-17 11 V
-17 10 V
-17 11 V
-16 11 V
-17 10 V
-16 11 V
-16 10 V
-16 11 V
-15 10 V
-16 11 V
-15 10 V
-15 10 V
-15 10 V
-15 10 V
-15 10 V
-14 10 V
-15 10 V
-14 10 V
-14 9 V
-14 9 V
-13 10 V
-14 9 V
-13 9 V
-14 8 V
-13 9 V
-13 8 V
-13 8 V
-13 8 V
-12 8 V
-13 7 V
-12 8 V
-12 7 V
-12 6 V
-12 7 V
-12 6 V
-12 6 V
-12 5 V
-11 5 V
-11 5 V
-12 4 V
-11 5 V
-11 3 V
-11 4 V
-11 3 V
-11 2 V
-10 3 V
-11 1 V
-10 2 V
-11 1 V
-10 1 V
-10 0 V
-10 0 V
-10 -1 V
-10 -1 V
-10 -1 V
-10 -2 V
-9 -2 V
-10 -2 V
-9 -3 V
-10 -3 V
-9 -3 V
-9 -4 V
-9 -4 V
-9 -4 V
-9 -5 V
-9 -5 V
-9 -5 V
-9 -5 V
-9 -6 V
-8 -6 V
-9 -6 V
-8 -6 V
-9 -6 V
-8 -6 V
-8 -7 V
-8 -7 V
-8 -6 V
-9 -7 V
-7 -7 V
-8 -7 V
-8 -7 V
-8 -7 V
-8 -7 V
-7 -8 V
-8 -7 V
-8 -7 V
-7 -7 V
-7 -7 V
-8 -7 V
-7 -8 V
-7 -7 V
-8 -7 V
-7 -7 V
-7 -7 V
-7 -7 V
-7 -7 V
-7 -7 V
-7 -6 V
-6 -7 V
-7 -7 V
-7 -7 V
-7 -6 V
-6 -7 V
-7 -6 V
-6 -7 V
-7 -6 V
-6 -7 V
-7 -6 V
-6 -6 V
-6 -7 V
-6 -6 V
-7 -6 V
-6 -6 V
-6 -6 V
-6 -6 V
-6 -6 V
-6 -6 V
-6 -6 V
-6 -6 V
-6 -5 V
-5 -6 V
-6 -6 V
-6 -5 V
-6 -6 V
-5 -5 V
-6 -6 V
-6 -5 V
-5 -6 V
-6 -5 V
-5 -5 V
-5 -6 V
-6 -5 V
-5 -5 V
-6 -5 V
1.000 UL
LT3
2537 1647 M
263 0 V
750 -619 R
-23 6 V
-42 12 V
-40 11 V
-40 12 V
-39 11 V
-39 12 V
-38 12 V
-37 12 V
-36 12 V
-36 12 V
-36 12 V
-34 13 V
-35 12 V
-33 12 V
-33 13 V
-33 13 V
-32 12 V
-32 13 V
-31 13 V
-30 13 V
-31 13 V
-29 13 V
-30 14 V
-28 13 V
-29 14 V
-28 13 V
-27 14 V
-27 13 V
-27 14 V
-27 14 V
-26 14 V
-25 14 V
-26 14 V
-25 14 V
-24 14 V
-24 14 V
-24 15 V
-24 14 V
-23 14 V
-23 15 V
-23 14 V
-22 15 V
-22 15 V
-22 14 V
-22 15 V
-21 15 V
-21 14 V
-21 15 V
-20 15 V
-20 15 V
-20 15 V
-20 14 V
-19 15 V
-19 15 V
-19 15 V
-19 14 V
-19 15 V
-18 14 V
-18 15 V
-18 14 V
-18 14 V
-17 14 V
-17 14 V
-17 14 V
-17 13 V
-17 13 V
-16 13 V
-17 12 V
-16 12 V
-16 11 V
-16 10 V
-15 10 V
-16 9 V
-15 9 V
-15 7 V
-15 7 V
-15 5 V
-14 5 V
-15 3 V
-14 3 V
-14 2 V
-14 1 V
-14 1 V
-14 1 V
-13 0 V
-14 0 V
-13 -1 V
-13 0 V
-13 0 V
-13 -1 V
-13 0 V
-13 0 V
-12 -1 V
-12 0 V
-13 0 V
-12 0 V
-12 -1 V
-12 0 V
-12 0 V
-11 0 V
-12 0 V
-11 0 V
-12 0 V
-11 -1 V
-11 0 V
-11 0 V
-11 0 V
-11 0 V
-11 0 V
-11 0 V
-10 0 V
-11 0 V
-10 0 V
-10 -1 V
-10 0 V
-11 0 V
-10 0 V
-9 0 V
-10 0 V
-10 0 V
-10 0 V
-9 0 V
-10 -1 V
-9 0 V
-10 0 V
-9 0 V
-9 0 V
-9 0 V
-9 0 V
-9 0 V
-9 -1 V
-9 0 V
-9 0 V
-8 0 V
-9 0 V
-8 0 V
-9 0 V
-8 0 V
-8 0 V
-9 -1 V
-8 0 V
-8 0 V
-8 0 V
-8 0 V
-8 0 V
-8 0 V
-8 -1 V
-7 0 V
-8 0 V
-8 0 V
-7 0 V
-8 0 V
-7 0 V
-8 0 V
-7 -1 V
-7 0 V
-8 0 V
-7 0 V
-7 0 V
-7 0 V
-7 0 V
-7 -1 V
-7 0 V
-7 0 V
-6 0 V
-7 0 V
-7 0 V
-7 0 V
-6 -1 V
-7 0 V
-6 0 V
-7 0 V
-6 0 V
-7 0 V
-6 0 V
-6 -1 V
-7 0 V
-6 0 V
-6 0 V
-6 0 V
-6 0 V
-6 0 V
-6 -1 V
-6 0 V
-6 0 V
-6 0 V
-6 0 V
-6 0 V
-5 -1 V
-6 0 V
stroke
grestore
end
showpage
}
\put(2850,1647){\makebox(0,0)[l]{$t = 100\tau$}}
\put(2850,1747){\makebox(0,0)[l]{$t =5\tau$}}
\put(2850,1847){\makebox(0,0)[l]{$t = \tau$}}
\put(2850,1947){\makebox(0,0)[l]{hole edge}}
\put(2000,150){\makebox(0,0){$r/R(0)$}}
\put(100,1230){\makebox(0,0)[b]{\shortstack{$\displaystyle{\Delta \over \lambda \mu}$}}}
\put(3550,300){\makebox(0,0){3}}
\put(2000,300){\makebox(0,0){2}}
\put(450,300){\makebox(0,0){1}}
\put(400,2060){\makebox(0,0)[r]{0.1}}
\put(400,400){\makebox(0,0)[r]{0}}
\end{picture}

%% file: highlambda.tex
\setlength{\unitlength}{0.1bp}
\special{!
/gnudict 120 dict def
gnudict begin
/Color false def
/Solid false def
/gnulinewidth 5.000 def
/userlinewidth gnulinewidth def
/vshift -33 def
/dl {10 mul} def
/hpt_ 31.5 def
/vpt_ 31.5 def
/hpt hpt_ def
/vpt vpt_ def
/M {moveto} bind def
/L {lineto} bind def
/R {rmoveto} bind def
/V {rlineto} bind def
/vpt2 vpt 2 mul def
/hpt2 hpt 2 mul def
/Lshow { currentpoint stroke M
  0 vshift R show } def
/Rshow { currentpoint stroke M
  dup stringwidth pop neg vshift R show } def
/Cshow { currentpoint stroke M
  dup stringwidth pop -2 div vshift R show } def
/UP { dup vpt_ mul /vpt exch def hpt_ mul /hpt exch def
  /hpt2 hpt 2 mul def /vpt2 vpt 2 mul def } def
/DL { Color {setrgbcolor Solid {pop []} if 0 setdash }
 {pop pop pop Solid {pop []} if 0 setdash} ifelse } def
/BL { stroke gnulinewidth 2 mul setlinewidth } def
/AL { stroke gnulinewidth 2 div setlinewidth } def
/UL { gnulinewidth mul /userlinewidth exch def } def
/PL { stroke userlinewidth setlinewidth } def
/LTb { BL [] 0 0 0 DL } def
/LTa { AL [1 dl 2 dl] 0 setdash 0 0 0 setrgbcolor } def
/LT0 { PL [] 0 1 0 DL } def
/LT1 { PL [4 dl 2 dl] 0 0 1 DL } def
/LT2 { PL [2 dl 3 dl] 1 0 0 DL } def
/LT3 { PL [1 dl 1.5 dl] 1 0 1 DL } def
/LT4 { PL [5 dl 2 dl 1 dl 2 dl] 0 1 1 DL } def
/LT5 { PL [4 dl 3 dl 1 dl 3 dl] 1 1 0 DL } def
/LT6 { PL [2 dl 2 dl 2 dl 4 dl] 0 0 0 DL } def
/LT7 { PL [2 dl 2 dl 2 dl 2 dl 2 dl 4 dl] 1 0.3 0 DL } def
/LT8 { PL [2 dl 2 dl 2 dl 2 dl 2 dl 2 dl 2 dl 4 dl] 0.5 0.5 0.5 DL } def
/Pnt { stroke [] 0 setdash
   gsave 1 setlinecap M 0 0 V stroke grestore } def
/Dia { stroke [] 0 setdash 2 copy vpt add M
  hpt neg vpt neg V hpt vpt neg V
  hpt vpt V hpt neg vpt V closepath stroke
  Pnt } def
/Pls { stroke [] 0 setdash vpt sub M 0 vpt2 V
  currentpoint stroke M
  hpt neg vpt neg R hpt2 0 V stroke
  } def
/Box { stroke [] 0 setdash 2 copy exch hpt sub exch vpt add M
  0 vpt2 neg V hpt2 0 V 0 vpt2 V
  hpt2 neg 0 V closepath stroke
  Pnt } def
/Crs { stroke [] 0 setdash exch hpt sub exch vpt add M
  hpt2 vpt2 neg V currentpoint stroke M
  hpt2 neg 0 R hpt2 vpt2 V stroke } def
/TriU { stroke [] 0 setdash 2 copy vpt 1.12 mul add M
  hpt neg vpt -1.62 mul V
  hpt 2 mul 0 V
  hpt neg vpt 1.62 mul V closepath stroke
  Pnt  } def
/Star { 2 copy Pls Crs } def
/BoxF { stroke [] 0 setdash exch hpt sub exch vpt add M
  0 vpt2 neg V  hpt2 0 V  0 vpt2 V
  hpt2 neg 0 V  closepath fill } def
/TriUF { stroke [] 0 setdash vpt 1.12 mul add M
  hpt neg vpt -1.62 mul V
  hpt 2 mul 0 V
  hpt neg vpt 1.62 mul V closepath fill } def
/TriD { stroke [] 0 setdash 2 copy vpt 1.12 mul sub M
  hpt neg vpt 1.62 mul V
  hpt 2 mul 0 V
  hpt neg vpt -1.62 mul V closepath stroke
  Pnt  } def
/TriDF { stroke [] 0 setdash vpt 1.12 mul sub M
  hpt neg vpt 1.62 mul V
  hpt 2 mul 0 V
  hpt neg vpt -1.62 mul V closepath fill} def
/DiaF { stroke [] 0 setdash vpt add M
  hpt neg vpt neg V hpt vpt neg V
  hpt vpt V hpt neg vpt V closepath fill } def
/Pent { stroke [] 0 setdash 2 copy gsave
  translate 0 hpt M 4 {72 rotate 0 hpt L} repeat
  closepath stroke grestore Pnt } def
/PentF { stroke [] 0 setdash gsave
  translate 0 hpt M 4 {72 rotate 0 hpt L} repeat
  closepath fill grestore } def
/Circle { stroke [] 0 setdash 2 copy
  hpt 0 360 arc stroke Pnt } def
/CircleF { stroke [] 0 setdash hpt 0 360 arc fill } def
/C0 { BL [] 0 setdash 2 copy moveto vpt 90 450  arc } bind def
/C1 { BL [] 0 setdash 2 copy        moveto
       2 copy  vpt 0 90 arc closepath fill
               vpt 0 360 arc closepath } bind def
/C2 { BL [] 0 setdash 2 copy moveto
       2 copy  vpt 90 180 arc closepath fill
               vpt 0 360 arc closepath } bind def
/C3 { BL [] 0 setdash 2 copy moveto
       2 copy  vpt 0 180 arc closepath fill
               vpt 0 360 arc closepath } bind def
/C4 { BL [] 0 setdash 2 copy moveto
       2 copy  vpt 180 270 arc closepath fill
               vpt 0 360 arc closepath } bind def
/C5 { BL [] 0 setdash 2 copy moveto
       2 copy  vpt 0 90 arc
       2 copy moveto
       2 copy  vpt 180 270 arc closepath fill
               vpt 0 360 arc } bind def
/C6 { BL [] 0 setdash 2 copy moveto
      2 copy  vpt 90 270 arc closepath fill
              vpt 0 360 arc closepath } bind def
/C7 { BL [] 0 setdash 2 copy moveto
      2 copy  vpt 0 270 arc closepath fill
              vpt 0 360 arc closepath } bind def
/C8 { BL [] 0 setdash 2 copy moveto
      2 copy vpt 270 360 arc closepath fill
              vpt 0 360 arc closepath } bind def
/C9 { BL [] 0 setdash 2 copy moveto
      2 copy  vpt 270 450 arc closepath fill
              vpt 0 360 arc closepath } bind def
/C10 { BL [] 0 setdash 2 copy 2 copy moveto vpt 270 360 arc closepath fill
       2 copy moveto
       2 copy vpt 90 180 arc closepath fill
               vpt 0 360 arc closepath } bind def
/C11 { BL [] 0 setdash 2 copy moveto
       2 copy  vpt 0 180 arc closepath fill
       2 copy moveto
       2 copy  vpt 270 360 arc closepath fill
               vpt 0 360 arc closepath } bind def
/C12 { BL [] 0 setdash 2 copy moveto
       2 copy  vpt 180 360 arc closepath fill
               vpt 0 360 arc closepath } bind def
/C13 { BL [] 0 setdash  2 copy moveto
       2 copy  vpt 0 90 arc closepath fill
       2 copy moveto
       2 copy  vpt 180 360 arc closepath fill
               vpt 0 360 arc closepath } bind def
/C14 { BL [] 0 setdash 2 copy moveto
       2 copy  vpt 90 360 arc closepath fill
               vpt 0 360 arc } bind def
/C15 { BL [] 0 setdash 2 copy vpt 0 360 arc closepath fill
               vpt 0 360 arc closepath } bind def
/Rec   { newpath 4 2 roll moveto 1 index 0 rlineto 0 exch rlineto
       neg 0 rlineto closepath } bind def
/Square { dup Rec } bind def
/Bsquare { vpt sub exch vpt sub exch vpt2 Square } bind def
/S0 { BL [] 0 setdash 2 copy moveto 0 vpt rlineto BL Bsquare } bind def
/S1 { BL [] 0 setdash 2 copy vpt Square fill Bsquare } bind def
/S2 { BL [] 0 setdash 2 copy exch vpt sub exch vpt Square fill Bsquare } bind def
/S3 { BL [] 0 setdash 2 copy exch vpt sub exch vpt2 vpt Rec fill Bsquare } bind def
/S4 { BL [] 0 setdash 2 copy exch vpt sub exch vpt sub vpt Square fill Bsquare } bind def
/S5 { BL [] 0 setdash 2 copy 2 copy vpt Square fill
       exch vpt sub exch vpt sub vpt Square fill Bsquare } bind def
/S6 { BL [] 0 setdash 2 copy exch vpt sub exch vpt sub vpt vpt2 Rec fill Bsquare } bind def
/S7 { BL [] 0 setdash 2 copy exch vpt sub exch vpt sub vpt vpt2 Rec fill
       2 copy vpt Square fill
       Bsquare } bind def
/S8 { BL [] 0 setdash 2 copy vpt sub vpt Square fill Bsquare } bind def
/S9 { BL [] 0 setdash 2 copy vpt sub vpt vpt2 Rec fill Bsquare } bind def
/S10 { BL [] 0 setdash 2 copy vpt sub vpt Square fill 2 copy exch vpt sub exch vpt Square fill
       Bsquare } bind def
/S11 { BL [] 0 setdash 2 copy vpt sub vpt Square fill 2 copy exch vpt sub exch vpt2 vpt Rec fill
       Bsquare } bind def
/S12 { BL [] 0 setdash 2 copy exch vpt sub exch vpt sub vpt2 vpt Rec fill Bsquare } bind def
/S13 { BL [] 0 setdash 2 copy exch vpt sub exch vpt sub vpt2 vpt Rec fill
       2 copy vpt Square fill Bsquare } bind def
/S14 { BL [] 0 setdash 2 copy exch vpt sub exch vpt sub vpt2 vpt Rec fill
       2 copy exch vpt sub exch vpt Square fill Bsquare } bind def
/S15 { BL [] 0 setdash 2 copy Bsquare fill Bsquare } bind def
/D0 { gsave translate 45 rotate 0 0 S0 stroke grestore } bind def
/D1 { gsave translate 45 rotate 0 0 S1 stroke grestore } bind def
/D2 { gsave translate 45 rotate 0 0 S2 stroke grestore } bind def
/D3 { gsave translate 45 rotate 0 0 S3 stroke grestore } bind def
/D4 { gsave translate 45 rotate 0 0 S4 stroke grestore } bind def
/D5 { gsave translate 45 rotate 0 0 S5 stroke grestore } bind def
/D6 { gsave translate 45 rotate 0 0 S6 stroke grestore } bind def
/D7 { gsave translate 45 rotate 0 0 S7 stroke grestore } bind def
/D8 { gsave translate 45 rotate 0 0 S8 stroke grestore } bind def
/D9 { gsave translate 45 rotate 0 0 S9 stroke grestore } bind def
/D10 { gsave translate 45 rotate 0 0 S10 stroke grestore } bind def
/D11 { gsave translate 45 rotate 0 0 S11 stroke grestore } bind def
/D12 { gsave translate 45 rotate 0 0 S12 stroke grestore } bind def
/D13 { gsave translate 45 rotate 0 0 S13 stroke grestore } bind def
/D14 { gsave translate 45 rotate 0 0 S14 stroke grestore } bind def
/D15 { gsave translate 45 rotate 0 0 S15 stroke grestore } bind def
/DiaE { stroke [] 0 setdash vpt add M
  hpt neg vpt neg V hpt vpt neg V
  hpt vpt V hpt neg vpt V closepath stroke } def
/BoxE { stroke [] 0 setdash exch hpt sub exch vpt add M
  0 vpt2 neg V hpt2 0 V 0 vpt2 V
  hpt2 neg 0 V closepath stroke } def
/TriUE { stroke [] 0 setdash vpt 1.12 mul add M
  hpt neg vpt -1.62 mul V
  hpt 2 mul 0 V
  hpt neg vpt 1.62 mul V closepath stroke } def
/TriDE { stroke [] 0 setdash vpt 1.12 mul sub M
  hpt neg vpt 1.62 mul V
  hpt 2 mul 0 V
  hpt neg vpt -1.62 mul V closepath stroke } def
/PentE { stroke [] 0 setdash gsave
  translate 0 hpt M 4 {72 rotate 0 hpt L} repeat
  closepath stroke grestore } def
/CircE { stroke [] 0 setdash 
  hpt 0 360 arc stroke } def
/BoxFill { gsave Rec 1 setgray fill grestore } def
end
}
\begin{picture}(3600,2160)(0,0)
\special{"
gnudict begin
gsave
0 0 translate
0.100 0.100 scale
0 setgray
newpath
LTb
450 400 M
63 0 V
3037 0 R
-63 0 V
450 1230 M
63 0 V
3037 0 R
-63 0 V
450 2060 M
63 0 V
3037 0 R
-63 0 V
450 400 M
0 63 V
0 1597 R
0 -63 V
2000 400 M
0 63 V
0 1597 R
0 -63 V
3550 400 M
0 63 V
0 1597 R
0 -63 V
LTb
450 400 M
3100 0 V
0 1660 V
-3100 0 V
450 400 L
1.000 UL
LT0
2587 1947 M
263 0 V
451 2060 M
36 -1 V
31 -10 V
28 -19 V
26 -25 V
24 -30 V
22 -32 V
21 -33 V
19 -33 V
19 -33 V
17 -31 V
17 -31 V
16 -29 V
15 -27 V
15 -26 V
14 -25 V
14 -22 V
13 -22 V
13 -20 V
12 -18 V
12 -18 V
12 -16 V
11 -15 V
11 -14 V
11 -13 V
10 -13 V
10 -11 V
10 -11 V
9 -10 V
10 -10 V
9 -9 V
9 -8 V
8 -8 V
9 -8 V
8 -7 V
8 -6 V
8 -7 V
7 -6 V
8 -6 V
7 -5 V
7 -5 V
7 -5 V
7 -5 V
7 -4 V
7 -5 V
6 -4 V
6 -4 V
7 -4 V
6 -3 V
6 -4 V
5 -3 V
6 -3 V
6 -4 V
5 -3 V
5 -3 V
6 -2 V
5 -3 V
5 -3 V
5 -2 V
5 -3 V
5 -2 V
4 -2 V
5 -3 V
4 -2 V
5 -2 V
4 -2 V
5 -2 V
4 -2 V
4 -2 V
4 -1 V
4 -2 V
4 -2 V
4 -1 V
3 -2 V
4 -2 V
4 -1 V
3 -2 V
4 -1 V
3 -2 V
4 -1 V
3 -1 V
3 -2 V
4 -1 V
3 -1 V
3 -1 V
3 -2 V
3 -1 V
3 -1 V
3 -1 V
3 -1 V
3 -1 V
2 -1 V
3 -1 V
3 -1 V
2 -1 V
3 -1 V
3 -1 V
2 -1 V
3 -1 V
2 -1 V
2 0 V
3 -1 V
2 -1 V
2 -1 V
3 -1 V
2 0 V
2 -1 V
2 -1 V
2 -1 V
2 0 V
2 -1 V
2 -1 V
2 0 V
2 -1 V
2 -1 V
2 0 V
2 -1 V
2 0 V
2 -1 V
1 0 V
2 -1 V
2 -1 V
1 0 V
2 -1 V
2 0 V
1 -1 V
2 0 V
2 -1 V
1 0 V
2 -1 V
1 0 V
2 -1 V
1 0 V
1 0 V
2 -1 V
1 0 V
2 -1 V
1 0 V
1 -1 V
1 0 V
2 0 V
1 -1 V
1 0 V
1 0 V
2 -1 V
1 0 V
1 0 V
1 -1 V
1 0 V
1 0 V
2 -1 V
1 0 V
1 0 V
1 -1 V
1 0 V
1 0 V
1 -1 V
1 0 V
1 0 V
1 0 V
1 -1 V
1 0 V
1 -1 V
1 0 V
1 0 V
1 0 V
1 0 V
1 -1 V
1 0 V
1 0 V
1 -1 V
1 0 V
1 0 V
1 0 V
1 -1 V
1 0 V
1 0 V
1 0 V
0 -1 V
1 0 V
1 0 V
1 0 V
0 -1 V
1 0 V
1 0 V
1 0 V
1 -1 V
1 0 V
1 0 V
1 0 V
1 -1 V
1 0 V
1 0 V
1 0 V
0 -1 V
1 0 V
1 0 V
1 0 V
1 0 V
0 -1 V
1 0 V
1 0 V
1 0 V
1 -1 V
1 0 V
1 0 V
1 0 V
1 -1 V
1 0 V
1 0 V
1 0 V
0 -1 V
1 0 V
1 0 V
1 0 V
1 0 V
0 -1 V
1 0 V
1 0 V
1 0 V
1 0 V
0 -1 V
1 0 V
1 0 V
1 0 V
1 0 V
0 -1 V
1 0 V
1 0 V
1 0 V
1 0 V
0 -1 V
1 0 V
1 0 V
1.000 UL
LT1
2587 1847 M
263 0 V
3550 585 M
-45 3 V
-45 4 V
-44 4 V
-43 3 V
-43 4 V
-42 4 V
-41 4 V
-40 4 V
-40 4 V
-38 4 V
-38 4 V
-38 4 V
-37 5 V
-36 4 V
-35 4 V
-35 4 V
-34 4 V
-34 5 V
-33 4 V
-33 5 V
-32 4 V
-31 5 V
-31 4 V
-31 5 V
-30 4 V
-29 5 V
-29 4 V
-29 5 V
-28 5 V
-28 5 V
-27 5 V
-27 4 V
-27 5 V
-26 5 V
-25 5 V
-26 5 V
-25 5 V
-24 5 V
-25 5 V
-24 6 V
-23 5 V
-23 5 V
-23 5 V
-23 5 V
-22 6 V
-22 5 V
-22 6 V
-21 5 V
-21 5 V
-21 6 V
-21 6 V
-20 5 V
-20 6 V
-20 5 V
-19 6 V
-19 6 V
-19 6 V
-19 5 V
-19 6 V
-18 6 V
-18 6 V
-18 6 V
-17 6 V
-18 6 V
-17 6 V
-17 6 V
-17 6 V
-16 7 V
-17 6 V
-16 6 V
-16 6 V
-16 7 V
-15 6 V
-16 6 V
-15 7 V
-15 6 V
-15 7 V
-15 6 V
-14 7 V
-15 7 V
-14 6 V
-14 7 V
-14 7 V
-14 6 V
-14 7 V
-13 7 V
-14 7 V
-13 7 V
-13 7 V
-13 7 V
-13 7 V
-12 7 V
-13 7 V
-12 7 V
-13 7 V
-12 8 V
-12 7 V
-12 7 V
-12 7 V
-11 8 V
-12 7 V
-11 8 V
-12 7 V
-11 8 V
-11 7 V
-11 8 V
-11 7 V
-11 8 V
-10 8 V
-11 7 V
-11 8 V
-10 8 V
-10 8 V
-10 8 V
-11 8 V
-10 8 V
-9 8 V
-10 8 V
-10 8 V
-10 8 V
-9 8 V
-10 8 V
-9 8 V
-9 8 V
-10 9 V
-9 8 V
-9 8 V
-9 9 V
-9 8 V
-9 9 V
-8 8 V
-9 9 V
-9 8 V
-8 9 V
-9 9 V
-8 8 V
-8 9 V
-9 9 V
-8 9 V
-8 8 V
-8 9 V
-8 9 V
-8 9 V
-8 9 V
-7 9 V
-8 9 V
-8 9 V
-7 9 V
-8 10 V
-7 9 V
-8 9 V
-7 9 V
-7 10 V
-8 9 V
-7 9 V
-7 10 V
-7 9 V
-7 10 V
-7 9 V
-7 10 V
-7 9 V
-7 10 V
-6 10 V
-7 10 V
-7 9 V
-6 10 V
-7 10 V
-6 10 V
-7 10 V
-6 10 V
-7 10 V
-6 10 V
-6 10 V
-6 10 V
-7 10 V
-6 10 V
-6 10 V
-6 10 V
-6 11 V
-6 10 V
-6 10 V
-6 11 V
-5 10 V
-6 11 V
-6 10 V
-6 11 V
-5 10 V
-6 11 V
-5 11 V
-6 10 V
-6 11 V
-5 11 V
-5 10 V
-6 11 V
-5 11 V
-6 11 V
-5 11 V
-5 11 V
-5 11 V
-6 11 V
1.000 UL
LT2
2587 1747 M
263 0 V
3550 780 M
-22 3 V
-35 6 V
-36 5 V
-34 6 V
-34 6 V
-34 6 V
-33 5 V
-32 6 V
-32 6 V
-32 6 V
-31 6 V
-31 6 V
-30 5 V
-30 6 V
-29 6 V
-29 6 V
-29 6 V
-28 6 V
-28 6 V
-27 6 V
-27 6 V
-27 6 V
-26 6 V
-26 7 V
-26 6 V
-25 6 V
-25 6 V
-25 6 V
-24 6 V
-24 6 V
-24 6 V
-24 6 V
-23 7 V
-23 6 V
-22 6 V
-23 6 V
-22 6 V
-21 6 V
-22 6 V
-21 6 V
-21 7 V
-21 6 V
-21 6 V
-20 6 V
-20 6 V
-20 6 V
-19 6 V
-20 6 V
-19 6 V
-19 6 V
-19 6 V
-18 6 V
-19 6 V
-18 6 V
-18 5 V
-17 6 V
-18 6 V
-17 6 V
-17 6 V
-17 5 V
-17 6 V
-17 6 V
-16 5 V
-16 6 V
-17 5 V
-16 6 V
-15 5 V
-16 6 V
-15 5 V
-16 5 V
-15 5 V
-15 6 V
-15 5 V
-14 5 V
-15 5 V
-14 5 V
-15 5 V
-14 5 V
-14 5 V
-14 4 V
-13 5 V
-14 5 V
-13 4 V
-14 5 V
-13 4 V
-13 5 V
-13 4 V
-13 4 V
-13 5 V
-12 4 V
-13 4 V
-12 4 V
-13 4 V
-12 4 V
-12 4 V
-12 4 V
-12 4 V
-11 3 V
-12 4 V
-11 4 V
-12 3 V
-11 4 V
-11 3 V
-12 4 V
-11 3 V
-11 3 V
-10 3 V
-11 4 V
-11 3 V
-10 3 V
-11 3 V
-10 3 V
-11 3 V
-10 3 V
-10 3 V
-10 2 V
-10 3 V
-10 3 V
-10 3 V
-10 2 V
-9 3 V
-10 3 V
-9 2 V
-10 3 V
-9 2 V
-9 3 V
-10 2 V
-9 3 V
-9 2 V
-9 2 V
-9 3 V
-9 2 V
-9 3 V
-8 2 V
-9 2 V
-9 2 V
-8 3 V
-9 2 V
-8 2 V
-8 2 V
-9 3 V
-8 2 V
-8 2 V
-8 2 V
-8 2 V
-8 3 V
-8 2 V
-8 2 V
-8 2 V
-7 2 V
-8 3 V
-8 2 V
-7 2 V
-8 2 V
-7 2 V
-8 2 V
-7 3 V
-7 2 V
-8 2 V
-7 2 V
-7 3 V
-7 2 V
-7 2 V
-7 2 V
-7 2 V
-7 3 V
-7 2 V
-7 2 V
-7 3 V
-7 2 V
-6 2 V
-7 2 V
1.000 UL
LT3
2587 1647 M
263 0 V
3550 890 M
-9 2 V
-33 6 V
-32 6 V
-32 6 V
-31 6 V
-31 7 V
-30 6 V
-30 6 V
-30 6 V
-29 6 V
-29 6 V
-29 7 V
-28 6 V
-28 6 V
-27 6 V
-27 6 V
-27 6 V
-26 6 V
-26 6 V
-26 6 V
-25 6 V
-25 6 V
-25 6 V
-25 6 V
-24 6 V
-24 6 V
-23 6 V
-24 5 V
-23 6 V
-23 6 V
-22 5 V
-22 6 V
-22 6 V
-22 5 V
-22 6 V
-21 5 V
-21 5 V
-21 6 V
-20 5 V
-20 5 V
-21 5 V
-19 5 V
-20 5 V
-19 5 V
-20 5 V
-19 5 V
-19 4 V
-18 5 V
-19 4 V
-18 5 V
-18 4 V
-18 4 V
-17 5 V
-18 4 V
-17 4 V
-17 4 V
-17 4 V
-17 3 V
-17 4 V
-16 4 V
-16 3 V
-16 4 V
-16 3 V
-16 3 V
-16 4 V
-15 3 V
-16 3 V
-15 3 V
-15 3 V
-15 2 V
-15 3 V
-14 3 V
-15 2 V
-14 3 V
-14 2 V
-14 2 V
-14 3 V
-14 2 V
-14 2 V
-13 2 V
-14 2 V
-13 2 V
-13 2 V
-14 1 V
-13 2 V
-12 2 V
-13 1 V
-13 2 V
-12 1 V
-13 2 V
-12 1 V
-12 2 V
-12 1 V
-12 1 V
-12 2 V
-12 1 V
-12 1 V
-11 1 V
-12 1 V
-11 1 V
-12 1 V
-11 1 V
-11 1 V
-11 1 V
-11 1 V
-11 1 V
-11 1 V
-10 1 V
-11 1 V
-10 1 V
-11 1 V
-10 1 V
-10 1 V
-11 0 V
-10 1 V
-10 1 V
-10 1 V
-10 1 V
-9 1 V
-10 0 V
-10 1 V
-9 1 V
-10 1 V
-9 1 V
-10 0 V
-9 1 V
-9 1 V
-9 1 V
-9 1 V
-9 0 V
-9 1 V
-9 1 V
-9 1 V
-9 1 V
-8 1 V
-9 0 V
-9 1 V
-8 1 V
-8 1 V
-9 1 V
-8 1 V
-8 0 V
-9 1 V
-8 1 V
-8 1 V
-8 1 V
-8 1 V
-8 1 V
-8 1 V
-7 0 V
-8 1 V
-8 1 V
-8 1 V
-7 1 V
-8 1 V
-7 1 V
-8 1 V
-7 1 V
-7 1 V
-8 1 V
1.000 UL
LT4
2587 1547 M
263 0 V
700 -530 R
-29 6 V
-29 6 V
-29 5 V
-29 6 V
-28 6 V
-28 5 V
-27 6 V
-27 5 V
-27 5 V
-26 6 V
-27 5 V
-25 5 V
-26 5 V
-25 5 V
-25 5 V
-25 5 V
-24 4 V
-24 5 V
-24 5 V
-24 4 V
-23 4 V
-23 5 V
-23 4 V
-22 4 V
-23 4 V
-22 4 V
-21 3 V
-22 4 V
-21 4 V
-21 3 V
-21 3 V
-21 4 V
-20 3 V
-21 3 V
-20 3 V
-19 3 V
-20 2 V
-19 3 V
-19 2 V
-19 3 V
-19 2 V
-19 2 V
-18 3 V
-18 2 V
-18 2 V
-18 2 V
-18 1 V
-18 2 V
-17 2 V
-17 1 V
-17 2 V
-17 1 V
-17 2 V
-16 1 V
-17 1 V
-16 1 V
-16 1 V
-16 1 V
-16 1 V
-15 1 V
-16 1 V
-15 1 V
-15 1 V
-15 1 V
-15 0 V
-15 1 V
-15 1 V
-14 0 V
-15 1 V
-14 0 V
-14 1 V
-14 0 V
-14 1 V
-14 0 V
-13 1 V
-14 0 V
-13 0 V
-14 1 V
-13 0 V
-13 1 V
-13 0 V
-13 0 V
-12 0 V
-13 1 V
-13 0 V
-12 0 V
-12 1 V
-13 0 V
-12 0 V
-12 0 V
-12 1 V
-12 0 V
-11 0 V
-12 0 V
-12 0 V
-11 1 V
-11 0 V
-12 0 V
-11 0 V
-11 0 V
-11 1 V
-11 0 V
-11 0 V
-11 0 V
-10 0 V
-11 1 V
-10 0 V
-11 0 V
-10 0 V
-11 0 V
-10 1 V
-10 0 V
-10 0 V
-10 0 V
-10 0 V
-10 1 V
-9 0 V
-10 0 V
-10 0 V
-9 0 V
-10 1 V
-9 0 V
-10 0 V
-9 0 V
-9 0 V
-9 1 V
-9 0 V
-9 0 V
-9 0 V
-9 0 V
-9 1 V
-9 0 V
-9 0 V
-8 0 V
-9 0 V
-8 1 V
-9 0 V
-8 0 V
-9 0 V
-8 0 V
-8 1 V
-8 0 V
-9 0 V
-8 0 V
stroke
grestore
end
showpage
}
\put(2900,1547){\makebox(0,0)[l]{$t = 25\tau$}}
\put(2900,1647){\makebox(0,0)[l]{$t = 10\tau$}}
\put(2900,1747){\makebox(0,0)[l]{$t =5\tau$}}
\put(2900,1847){\makebox(0,0)[l]{$t = 0$}}
\put(2900,1947){\makebox(0,0)[l]{hole edge}}
\put(2000,150){\makebox(0,0){$r/R(0)$}}
\put(100,1230){\makebox(0,0)[b]{\shortstack{$\displaystyle{s \over \mu}$}}}
\put(3550,300){\makebox(0,0){3}}
\put(2000,300){\makebox(0,0){2}}
\put(450,300){\makebox(0,0){1}}
\put(400,2060){\makebox(0,0)[r]{0.2}}
\put(400,1230){\makebox(0,0)[r]{0.1}}
\put(400,400){\makebox(0,0)[r]{0}}
\end{picture}

%% file: Rfinal.tex
\setlength{\unitlength}{0.1bp}
\special{!
/gnudict 120 dict def
gnudict begin
/Color false def
/Solid false def
/gnulinewidth 5.000 def
/userlinewidth gnulinewidth def
/vshift -33 def
/dl {10 mul} def
/hpt_ 31.5 def
/vpt_ 31.5 def
/hpt hpt_ def
/vpt vpt_ def
/M {moveto} bind def
/L {lineto} bind def
/R {rmoveto} bind def
/V {rlineto} bind def
/vpt2 vpt 2 mul def
/hpt2 hpt 2 mul def
/Lshow { currentpoint stroke M
  0 vshift R show } def
/Rshow { currentpoint stroke M
  dup stringwidth pop neg vshift R show } def
/Cshow { currentpoint stroke M
  dup stringwidth pop -2 div vshift R show } def
/UP { dup vpt_ mul /vpt exch def hpt_ mul /hpt exch def
  /hpt2 hpt 2 mul def /vpt2 vpt 2 mul def } def
/DL { Color {setrgbcolor Solid {pop []} if 0 setdash }
 {pop pop pop Solid {pop []} if 0 setdash} ifelse } def
/BL { stroke gnulinewidth 2 mul setlinewidth } def
/AL { stroke gnulinewidth 2 div setlinewidth } def
/UL { gnulinewidth mul /userlinewidth exch def } def
/PL { stroke userlinewidth setlinewidth } def
/LTb { BL [] 0 0 0 DL } def
/LTa { AL [1 dl 2 dl] 0 setdash 0 0 0 setrgbcolor } def
/LT0 { PL [] 0 1 0 DL } def
/LT1 { PL [4 dl 2 dl] 0 0 1 DL } def
/LT2 { PL [2 dl 3 dl] 1 0 0 DL } def
/LT3 { PL [1 dl 1.5 dl] 1 0 1 DL } def
/LT4 { PL [5 dl 2 dl 1 dl 2 dl] 0 1 1 DL } def
/LT5 { PL [4 dl 3 dl 1 dl 3 dl] 1 1 0 DL } def
/LT6 { PL [2 dl 2 dl 2 dl 4 dl] 0 0 0 DL } def
/LT7 { PL [2 dl 2 dl 2 dl 2 dl 2 dl 4 dl] 1 0.3 0 DL } def
/LT8 { PL [2 dl 2 dl 2 dl 2 dl 2 dl 2 dl 2 dl 4 dl] 0.5 0.5 0.5 DL } def
/Pnt { stroke [] 0 setdash
   gsave 1 setlinecap M 0 0 V stroke grestore } def
/Dia { stroke [] 0 setdash 2 copy vpt add M
  hpt neg vpt neg V hpt vpt neg V
  hpt vpt V hpt neg vpt V closepath stroke
  Pnt } def
/Pls { stroke [] 0 setdash vpt sub M 0 vpt2 V
  currentpoint stroke M
  hpt neg vpt neg R hpt2 0 V stroke
  } def
/Box { stroke [] 0 setdash 2 copy exch hpt sub exch vpt add M
  0 vpt2 neg V hpt2 0 V 0 vpt2 V
  hpt2 neg 0 V closepath stroke
  Pnt } def
/Crs { stroke [] 0 setdash exch hpt sub exch vpt add M
  hpt2 vpt2 neg V currentpoint stroke M
  hpt2 neg 0 R hpt2 vpt2 V stroke } def
/TriU { stroke [] 0 setdash 2 copy vpt 1.12 mul add M
  hpt neg vpt -1.62 mul V
  hpt 2 mul 0 V
  hpt neg vpt 1.62 mul V closepath stroke
  Pnt  } def
/Star { 2 copy Pls Crs } def
/BoxF { stroke [] 0 setdash exch hpt sub exch vpt add M
  0 vpt2 neg V  hpt2 0 V  0 vpt2 V
  hpt2 neg 0 V  closepath fill } def
/TriUF { stroke [] 0 setdash vpt 1.12 mul add M
  hpt neg vpt -1.62 mul V
  hpt 2 mul 0 V
  hpt neg vpt 1.62 mul V closepath fill } def
/TriD { stroke [] 0 setdash 2 copy vpt 1.12 mul sub M
  hpt neg vpt 1.62 mul V
  hpt 2 mul 0 V
  hpt neg vpt -1.62 mul V closepath stroke
  Pnt  } def
/TriDF { stroke [] 0 setdash vpt 1.12 mul sub M
  hpt neg vpt 1.62 mul V
  hpt 2 mul 0 V
  hpt neg vpt -1.62 mul V closepath fill} def
/DiaF { stroke [] 0 setdash vpt add M
  hpt neg vpt neg V hpt vpt neg V
  hpt vpt V hpt neg vpt V closepath fill } def
/Pent { stroke [] 0 setdash 2 copy gsave
  translate 0 hpt M 4 {72 rotate 0 hpt L} repeat
  closepath stroke grestore Pnt } def
/PentF { stroke [] 0 setdash gsave
  translate 0 hpt M 4 {72 rotate 0 hpt L} repeat
  closepath fill grestore } def
/Circle { stroke [] 0 setdash 2 copy
  hpt 0 360 arc stroke Pnt } def
/CircleF { stroke [] 0 setdash hpt 0 360 arc fill } def
/C0 { BL [] 0 setdash 2 copy moveto vpt 90 450  arc } bind def
/C1 { BL [] 0 setdash 2 copy        moveto
       2 copy  vpt 0 90 arc closepath fill
               vpt 0 360 arc closepath } bind def
/C2 { BL [] 0 setdash 2 copy moveto
       2 copy  vpt 90 180 arc closepath fill
               vpt 0 360 arc closepath } bind def
/C3 { BL [] 0 setdash 2 copy moveto
       2 copy  vpt 0 180 arc closepath fill
               vpt 0 360 arc closepath } bind def
/C4 { BL [] 0 setdash 2 copy moveto
       2 copy  vpt 180 270 arc closepath fill
               vpt 0 360 arc closepath } bind def
/C5 { BL [] 0 setdash 2 copy moveto
       2 copy  vpt 0 90 arc
       2 copy moveto
       2 copy  vpt 180 270 arc closepath fill
               vpt 0 360 arc } bind def
/C6 { BL [] 0 setdash 2 copy moveto
      2 copy  vpt 90 270 arc closepath fill
              vpt 0 360 arc closepath } bind def
/C7 { BL [] 0 setdash 2 copy moveto
      2 copy  vpt 0 270 arc closepath fill
              vpt 0 360 arc closepath } bind def
/C8 { BL [] 0 setdash 2 copy moveto
      2 copy vpt 270 360 arc closepath fill
              vpt 0 360 arc closepath } bind def
/C9 { BL [] 0 setdash 2 copy moveto
      2 copy  vpt 270 450 arc closepath fill
              vpt 0 360 arc closepath } bind def
/C10 { BL [] 0 setdash 2 copy 2 copy moveto vpt 270 360 arc closepath fill
       2 copy moveto
       2 copy vpt 90 180 arc closepath fill
               vpt 0 360 arc closepath } bind def
/C11 { BL [] 0 setdash 2 copy moveto
       2 copy  vpt 0 180 arc closepath fill
       2 copy moveto
       2 copy  vpt 270 360 arc closepath fill
               vpt 0 360 arc closepath } bind def
/C12 { BL [] 0 setdash 2 copy moveto
       2 copy  vpt 180 360 arc closepath fill
               vpt 0 360 arc closepath } bind def
/C13 { BL [] 0 setdash  2 copy moveto
       2 copy  vpt 0 90 arc closepath fill
       2 copy moveto
       2 copy  vpt 180 360 arc closepath fill
               vpt 0 360 arc closepath } bind def
/C14 { BL [] 0 setdash 2 copy moveto
       2 copy  vpt 90 360 arc closepath fill
               vpt 0 360 arc } bind def
/C15 { BL [] 0 setdash 2 copy vpt 0 360 arc closepath fill
               vpt 0 360 arc closepath } bind def
/Rec   { newpath 4 2 roll moveto 1 index 0 rlineto 0 exch rlineto
       neg 0 rlineto closepath } bind def
/Square { dup Rec } bind def
/Bsquare { vpt sub exch vpt sub exch vpt2 Square } bind def
/S0 { BL [] 0 setdash 2 copy moveto 0 vpt rlineto BL Bsquare } bind def
/S1 { BL [] 0 setdash 2 copy vpt Square fill Bsquare } bind def
/S2 { BL [] 0 setdash 2 copy exch vpt sub exch vpt Square fill Bsquare } bind def
/S3 { BL [] 0 setdash 2 copy exch vpt sub exch vpt2 vpt Rec fill Bsquare } bind def
/S4 { BL [] 0 setdash 2 copy exch vpt sub exch vpt sub vpt Square fill Bsquare } bind def
/S5 { BL [] 0 setdash 2 copy 2 copy vpt Square fill
       exch vpt sub exch vpt sub vpt Square fill Bsquare } bind def
/S6 { BL [] 0 setdash 2 copy exch vpt sub exch vpt sub vpt vpt2 Rec fill Bsquare } bind def
/S7 { BL [] 0 setdash 2 copy exch vpt sub exch vpt sub vpt vpt2 Rec fill
       2 copy vpt Square fill
       Bsquare } bind def
/S8 { BL [] 0 setdash 2 copy vpt sub vpt Square fill Bsquare } bind def
/S9 { BL [] 0 setdash 2 copy vpt sub vpt vpt2 Rec fill Bsquare } bind def
/S10 { BL [] 0 setdash 2 copy vpt sub vpt Square fill 2 copy exch vpt sub exch vpt Square fill
       Bsquare } bind def
/S11 { BL [] 0 setdash 2 copy vpt sub vpt Square fill 2 copy exch vpt sub exch vpt2 vpt Rec fill
       Bsquare } bind def
/S12 { BL [] 0 setdash 2 copy exch vpt sub exch vpt sub vpt2 vpt Rec fill Bsquare } bind def
/S13 { BL [] 0 setdash 2 copy exch vpt sub exch vpt sub vpt2 vpt Rec fill
       2 copy vpt Square fill Bsquare } bind def
/S14 { BL [] 0 setdash 2 copy exch vpt sub exch vpt sub vpt2 vpt Rec fill
       2 copy exch vpt sub exch vpt Square fill Bsquare } bind def
/S15 { BL [] 0 setdash 2 copy Bsquare fill Bsquare } bind def
/D0 { gsave translate 45 rotate 0 0 S0 stroke grestore } bind def
/D1 { gsave translate 45 rotate 0 0 S1 stroke grestore } bind def
/D2 { gsave translate 45 rotate 0 0 S2 stroke grestore } bind def
/D3 { gsave translate 45 rotate 0 0 S3 stroke grestore } bind def
/D4 { gsave translate 45 rotate 0 0 S4 stroke grestore } bind def
/D5 { gsave translate 45 rotate 0 0 S5 stroke grestore } bind def
/D6 { gsave translate 45 rotate 0 0 S6 stroke grestore } bind def
/D7 { gsave translate 45 rotate 0 0 S7 stroke grestore } bind def
/D8 { gsave translate 45 rotate 0 0 S8 stroke grestore } bind def
/D9 { gsave translate 45 rotate 0 0 S9 stroke grestore } bind def
/D10 { gsave translate 45 rotate 0 0 S10 stroke grestore } bind def
/D11 { gsave translate 45 rotate 0 0 S11 stroke grestore } bind def
/D12 { gsave translate 45 rotate 0 0 S12 stroke grestore } bind def
/D13 { gsave translate 45 rotate 0 0 S13 stroke grestore } bind def
/D14 { gsave translate 45 rotate 0 0 S14 stroke grestore } bind def
/D15 { gsave translate 45 rotate 0 0 S15 stroke grestore } bind def
/DiaE { stroke [] 0 setdash vpt add M
  hpt neg vpt neg V hpt vpt neg V
  hpt vpt V hpt neg vpt V closepath stroke } def
/BoxE { stroke [] 0 setdash exch hpt sub exch vpt add M
  0 vpt2 neg V hpt2 0 V 0 vpt2 V
  hpt2 neg 0 V closepath stroke } def
/TriUE { stroke [] 0 setdash vpt 1.12 mul add M
  hpt neg vpt -1.62 mul V
  hpt 2 mul 0 V
  hpt neg vpt 1.62 mul V closepath stroke } def
/TriDE { stroke [] 0 setdash vpt 1.12 mul sub M
  hpt neg vpt 1.62 mul V
  hpt 2 mul 0 V
  hpt neg vpt -1.62 mul V closepath stroke } def
/PentE { stroke [] 0 setdash gsave
  translate 0 hpt M 4 {72 rotate 0 hpt L} repeat
  closepath stroke grestore } def
/CircE { stroke [] 0 setdash 
  hpt 0 360 arc stroke } def
/BoxFill { gsave Rec 1 setgray fill grestore } def
end
}
\begin{picture}(3600,2160)(0,0)
\special{"
gnudict begin
gsave
0 0 translate
0.100 0.100 scale
0 setgray
newpath
LTb
350 400 M
63 0 V
3137 0 R
-63 0 V
350 1064 M
63 0 V
3137 0 R
-63 0 V
350 1728 M
63 0 V
3137 0 R
-63 0 V
350 400 M
0 63 V
0 1597 R
0 -63 V
807 400 M
0 63 V
0 1597 R
0 -63 V
1264 400 M
0 63 V
0 1597 R
0 -63 V
1721 400 M
0 63 V
0 1597 R
0 -63 V
2179 400 M
0 63 V
0 1597 R
0 -63 V
2636 400 M
0 63 V
0 1597 R
0 -63 V
3093 400 M
0 63 V
0 1597 R
0 -63 V
3550 400 M
0 63 V
0 1597 R
0 -63 V
LTb
350 400 M
3200 0 V
0 1660 V
-3200 0 V
350 400 L
1.000 UL
LT0
773 1728 M
263 0 V
2065 472 M
13 1 V
14 2 V
13 2 V
13 1 V
14 2 V
13 2 V
13 2 V
14 2 V
13 1 V
13 2 V
14 2 V
13 2 V
14 2 V
13 2 V
13 2 V
14 3 V
13 2 V
13 2 V
14 3 V
13 2 V
14 2 V
13 3 V
13 3 V
14 2 V
13 3 V
13 3 V
14 3 V
13 3 V
13 3 V
14 3 V
13 3 V
14 3 V
13 4 V
13 3 V
14 3 V
13 4 V
13 4 V
14 4 V
13 3 V
14 5 V
13 4 V
13 4 V
14 4 V
13 5 V
13 4 V
14 5 V
13 5 V
13 5 V
14 5 V
13 6 V
14 5 V
13 6 V
13 6 V
14 6 V
13 7 V
13 6 V
14 7 V
13 7 V
14 8 V
13 7 V
13 8 V
14 9 V
13 8 V
13 9 V
14 10 V
13 10 V
13 10 V
14 10 V
13 12 V
14 11 V
13 13 V
13 12 V
14 14 V
13 14 V
13 15 V
14 16 V
13 16 V
14 18 V
13 19 V
13 20 V
14 21 V
13 23 V
13 25 V
14 26 V
13 28 V
14 31 V
13 33 V
13 37 V
14 40 V
13 45 V
13 49 V
14 56 V
13 62 V
13 69 V
14 83 V
13 104 V
14 124 V
13 136 V
12 127 V
1.000 UL
LT1
773 1628 M
263 0 V
1831 526 M
16 2 V
15 2 V
15 2 V
15 2 V
15 2 V
16 2 V
15 2 V
15 3 V
15 2 V
15 2 V
15 2 V
16 3 V
15 2 V
15 3 V
15 2 V
15 3 V
16 2 V
15 3 V
15 3 V
15 3 V
15 2 V
15 3 V
16 3 V
15 3 V
15 4 V
15 3 V
15 3 V
16 3 V
15 4 V
15 3 V
15 4 V
15 4 V
16 3 V
15 4 V
15 4 V
15 4 V
15 4 V
15 4 V
16 5 V
15 4 V
15 5 V
15 5 V
15 4 V
16 5 V
15 5 V
15 6 V
15 5 V
15 6 V
15 5 V
16 6 V
15 6 V
15 6 V
15 7 V
15 6 V
16 7 V
15 7 V
15 7 V
15 8 V
15 8 V
15 8 V
16 8 V
15 9 V
15 9 V
15 9 V
15 10 V
16 10 V
15 11 V
15 11 V
15 11 V
15 12 V
15 12 V
16 13 V
15 14 V
15 14 V
15 15 V
15 15 V
16 16 V
15 18 V
15 18 V
15 20 V
15 21 V
15 23 V
16 24 V
15 25 V
15 27 V
15 28 V
15 30 V
16 32 V
15 33 V
15 36 V
15 42 V
15 49 V
15 57 V
16 67 V
15 75 V
15 82 V
15 87 V
15 90 V
16 91 V
1.000 UL
LT2
773 1528 M
263 0 V
350 519 M
8 1 V
27 3 V
26 3 V
26 3 V
26 4 V
26 3 V
26 3 V
26 3 V
27 4 V
26 3 V
26 4 V
26 3 V
26 4 V
26 3 V
27 4 V
26 4 V
26 4 V
26 4 V
26 4 V
26 4 V
26 4 V
27 4 V
26 4 V
26 5 V
26 4 V
26 5 V
26 4 V
27 5 V
26 5 V
26 5 V
26 4 V
26 5 V
26 6 V
26 5 V
27 5 V
26 6 V
26 5 V
26 6 V
26 5 V
26 6 V
27 6 V
26 7 V
26 6 V
26 6 V
26 7 V
26 6 V
26 7 V
27 7 V
26 8 V
26 7 V
26 8 V
26 7 V
26 8 V
26 9 V
27 8 V
26 9 V
26 9 V
26 9 V
26 9 V
26 10 V
27 10 V
26 10 V
26 10 V
26 11 V
26 11 V
26 12 V
26 12 V
27 12 V
26 13 V
26 13 V
26 13 V
26 14 V
26 15 V
27 16 V
26 16 V
26 17 V
26 17 V
26 19 V
26 19 V
26 21 V
27 21 V
26 22 V
26 22 V
26 24 V
26 25 V
26 25 V
26 27 V
27 29 V
26 31 V
26 34 V
26 38 V
26 41 V
26 46 V
27 50 V
26 54 V
26 57 V
26 59 V
26 61 V
26 61 V
stroke
grestore
end
showpage
}
\put(1086,1528){\makebox(0,0)[l]{$\lambda s_y = 0.5$}}
\put(1086,1628){\makebox(0,0)[l]{$\lambda s_y = 0.05$}}
\put(1086,1728){\makebox(0,0)[l]{$\lambda s_y = 0.005$}}
\put(1950,150){\makebox(0,0){$\sigma_\infty/\sigma_\infty^{th}$}}
\put(100,1230){\makebox(0,0)[b]{\shortstack{$\displaystyle{R_{\rm final} \over R(0)}$}}}
\put(3550,300){\makebox(0,0){1}}
\put(3093,300){\makebox(0,0){0.9}}
\put(2636,300){\makebox(0,0){0.8}}
\put(2179,300){\makebox(0,0){0.7}}
\put(1721,300){\makebox(0,0){0.6}}
\put(1264,300){\makebox(0,0){0.5}}
\put(807,300){\makebox(0,0){0.4}}
\put(350,300){\makebox(0,0){0.3}}
\put(300,1728){\makebox(0,0)[r]{3}}
\put(300,1064){\makebox(0,0)[r]{2}}
\put(300,400){\makebox(0,0)[r]{1}}
\end{picture}

%% file: lam_thresh.tex
\setlength{\unitlength}{0.1bp}
\special{!
/gnudict 120 dict def
gnudict begin
/Color false def
/Solid false def
/gnulinewidth 5.000 def
/userlinewidth gnulinewidth def
/vshift -33 def
/dl {10 mul} def
/hpt_ 31.5 def
/vpt_ 31.5 def
/hpt hpt_ def
/vpt vpt_ def
/M {moveto} bind def
/L {lineto} bind def
/R {rmoveto} bind def
/V {rlineto} bind def
/vpt2 vpt 2 mul def
/hpt2 hpt 2 mul def
/Lshow { currentpoint stroke M
  0 vshift R show } def
/Rshow { currentpoint stroke M
  dup stringwidth pop neg vshift R show } def
/Cshow { currentpoint stroke M
  dup stringwidth pop -2 div vshift R show } def
/UP { dup vpt_ mul /vpt exch def hpt_ mul /hpt exch def
  /hpt2 hpt 2 mul def /vpt2 vpt 2 mul def } def
/DL { Color {setrgbcolor Solid {pop []} if 0 setdash }
 {pop pop pop Solid {pop []} if 0 setdash} ifelse } def
/BL { stroke gnulinewidth 2 mul setlinewidth } def
/AL { stroke gnulinewidth 2 div setlinewidth } def
/UL { gnulinewidth mul /userlinewidth exch def } def
/PL { stroke userlinewidth setlinewidth } def
/LTb { BL [] 0 0 0 DL } def
/LTa { AL [1 dl 2 dl] 0 setdash 0 0 0 setrgbcolor } def
/LT0 { PL [] 0 1 0 DL } def
/LT1 { PL [4 dl 2 dl] 0 0 1 DL } def
/LT2 { PL [2 dl 3 dl] 1 0 0 DL } def
/LT3 { PL [1 dl 1.5 dl] 1 0 1 DL } def
/LT4 { PL [5 dl 2 dl 1 dl 2 dl] 0 1 1 DL } def
/LT5 { PL [4 dl 3 dl 1 dl 3 dl] 1 1 0 DL } def
/LT6 { PL [2 dl 2 dl 2 dl 4 dl] 0 0 0 DL } def
/LT7 { PL [2 dl 2 dl 2 dl 2 dl 2 dl 4 dl] 1 0.3 0 DL } def
/LT8 { PL [2 dl 2 dl 2 dl 2 dl 2 dl 2 dl 2 dl 4 dl] 0.5 0.5 0.5 DL } def
/Pnt { stroke [] 0 setdash
   gsave 1 setlinecap M 0 0 V stroke grestore } def
/Dia { stroke [] 0 setdash 2 copy vpt add M
  hpt neg vpt neg V hpt vpt neg V
  hpt vpt V hpt neg vpt V closepath stroke
  Pnt } def
/Pls { stroke [] 0 setdash vpt sub M 0 vpt2 V
  currentpoint stroke M
  hpt neg vpt neg R hpt2 0 V stroke
  } def
/Box { stroke [] 0 setdash 2 copy exch hpt sub exch vpt add M
  0 vpt2 neg V hpt2 0 V 0 vpt2 V
  hpt2 neg 0 V closepath stroke
  Pnt } def
/Crs { stroke [] 0 setdash exch hpt sub exch vpt add M
  hpt2 vpt2 neg V currentpoint stroke M
  hpt2 neg 0 R hpt2 vpt2 V stroke } def
/TriU { stroke [] 0 setdash 2 copy vpt 1.12 mul add M
  hpt neg vpt -1.62 mul V
  hpt 2 mul 0 V
  hpt neg vpt 1.62 mul V closepath stroke
  Pnt  } def
/Star { 2 copy Pls Crs } def
/BoxF { stroke [] 0 setdash exch hpt sub exch vpt add M
  0 vpt2 neg V  hpt2 0 V  0 vpt2 V
  hpt2 neg 0 V  closepath fill } def
/TriUF { stroke [] 0 setdash vpt 1.12 mul add M
  hpt neg vpt -1.62 mul V
  hpt 2 mul 0 V
  hpt neg vpt 1.62 mul V closepath fill } def
/TriD { stroke [] 0 setdash 2 copy vpt 1.12 mul sub M
  hpt neg vpt 1.62 mul V
  hpt 2 mul 0 V
  hpt neg vpt -1.62 mul V closepath stroke
  Pnt  } def
/TriDF { stroke [] 0 setdash vpt 1.12 mul sub M
  hpt neg vpt 1.62 mul V
  hpt 2 mul 0 V
  hpt neg vpt -1.62 mul V closepath fill} def
/DiaF { stroke [] 0 setdash vpt add M
  hpt neg vpt neg V hpt vpt neg V
  hpt vpt V hpt neg vpt V closepath fill } def
/Pent { stroke [] 0 setdash 2 copy gsave
  translate 0 hpt M 4 {72 rotate 0 hpt L} repeat
  closepath stroke grestore Pnt } def
/PentF { stroke [] 0 setdash gsave
  translate 0 hpt M 4 {72 rotate 0 hpt L} repeat
  closepath fill grestore } def
/Circle { stroke [] 0 setdash 2 copy
  hpt 0 360 arc stroke Pnt } def
/CircleF { stroke [] 0 setdash hpt 0 360 arc fill } def
/C0 { BL [] 0 setdash 2 copy moveto vpt 90 450  arc } bind def
/C1 { BL [] 0 setdash 2 copy        moveto
       2 copy  vpt 0 90 arc closepath fill
               vpt 0 360 arc closepath } bind def
/C2 { BL [] 0 setdash 2 copy moveto
       2 copy  vpt 90 180 arc closepath fill
               vpt 0 360 arc closepath } bind def
/C3 { BL [] 0 setdash 2 copy moveto
       2 copy  vpt 0 180 arc closepath fill
               vpt 0 360 arc closepath } bind def
/C4 { BL [] 0 setdash 2 copy moveto
       2 copy  vpt 180 270 arc closepath fill
               vpt 0 360 arc closepath } bind def
/C5 { BL [] 0 setdash 2 copy moveto
       2 copy  vpt 0 90 arc
       2 copy moveto
       2 copy  vpt 180 270 arc closepath fill
               vpt 0 360 arc } bind def
/C6 { BL [] 0 setdash 2 copy moveto
      2 copy  vpt 90 270 arc closepath fill
              vpt 0 360 arc closepath } bind def
/C7 { BL [] 0 setdash 2 copy moveto
      2 copy  vpt 0 270 arc closepath fill
              vpt 0 360 arc closepath } bind def
/C8 { BL [] 0 setdash 2 copy moveto
      2 copy vpt 270 360 arc closepath fill
              vpt 0 360 arc closepath } bind def
/C9 { BL [] 0 setdash 2 copy moveto
      2 copy  vpt 270 450 arc closepath fill
              vpt 0 360 arc closepath } bind def
/C10 { BL [] 0 setdash 2 copy 2 copy moveto vpt 270 360 arc closepath fill
       2 copy moveto
       2 copy vpt 90 180 arc closepath fill
               vpt 0 360 arc closepath } bind def
/C11 { BL [] 0 setdash 2 copy moveto
       2 copy  vpt 0 180 arc closepath fill
       2 copy moveto
       2 copy  vpt 270 360 arc closepath fill
               vpt 0 360 arc closepath } bind def
/C12 { BL [] 0 setdash 2 copy moveto
       2 copy  vpt 180 360 arc closepath fill
               vpt 0 360 arc closepath } bind def
/C13 { BL [] 0 setdash  2 copy moveto
       2 copy  vpt 0 90 arc closepath fill
       2 copy moveto
       2 copy  vpt 180 360 arc closepath fill
               vpt 0 360 arc closepath } bind def
/C14 { BL [] 0 setdash 2 copy moveto
       2 copy  vpt 90 360 arc closepath fill
               vpt 0 360 arc } bind def
/C15 { BL [] 0 setdash 2 copy vpt 0 360 arc closepath fill
               vpt 0 360 arc closepath } bind def
/Rec   { newpath 4 2 roll moveto 1 index 0 rlineto 0 exch rlineto
       neg 0 rlineto closepath } bind def
/Square { dup Rec } bind def
/Bsquare { vpt sub exch vpt sub exch vpt2 Square } bind def
/S0 { BL [] 0 setdash 2 copy moveto 0 vpt rlineto BL Bsquare } bind def
/S1 { BL [] 0 setdash 2 copy vpt Square fill Bsquare } bind def
/S2 { BL [] 0 setdash 2 copy exch vpt sub exch vpt Square fill Bsquare } bind def
/S3 { BL [] 0 setdash 2 copy exch vpt sub exch vpt2 vpt Rec fill Bsquare } bind def
/S4 { BL [] 0 setdash 2 copy exch vpt sub exch vpt sub vpt Square fill Bsquare } bind def
/S5 { BL [] 0 setdash 2 copy 2 copy vpt Square fill
       exch vpt sub exch vpt sub vpt Square fill Bsquare } bind def
/S6 { BL [] 0 setdash 2 copy exch vpt sub exch vpt sub vpt vpt2 Rec fill Bsquare } bind def
/S7 { BL [] 0 setdash 2 copy exch vpt sub exch vpt sub vpt vpt2 Rec fill
       2 copy vpt Square fill
       Bsquare } bind def
/S8 { BL [] 0 setdash 2 copy vpt sub vpt Square fill Bsquare } bind def
/S9 { BL [] 0 setdash 2 copy vpt sub vpt vpt2 Rec fill Bsquare } bind def
/S10 { BL [] 0 setdash 2 copy vpt sub vpt Square fill 2 copy exch vpt sub exch vpt Square fill
       Bsquare } bind def
/S11 { BL [] 0 setdash 2 copy vpt sub vpt Square fill 2 copy exch vpt sub exch vpt2 vpt Rec fill
       Bsquare } bind def
/S12 { BL [] 0 setdash 2 copy exch vpt sub exch vpt sub vpt2 vpt Rec fill Bsquare } bind def
/S13 { BL [] 0 setdash 2 copy exch vpt sub exch vpt sub vpt2 vpt Rec fill
       2 copy vpt Square fill Bsquare } bind def
/S14 { BL [] 0 setdash 2 copy exch vpt sub exch vpt sub vpt2 vpt Rec fill
       2 copy exch vpt sub exch vpt Square fill Bsquare } bind def
/S15 { BL [] 0 setdash 2 copy Bsquare fill Bsquare } bind def
/D0 { gsave translate 45 rotate 0 0 S0 stroke grestore } bind def
/D1 { gsave translate 45 rotate 0 0 S1 stroke grestore } bind def
/D2 { gsave translate 45 rotate 0 0 S2 stroke grestore } bind def
/D3 { gsave translate 45 rotate 0 0 S3 stroke grestore } bind def
/D4 { gsave translate 45 rotate 0 0 S4 stroke grestore } bind def
/D5 { gsave translate 45 rotate 0 0 S5 stroke grestore } bind def
/D6 { gsave translate 45 rotate 0 0 S6 stroke grestore } bind def
/D7 { gsave translate 45 rotate 0 0 S7 stroke grestore } bind def
/D8 { gsave translate 45 rotate 0 0 S8 stroke grestore } bind def
/D9 { gsave translate 45 rotate 0 0 S9 stroke grestore } bind def
/D10 { gsave translate 45 rotate 0 0 S10 stroke grestore } bind def
/D11 { gsave translate 45 rotate 0 0 S11 stroke grestore } bind def
/D12 { gsave translate 45 rotate 0 0 S12 stroke grestore } bind def
/D13 { gsave translate 45 rotate 0 0 S13 stroke grestore } bind def
/D14 { gsave translate 45 rotate 0 0 S14 stroke grestore } bind def
/D15 { gsave translate 45 rotate 0 0 S15 stroke grestore } bind def
/DiaE { stroke [] 0 setdash vpt add M
  hpt neg vpt neg V hpt vpt neg V
  hpt vpt V hpt neg vpt V closepath stroke } def
/BoxE { stroke [] 0 setdash exch hpt sub exch vpt add M
  0 vpt2 neg V hpt2 0 V 0 vpt2 V
  hpt2 neg 0 V closepath stroke } def
/TriUE { stroke [] 0 setdash vpt 1.12 mul add M
  hpt neg vpt -1.62 mul V
  hpt 2 mul 0 V
  hpt neg vpt 1.62 mul V closepath stroke } def
/TriDE { stroke [] 0 setdash vpt 1.12 mul sub M
  hpt neg vpt 1.62 mul V
  hpt 2 mul 0 V
  hpt neg vpt -1.62 mul V closepath stroke } def
/PentE { stroke [] 0 setdash gsave
  translate 0 hpt M 4 {72 rotate 0 hpt L} repeat
  closepath stroke grestore } def
/CircE { stroke [] 0 setdash 
  hpt 0 360 arc stroke } def
/BoxFill { gsave Rec 1 setgray fill grestore } def
end
}
\begin{picture}(3600,2160)(0,0)
\special{"
gnudict begin
gsave
0 0 translate
0.100 0.100 scale
0 setgray
newpath
LTb
350 400 M
63 0 V
3137 0 R
-63 0 V
350 953 M
63 0 V
3137 0 R
-63 0 V
350 1507 M
63 0 V
3137 0 R
-63 0 V
350 2060 M
63 0 V
3137 0 R
-63 0 V
347 400 M
0 63 V
0 1597 R
0 -63 V
987 400 M
0 63 V
0 1597 R
0 -63 V
1628 400 M
0 63 V
0 1597 R
0 -63 V
2269 400 M
0 63 V
0 1597 R
0 -63 V
2909 400 M
0 63 V
0 1597 R
0 -63 V
3550 400 M
0 63 V
0 1597 R
0 -63 V
LTb
350 400 M
3200 0 V
0 1660 V
-3200 0 V
350 400 L
1.000 UL
LT0
350 2057 M
3 -4 V
7 -9 V
6 -10 V
6 -9 V
7 -9 V
32 -42 V
32 -40 V
32 -37 V
32 -34 V
32 -33 V
32 -30 V
32 -29 V
32 -28 V
32 -26 V
64 -49 V
64 -44 V
64 -42 V
64 -38 V
64 -36 V
129 -66 V
128 -58 V
128 -53 V
128 -48 V
128 -44 V
320 -94 V
321 -77 V
320 -64 V
320 -54 V
321 -46 V
320 -39 V
1.000 UL
LT1
350 953 M
32 0 V
33 0 V
32 0 V
32 0 V
33 0 V
32 0 V
32 0 V
33 0 V
32 0 V
32 0 V
33 0 V
32 0 V
32 0 V
33 0 V
32 0 V
32 0 V
32 0 V
33 0 V
32 0 V
32 0 V
33 0 V
32 0 V
32 0 V
33 0 V
32 0 V
32 0 V
33 0 V
32 0 V
32 0 V
33 0 V
32 0 V
32 0 V
33 0 V
32 0 V
32 0 V
33 0 V
32 0 V
32 0 V
33 0 V
32 0 V
32 0 V
33 0 V
32 0 V
32 0 V
33 0 V
32 0 V
32 0 V
33 0 V
32 0 V
32 0 V
32 0 V
33 0 V
32 0 V
32 0 V
33 0 V
32 0 V
32 0 V
33 0 V
32 0 V
32 0 V
33 0 V
32 0 V
32 0 V
33 0 V
32 0 V
32 0 V
33 0 V
32 0 V
32 0 V
33 0 V
32 0 V
32 0 V
33 0 V
32 0 V
32 0 V
33 0 V
32 0 V
32 0 V
33 0 V
32 0 V
32 0 V
33 0 V
32 0 V
32 0 V
32 0 V
33 0 V
32 0 V
32 0 V
33 0 V
32 0 V
32 0 V
33 0 V
32 0 V
32 0 V
33 0 V
32 0 V
32 0 V
33 0 V
32 0 V
stroke
grestore
end
showpage
}
\put(1950,150){\makebox(0,0){$\lambda s_y$}}
\put(100,1230){\makebox(0,0)[b]{\shortstack{$\displaystyle{\sigma_\infty^{th} \over s_y}$}}}
\put(3550,300){\makebox(0,0){0.5}}
\put(2909,300){\makebox(0,0){0.4}}
\put(2269,300){\makebox(0,0){0.3}}
\put(1628,300){\makebox(0,0){0.2}}
\put(987,300){\makebox(0,0){0.1}}
\put(347,300){\makebox(0,0){0}}
\put(300,2060){\makebox(0,0)[r]{3}}
\put(300,1507){\makebox(0,0)[r]{2}}
\put(300,953){\makebox(0,0)[r]{1}}
\put(300,400){\makebox(0,0)[r]{0}}
\end{picture}

%% file: sim_omega.tex
\setlength{\unitlength}{0.1bp}
\special{!
/gnudict 120 dict def
gnudict begin
/Color false def
/Solid false def
/gnulinewidth 5.000 def
/userlinewidth gnulinewidth def
/vshift -33 def
/dl {10 mul} def
/hpt_ 31.5 def
/vpt_ 31.5 def
/hpt hpt_ def
/vpt vpt_ def
/M {moveto} bind def
/L {lineto} bind def
/R {rmoveto} bind def
/V {rlineto} bind def
/vpt2 vpt 2 mul def
/hpt2 hpt 2 mul def
/Lshow { currentpoint stroke M
  0 vshift R show } def
/Rshow { currentpoint stroke M
  dup stringwidth pop neg vshift R show } def
/Cshow { currentpoint stroke M
  dup stringwidth pop -2 div vshift R show } def
/UP { dup vpt_ mul /vpt exch def hpt_ mul /hpt exch def
  /hpt2 hpt 2 mul def /vpt2 vpt 2 mul def } def
/DL { Color {setrgbcolor Solid {pop []} if 0 setdash }
 {pop pop pop Solid {pop []} if 0 setdash} ifelse } def
/BL { stroke gnulinewidth 2 mul setlinewidth } def
/AL { stroke gnulinewidth 2 div setlinewidth } def
/UL { gnulinewidth mul /userlinewidth exch def } def
/PL { stroke userlinewidth setlinewidth } def
/LTb { BL [] 0 0 0 DL } def
/LTa { AL [1 dl 2 dl] 0 setdash 0 0 0 setrgbcolor } def
/LT0 { PL [] 0 1 0 DL } def
/LT1 { PL [4 dl 2 dl] 0 0 1 DL } def
/LT2 { PL [2 dl 3 dl] 1 0 0 DL } def
/LT3 { PL [1 dl 1.5 dl] 1 0 1 DL } def
/LT4 { PL [5 dl 2 dl 1 dl 2 dl] 0 1 1 DL } def
/LT5 { PL [4 dl 3 dl 1 dl 3 dl] 1 1 0 DL } def
/LT6 { PL [2 dl 2 dl 2 dl 4 dl] 0 0 0 DL } def
/LT7 { PL [2 dl 2 dl 2 dl 2 dl 2 dl 4 dl] 1 0.3 0 DL } def
/LT8 { PL [2 dl 2 dl 2 dl 2 dl 2 dl 2 dl 2 dl 4 dl] 0.5 0.5 0.5 DL } def
/Pnt { stroke [] 0 setdash
   gsave 1 setlinecap M 0 0 V stroke grestore } def
/Dia { stroke [] 0 setdash 2 copy vpt add M
  hpt neg vpt neg V hpt vpt neg V
  hpt vpt V hpt neg vpt V closepath stroke
  Pnt } def
/Pls { stroke [] 0 setdash vpt sub M 0 vpt2 V
  currentpoint stroke M
  hpt neg vpt neg R hpt2 0 V stroke
  } def
/Box { stroke [] 0 setdash 2 copy exch hpt sub exch vpt add M
  0 vpt2 neg V hpt2 0 V 0 vpt2 V
  hpt2 neg 0 V closepath stroke
  Pnt } def
/Crs { stroke [] 0 setdash exch hpt sub exch vpt add M
  hpt2 vpt2 neg V currentpoint stroke M
  hpt2 neg 0 R hpt2 vpt2 V stroke } def
/TriU { stroke [] 0 setdash 2 copy vpt 1.12 mul add M
  hpt neg vpt -1.62 mul V
  hpt 2 mul 0 V
  hpt neg vpt 1.62 mul V closepath stroke
  Pnt  } def
/Star { 2 copy Pls Crs } def
/BoxF { stroke [] 0 setdash exch hpt sub exch vpt add M
  0 vpt2 neg V  hpt2 0 V  0 vpt2 V
  hpt2 neg 0 V  closepath fill } def
/TriUF { stroke [] 0 setdash vpt 1.12 mul add M
  hpt neg vpt -1.62 mul V
  hpt 2 mul 0 V
  hpt neg vpt 1.62 mul V closepath fill } def
/TriD { stroke [] 0 setdash 2 copy vpt 1.12 mul sub M
  hpt neg vpt 1.62 mul V
  hpt 2 mul 0 V
  hpt neg vpt -1.62 mul V closepath stroke
  Pnt  } def
/TriDF { stroke [] 0 setdash vpt 1.12 mul sub M
  hpt neg vpt 1.62 mul V
  hpt 2 mul 0 V
  hpt neg vpt -1.62 mul V closepath fill} def
/DiaF { stroke [] 0 setdash vpt add M
  hpt neg vpt neg V hpt vpt neg V
  hpt vpt V hpt neg vpt V closepath fill } def
/Pent { stroke [] 0 setdash 2 copy gsave
  translate 0 hpt M 4 {72 rotate 0 hpt L} repeat
  closepath stroke grestore Pnt } def
/PentF { stroke [] 0 setdash gsave
  translate 0 hpt M 4 {72 rotate 0 hpt L} repeat
  closepath fill grestore } def
/Circle { stroke [] 0 setdash 2 copy
  hpt 0 360 arc stroke Pnt } def
/CircleF { stroke [] 0 setdash hpt 0 360 arc fill } def
/C0 { BL [] 0 setdash 2 copy moveto vpt 90 450  arc } bind def
/C1 { BL [] 0 setdash 2 copy        moveto
       2 copy  vpt 0 90 arc closepath fill
               vpt 0 360 arc closepath } bind def
/C2 { BL [] 0 setdash 2 copy moveto
       2 copy  vpt 90 180 arc closepath fill
               vpt 0 360 arc closepath } bind def
/C3 { BL [] 0 setdash 2 copy moveto
       2 copy  vpt 0 180 arc closepath fill
               vpt 0 360 arc closepath } bind def
/C4 { BL [] 0 setdash 2 copy moveto
       2 copy  vpt 180 270 arc closepath fill
               vpt 0 360 arc closepath } bind def
/C5 { BL [] 0 setdash 2 copy moveto
       2 copy  vpt 0 90 arc
       2 copy moveto
       2 copy  vpt 180 270 arc closepath fill
               vpt 0 360 arc } bind def
/C6 { BL [] 0 setdash 2 copy moveto
      2 copy  vpt 90 270 arc closepath fill
              vpt 0 360 arc closepath } bind def
/C7 { BL [] 0 setdash 2 copy moveto
      2 copy  vpt 0 270 arc closepath fill
              vpt 0 360 arc closepath } bind def
/C8 { BL [] 0 setdash 2 copy moveto
      2 copy vpt 270 360 arc closepath fill
              vpt 0 360 arc closepath } bind def
/C9 { BL [] 0 setdash 2 copy moveto
      2 copy  vpt 270 450 arc closepath fill
              vpt 0 360 arc closepath } bind def
/C10 { BL [] 0 setdash 2 copy 2 copy moveto vpt 270 360 arc closepath fill
       2 copy moveto
       2 copy vpt 90 180 arc closepath fill
               vpt 0 360 arc closepath } bind def
/C11 { BL [] 0 setdash 2 copy moveto
       2 copy  vpt 0 180 arc closepath fill
       2 copy moveto
       2 copy  vpt 270 360 arc closepath fill
               vpt 0 360 arc closepath } bind def
/C12 { BL [] 0 setdash 2 copy moveto
       2 copy  vpt 180 360 arc closepath fill
               vpt 0 360 arc closepath } bind def
/C13 { BL [] 0 setdash  2 copy moveto
       2 copy  vpt 0 90 arc closepath fill
       2 copy moveto
       2 copy  vpt 180 360 arc closepath fill
               vpt 0 360 arc closepath } bind def
/C14 { BL [] 0 setdash 2 copy moveto
       2 copy  vpt 90 360 arc closepath fill
               vpt 0 360 arc } bind def
/C15 { BL [] 0 setdash 2 copy vpt 0 360 arc closepath fill
               vpt 0 360 arc closepath } bind def
/Rec   { newpath 4 2 roll moveto 1 index 0 rlineto 0 exch rlineto
       neg 0 rlineto closepath } bind def
/Square { dup Rec } bind def
/Bsquare { vpt sub exch vpt sub exch vpt2 Square } bind def
/S0 { BL [] 0 setdash 2 copy moveto 0 vpt rlineto BL Bsquare } bind def
/S1 { BL [] 0 setdash 2 copy vpt Square fill Bsquare } bind def
/S2 { BL [] 0 setdash 2 copy exch vpt sub exch vpt Square fill Bsquare } bind def
/S3 { BL [] 0 setdash 2 copy exch vpt sub exch vpt2 vpt Rec fill Bsquare } bind def
/S4 { BL [] 0 setdash 2 copy exch vpt sub exch vpt sub vpt Square fill Bsquare } bind def
/S5 { BL [] 0 setdash 2 copy 2 copy vpt Square fill
       exch vpt sub exch vpt sub vpt Square fill Bsquare } bind def
/S6 { BL [] 0 setdash 2 copy exch vpt sub exch vpt sub vpt vpt2 Rec fill Bsquare } bind def
/S7 { BL [] 0 setdash 2 copy exch vpt sub exch vpt sub vpt vpt2 Rec fill
       2 copy vpt Square fill
       Bsquare } bind def
/S8 { BL [] 0 setdash 2 copy vpt sub vpt Square fill Bsquare } bind def
/S9 { BL [] 0 setdash 2 copy vpt sub vpt vpt2 Rec fill Bsquare } bind def
/S10 { BL [] 0 setdash 2 copy vpt sub vpt Square fill 2 copy exch vpt sub exch vpt Square fill
       Bsquare } bind def
/S11 { BL [] 0 setdash 2 copy vpt sub vpt Square fill 2 copy exch vpt sub exch vpt2 vpt Rec fill
       Bsquare } bind def
/S12 { BL [] 0 setdash 2 copy exch vpt sub exch vpt sub vpt2 vpt Rec fill Bsquare } bind def
/S13 { BL [] 0 setdash 2 copy exch vpt sub exch vpt sub vpt2 vpt Rec fill
       2 copy vpt Square fill Bsquare } bind def
/S14 { BL [] 0 setdash 2 copy exch vpt sub exch vpt sub vpt2 vpt Rec fill
       2 copy exch vpt sub exch vpt Square fill Bsquare } bind def
/S15 { BL [] 0 setdash 2 copy Bsquare fill Bsquare } bind def
/D0 { gsave translate 45 rotate 0 0 S0 stroke grestore } bind def
/D1 { gsave translate 45 rotate 0 0 S1 stroke grestore } bind def
/D2 { gsave translate 45 rotate 0 0 S2 stroke grestore } bind def
/D3 { gsave translate 45 rotate 0 0 S3 stroke grestore } bind def
/D4 { gsave translate 45 rotate 0 0 S4 stroke grestore } bind def
/D5 { gsave translate 45 rotate 0 0 S5 stroke grestore } bind def
/D6 { gsave translate 45 rotate 0 0 S6 stroke grestore } bind def
/D7 { gsave translate 45 rotate 0 0 S7 stroke grestore } bind def
/D8 { gsave translate 45 rotate 0 0 S8 stroke grestore } bind def
/D9 { gsave translate 45 rotate 0 0 S9 stroke grestore } bind def
/D10 { gsave translate 45 rotate 0 0 S10 stroke grestore } bind def
/D11 { gsave translate 45 rotate 0 0 S11 stroke grestore } bind def
/D12 { gsave translate 45 rotate 0 0 S12 stroke grestore } bind def
/D13 { gsave translate 45 rotate 0 0 S13 stroke grestore } bind def
/D14 { gsave translate 45 rotate 0 0 S14 stroke grestore } bind def
/D15 { gsave translate 45 rotate 0 0 S15 stroke grestore } bind def
/DiaE { stroke [] 0 setdash vpt add M
  hpt neg vpt neg V hpt vpt neg V
  hpt vpt V hpt neg vpt V closepath stroke } def
/BoxE { stroke [] 0 setdash exch hpt sub exch vpt add M
  0 vpt2 neg V hpt2 0 V 0 vpt2 V
  hpt2 neg 0 V closepath stroke } def
/TriUE { stroke [] 0 setdash vpt 1.12 mul add M
  hpt neg vpt -1.62 mul V
  hpt 2 mul 0 V
  hpt neg vpt 1.62 mul V closepath stroke } def
/TriDE { stroke [] 0 setdash vpt 1.12 mul sub M
  hpt neg vpt 1.62 mul V
  hpt 2 mul 0 V
  hpt neg vpt -1.62 mul V closepath stroke } def
/PentE { stroke [] 0 setdash gsave
  translate 0 hpt M 4 {72 rotate 0 hpt L} repeat
  closepath stroke grestore } def
/CircE { stroke [] 0 setdash 
  hpt 0 360 arc stroke } def
/BoxFill { gsave Rec 1 setgray fill grestore } def
end
}
\begin{picture}(3600,2160)(0,0)
\special{"
gnudict begin
gsave
0 0 translate
0.100 0.100 scale
0 setgray
newpath
LTb
450 400 M
63 0 V
3037 0 R
-63 0 V
450 1064 M
63 0 V
3037 0 R
-63 0 V
450 1728 M
63 0 V
3037 0 R
-63 0 V
450 400 M
0 63 V
0 1597 R
0 -63 V
1225 400 M
0 63 V
0 1597 R
0 -63 V
2000 400 M
0 63 V
0 1597 R
0 -63 V
2775 400 M
0 63 V
0 1597 R
0 -63 V
3550 400 M
0 63 V
0 1597 R
0 -63 V
LTb
450 400 M
3100 0 V
0 1660 V
-3100 0 V
450 400 L
1.000 UL
LT0
2125 1728 M
263 0 V
3550 584 M
-19 2 V
-19 2 V
-19 1 V
-19 2 V
-19 2 V
-18 2 V
-18 2 V
-18 2 V
-18 2 V
-18 2 V
-17 2 V
-17 2 V
-18 1 V
-17 2 V
-16 2 V
-17 2 V
-17 2 V
-16 2 V
-16 2 V
-16 2 V
-16 2 V
-16 2 V
-16 2 V
-15 2 V
-16 2 V
-15 3 V
-15 2 V
-15 2 V
-15 2 V
-14 2 V
-15 2 V
-14 2 V
-15 2 V
-14 2 V
-14 2 V
-14 3 V
-14 2 V
-14 2 V
-13 2 V
-14 2 V
-13 2 V
-14 3 V
-13 2 V
-13 2 V
-13 2 V
-13 3 V
-12 2 V
-13 2 V
-13 2 V
-12 3 V
-12 2 V
-13 2 V
-12 2 V
-12 3 V
-12 2 V
-12 2 V
-11 3 V
-12 2 V
-12 2 V
-11 3 V
-12 2 V
-11 2 V
-11 3 V
-11 2 V
-11 2 V
-11 3 V
-11 2 V
-11 3 V
-11 2 V
-11 3 V
-10 2 V
-11 2 V
-10 3 V
-11 2 V
-10 3 V
-10 2 V
-10 3 V
-10 2 V
-10 3 V
-10 2 V
-10 3 V
-10 2 V
-9 3 V
-10 2 V
-10 3 V
-9 2 V
-10 3 V
-9 3 V
-9 2 V
-10 3 V
-9 2 V
-9 3 V
-9 3 V
-9 2 V
-9 3 V
-9 2 V
-9 3 V
-8 3 V
-9 2 V
-9 3 V
-8 3 V
-9 2 V
-8 3 V
-9 3 V
-8 2 V
-8 3 V
-9 3 V
-8 2 V
-8 3 V
-8 3 V
-8 3 V
-8 2 V
-8 3 V
-8 3 V
-8 2 V
-7 3 V
-8 3 V
-8 3 V
-8 2 V
-7 3 V
-8 3 V
-7 3 V
-8 3 V
-7 2 V
-7 3 V
-8 3 V
-7 3 V
-7 3 V
-7 2 V
-8 3 V
-7 3 V
-7 3 V
-7 3 V
-7 2 V
-7 3 V
-6 3 V
-7 3 V
-7 3 V
-7 3 V
-7 2 V
-6 3 V
-7 3 V
-7 3 V
-6 3 V
-7 3 V
-6 2 V
-7 3 V
-6 3 V
-6 3 V
-7 3 V
-6 3 V
-6 2 V
-7 3 V
-6 3 V
-6 3 V
-6 3 V
-6 2 V
-6 3 V
-6 3 V
-6 3 V
-6 2 V
-6 3 V
-6 3 V
-6 3 V
-6 2 V
-6 3 V
-5 3 V
-6 2 V
-6 3 V
-5 3 V
-6 2 V
-6 3 V
-5 2 V
-6 3 V
-5 2 V
-6 3 V
-5 2 V
-6 2 V
-5 3 V
-6 2 V
-5 2 V
-5 2 V
-6 3 V
-5 2 V
-5 2 V
-5 2 V
-5 1 V
-6 2 V
-5 2 V
-5 2 V
-5 1 V
-5 2 V
-5 1 V
-5 2 V
-5 1 V
-5 1 V
-5 1 V
-5 1 V
-5 1 V
-4 1 V
-5 0 V
-5 1 V
-5 1 V
-5 0 V
-4 1 V
-5 0 V
-5 0 V
-4 1 V
-5 0 V
-5 0 V
-4 0 V
-5 0 V
-4 1 V
-5 0 V
-5 0 V
-4 0 V
-4 0 V
-5 0 V
-4 0 V
-5 0 V
-4 0 V
-5 0 V
-4 0 V
-4 0 V
-4 0 V
-5 0 V
-4 0 V
-4 0 V
-4 0 V
-5 0 V
-4 0 V
-4 0 V
-4 0 V
-4 0 V
-4 0 V
-4 0 V
-5 0 V
-4 0 V
-4 0 V
-4 0 V
-4 0 V
-4 0 V
-4 0 V
-4 0 V
-3 0 V
-4 0 V
-4 0 V
-4 0 V
-4 0 V
-4 0 V
-4 0 V
-3 0 V
-4 0 V
-4 0 V
-4 0 V
-3 0 V
-4 0 V
-4 0 V
-4 0 V
-3 0 V
-4 1 V
-4 0 V
-3 0 V
-4 0 V
-3 0 V
-4 0 V
-3 0 V
-4 0 V
-4 0 V
-3 0 V
-4 0 V
-3 0 V
-4 0 V
-3 0 V
-3 0 V
-4 0 V
-3 0 V
-4 0 V
-3 0 V
-3 0 V
-4 0 V
-3 0 V
-3 0 V
-4 0 V
-3 0 V
-3 0 V
-4 0 V
-3 0 V
-3 0 V
-3 0 V
-4 0 V
-3 0 V
-3 0 V
-3 0 V
-3 0 V
-3 0 V
-4 0 V
-3 0 V
-3 0 V
-3 0 V
-3 0 V
-3 0 V
-3 0 V
-3 0 V
-3 0 V
-3 0 V
-3 0 V
-3 0 V
-3 0 V
-3 0 V
-3 0 V
-3 0 V
-3 0 V
-3 0 V
-3 0 V
-3 0 V
-3 0 V
-3 0 V
-3 0 V
-3 0 V
-2 0 V
-3 0 V
-3 0 V
-3 0 V
-3 0 V
-3 0 V
-2 0 V
-3 0 V
-3 0 V
-3 0 V
-2 0 V
-3 0 V
-3 0 V
-3 0 V
-2 0 V
-3 0 V
-3 0 V
-2 0 V
-3 0 V
-3 0 V
-2 0 V
-3 0 V
-3 0 V
-2 0 V
-3 0 V
-3 0 V
-2 0 V
-3 0 V
-2 0 V
-3 0 V
-2 0 V
-3 0 V
-3 0 V
-2 0 V
-3 0 V
-2 0 V
-3 0 V
-2 0 V
-3 0 V
-2 0 V
-3 0 V
-2 0 V
-2 0 V
-3 0 V
-2 0 V
-3 0 V
-2 0 V
-3 0 V
-2 0 V
-2 0 V
-3 0 V
-2 0 V
-2 0 V
-3 0 V
-2 0 V
-3 0 V
-2 0 V
-2 0 V
-2 0 V
-3 0 V
-2 0 V
-2 0 V
-3 0 V
-2 0 V
-2 0 V
-2 0 V
-3 0 V
-2 0 V
-2 0 V
-2 0 V
-3 0 V
-2 0 V
-2 0 V
-2 0 V
-2 0 V
-3 0 V
-2 0 V
currentpoint stroke M
-2 1 V
-2 0 V
-2 0 V
-2 0 V
-3 0 V
-2 0 V
-2 0 V
-2 0 V
-2 0 V
-2 0 V
-2 0 V
-2 0 V
-3 0 V
-2 0 V
-2 0 V
-2 0 V
-2 0 V
-2 0 V
-2 0 V
-2 0 V
-2 0 V
-2 0 V
-2 0 V
-2 0 V
-2 0 V
-2 0 V
-2 0 V
-2 0 V
-2 0 V
-2 0 V
-2 0 V
-2 0 V
-2 0 V
-2 0 V
-2 0 V
-2 0 V
-2 0 V
-2 0 V
-1 0 V
-2 0 V
-2 0 V
-2 0 V
-2 0 V
-2 0 V
-2 0 V
-2 0 V
-2 0 V
-1 0 V
-2 0 V
-2 0 V
-2 0 V
-2 0 V
-2 0 V
-2 0 V
-1 0 V
-2 0 V
-2 0 V
-2 0 V
-2 0 V
-1 0 V
-2 0 V
-2 0 V
-2 0 V
-2 0 V
-1 0 V
-2 0 V
-2 0 V
-2 0 V
-1 0 V
-2 0 V
-2 0 V
-2 0 V
-1 0 V
-2 0 V
-2 0 V
-1 0 V
-2 0 V
-2 0 V
-2 0 V
-1 0 V
-2 0 V
-2 0 V
-1 0 V
-2 0 V
-2 0 V
-1 0 V
-2 0 V
-2 0 V
-1 0 V
-2 0 V
-2 0 V
-1 0 V
-2 0 V
-1 0 V
-2 0 V
-2 0 V
-1 0 V
-2 0 V
-1 0 V
-2 0 V
-2 0 V
-1 0 V
-2 0 V
-1 0 V
-2 0 V
-2 0 V
-1 0 V
-2 0 V
-1 0 V
-2 0 V
-1 1 V
-2 0 V
-1 0 V
-2 0 V
-2 0 V
-1 0 V
-2 0 V
-1 0 V
-2 0 V
-1 0 V
-2 0 V
-1 0 V
-2 0 V
-1 0 V
-2 0 V
-1 0 V
-2 0 V
-1 0 V
-2 0 V
-1 0 V
-1 0 V
-2 0 V
-1 0 V
-2 0 V
-1 0 V
-2 0 V
-1 0 V
-2 0 V
-1 0 V
-1 0 V
-2 0 V
-1 0 V
-2 0 V
-1 0 V
-2 0 V
-1 0 V
-1 0 V
-2 0 V
-1 0 V
-2 0 V
-1 0 V
-1 0 V
-2 0 V
-1 0 V
-1 0 V
-2 0 V
-1 0 V
-2 0 V
-1 0 V
-1 0 V
-2 0 V
-1 0 V
-1 0 V
-2 0 V
-1 0 V
-1 0 V
-2 0 V
-1 0 V
-1 0 V
-2 0 V
-1 0 V
-1 0 V
-2 0 V
-1 0 V
-1 0 V
-1 0 V
-2 0 V
-1 0 V
-1 0 V
-2 0 V
-1 0 V
-1 0 V
-1 0 V
-2 0 V
-1 0 V
-1 0 V
-1 0 V
-2 0 V
-1 0 V
-1 0 V
-1 0 V
-2 0 V
-1 0 V
-1 0 V
-1 0 V
-2 0 V
-1 0 V
-1 0 V
-1 0 V
-2 0 V
-1 0 V
-1 0 V
-1 0 V
-1 0 V
-2 0 V
-1 0 V
-1 1 V
-1 0 V
-1 0 V
-2 0 V
-1 0 V
-1 0 V
-1 0 V
-1 0 V
-2 0 V
-1 0 V
-1 0 V
-1 0 V
-1 0 V
-1 0 V
-2 0 V
-1 0 V
-1 0 V
-1 0 V
-1 0 V
-1 0 V
-1 0 V
-2 0 V
-1 0 V
-1 0 V
-1 0 V
-1 0 V
-1 0 V
-1 0 V
-2 0 V
-1 0 V
-1 0 V
-1 0 V
-1 0 V
-1 0 V
-1 0 V
-1 0 V
-1 0 V
-1 0 V
-2 0 V
-1 0 V
-1 0 V
-1 0 V
-1 0 V
-1 0 V
-1 0 V
-1 0 V
-1 0 V
-1 0 V
-1 0 V
-1 0 V
-2 0 V
-1 0 V
-1 0 V
-1 0 V
-1 0 V
-1 0 V
-1 0 V
-1 0 V
-1 0 V
-1 0 V
-1 0 V
-1 0 V
-1 0 V
-1 0 V
-1 0 V
-1 0 V
-1 0 V
-1 0 V
-1 0 V
-1 0 V
-1 0 V
-1 0 V
-1 0 V
-1 0 V
-1 0 V
-1 0 V
-1 0 V
-1 0 V
-1 0 V
-1 0 V
-1 0 V
-1 0 V
-1 0 V
-1 0 V
-1 0 V
-1 1 V
-1 0 V
-1 0 V
-1 0 V
-1 0 V
-1 0 V
-1 0 V
-1 0 V
-1 0 V
-1 0 V
-1 0 V
-1 0 V
-1 0 V
-1 0 V
-1 0 V
-1 0 V
-1 0 V
-1 0 V
-1 0 V
-1 0 V
-1 0 V
-1 0 V
-1 0 V
-1 0 V
-1 0 V
-1 0 V
-1 0 V
-1 0 V
-1 0 V
-1 0 V
-1 0 V
-1 0 V
-1 0 V
-1 0 V
-1 0 V
-1 0 V
-1 0 V
-1 0 V
-1 0 V
-1 0 V
-1 0 V
-1 0 V
-1 0 V
-1 0 V
-1 0 V
-1 0 V
-1 0 V
-1 0 V
-1 0 V
-1 0 V
-1 0 V
-1 0 V
-1 0 V
-1 0 V
-1 0 V
-1 0 V
-1 0 V
-1 0 V
-1 0 V
-1 0 V
-1 0 V
-1 0 V
-1 0 V
-1 0 V
-1 0 V
-1 0 V
-1 0 V
-1 0 V
-1 0 V
-1 0 V
-1 1 V
-1 0 V
-1 0 V
-1 0 V
-1 0 V
-1 0 V
-1 0 V
-1 0 V
-1 0 V
-1 0 V
-1 0 V
-1 0 V
-1 0 V
-1 0 V
-1 0 V
-1 0 V
-1 0 V
-1 0 V
-1 0 V
-1 0 V
-1 0 V
-1 0 V
-1 0 V
-1 0 V
-1 0 V
-1 0 V
-1 0 V
-1 0 V
1.000 UL
LT1
2125 1628 M
263 0 V
3550 588 M
-19 2 V
-19 2 V
-19 2 V
-19 2 V
-19 2 V
-18 2 V
-18 2 V
-18 1 V
-18 2 V
-18 2 V
-17 2 V
-17 2 V
-18 2 V
-17 2 V
-16 2 V
-17 2 V
-17 2 V
-16 2 V
-16 2 V
-16 2 V
-16 3 V
-16 2 V
-16 2 V
-15 2 V
-16 2 V
-15 2 V
-15 2 V
-15 2 V
-15 2 V
-14 2 V
-15 3 V
-14 2 V
-15 2 V
-14 2 V
-14 2 V
-14 3 V
-14 2 V
-14 2 V
-13 2 V
-14 2 V
-13 3 V
-14 2 V
-13 2 V
-13 2 V
-13 3 V
-13 2 V
-12 2 V
-13 3 V
-13 2 V
-12 2 V
-12 2 V
-13 3 V
-12 2 V
-12 3 V
-12 2 V
-12 2 V
-11 3 V
-12 2 V
-12 2 V
-11 3 V
-12 2 V
-11 3 V
-11 2 V
-11 2 V
-11 3 V
-11 2 V
-11 3 V
-11 2 V
-11 3 V
-11 2 V
-10 3 V
-11 2 V
-10 3 V
-11 2 V
-10 3 V
-10 2 V
-10 3 V
-10 2 V
-10 3 V
-10 3 V
-10 2 V
-10 3 V
-9 2 V
-10 3 V
-10 3 V
-9 2 V
-10 3 V
-9 2 V
-9 3 V
-10 3 V
-9 2 V
-9 3 V
-9 3 V
-9 2 V
-9 3 V
-9 3 V
-9 2 V
-8 3 V
-9 3 V
-9 2 V
-8 3 V
-9 3 V
-8 3 V
-9 2 V
-8 3 V
-8 3 V
-9 3 V
-8 2 V
-8 3 V
-8 3 V
-8 3 V
-8 3 V
-8 2 V
-8 3 V
-8 3 V
-7 3 V
-8 3 V
-8 2 V
-8 3 V
-7 3 V
-8 3 V
-7 3 V
-8 3 V
-7 3 V
-7 2 V
-8 3 V
-7 3 V
-7 3 V
-7 3 V
-8 3 V
-7 3 V
-7 3 V
-7 2 V
-7 3 V
-7 3 V
-6 3 V
-7 3 V
-7 3 V
-7 3 V
-7 3 V
-6 3 V
-7 3 V
-7 3 V
-6 2 V
-7 3 V
-6 3 V
-7 3 V
-6 3 V
-6 3 V
-7 3 V
-6 3 V
-6 3 V
-7 3 V
-6 3 V
-6 3 V
-6 2 V
-6 3 V
-6 3 V
-6 3 V
-6 3 V
-6 3 V
-6 3 V
-6 3 V
-6 3 V
-6 2 V
-6 3 V
-5 3 V
-6 3 V
-6 3 V
-5 3 V
-6 2 V
-6 3 V
-5 3 V
-6 3 V
-5 2 V
-6 3 V
-5 3 V
-6 2 V
-5 3 V
-6 3 V
-5 2 V
-5 3 V
-6 2 V
-5 3 V
-5 3 V
-5 2 V
-5 2 V
-6 3 V
-5 2 V
-5 3 V
-5 2 V
-5 2 V
-5 2 V
-5 3 V
-5 2 V
-5 2 V
-5 2 V
-5 2 V
-5 2 V
-4 2 V
-5 2 V
-5 2 V
-5 1 V
-5 2 V
-4 2 V
-5 2 V
-5 1 V
-4 2 V
-5 1 V
-5 2 V
-4 1 V
-5 1 V
-4 2 V
-5 1 V
-5 1 V
-4 1 V
-4 1 V
-5 1 V
-4 1 V
-5 1 V
-4 1 V
-5 1 V
-4 1 V
-4 0 V
-4 1 V
-5 1 V
-4 0 V
-4 1 V
-4 1 V
-5 0 V
-4 1 V
-4 0 V
-4 1 V
-4 0 V
-4 1 V
-4 0 V
-5 1 V
-4 0 V
-4 0 V
-4 1 V
-4 0 V
-4 1 V
-4 0 V
-4 0 V
-3 1 V
-4 0 V
-4 0 V
-4 1 V
-4 0 V
-4 0 V
-4 0 V
-3 1 V
-4 0 V
-4 0 V
-4 1 V
-3 0 V
-4 0 V
-4 1 V
-4 0 V
-3 0 V
-4 1 V
-4 0 V
-3 0 V
-4 0 V
-3 1 V
-4 0 V
-3 0 V
-4 1 V
-4 0 V
-3 0 V
-4 1 V
-3 0 V
-4 0 V
-3 1 V
-3 0 V
-4 0 V
-3 1 V
-4 0 V
-3 0 V
-3 1 V
-4 0 V
-3 0 V
-3 1 V
-4 0 V
-3 0 V
-3 1 V
-4 0 V
-3 0 V
-3 1 V
-3 0 V
-4 0 V
-3 1 V
-3 0 V
-3 0 V
-3 1 V
-3 0 V
-4 0 V
-3 1 V
-3 0 V
-3 0 V
-3 1 V
-3 0 V
-3 1 V
-3 0 V
-3 0 V
-3 1 V
-3 0 V
-3 0 V
-3 1 V
-3 0 V
-3 0 V
-3 1 V
-3 0 V
-3 1 V
-3 0 V
-3 0 V
-3 1 V
-3 0 V
-3 0 V
-3 1 V
-2 0 V
-3 1 V
-3 0 V
-3 0 V
-3 1 V
-3 0 V
-2 1 V
-3 0 V
-3 0 V
-3 1 V
-2 0 V
-3 1 V
-3 0 V
-3 0 V
-2 1 V
-3 0 V
-3 0 V
-2 1 V
-3 0 V
-3 1 V
-2 0 V
-3 0 V
-3 1 V
-2 0 V
-3 1 V
-3 0 V
-2 0 V
-3 1 V
-2 0 V
-3 1 V
-2 0 V
-3 1 V
-3 0 V
-2 0 V
-3 1 V
-2 0 V
-3 1 V
-2 0 V
-3 0 V
-2 1 V
-3 0 V
-2 1 V
-2 0 V
-3 1 V
-2 0 V
-3 0 V
-2 1 V
-3 0 V
-2 1 V
-2 0 V
-3 0 V
-2 1 V
-2 0 V
-3 1 V
-2 0 V
-3 1 V
-2 0 V
-2 1 V
-2 0 V
-3 0 V
-2 1 V
-2 0 V
-3 1 V
-2 0 V
-2 1 V
-2 0 V
-3 0 V
-2 1 V
-2 0 V
-2 1 V
-3 0 V
-2 1 V
-2 0 V
-2 1 V
-2 0 V
-3 0 V
-2 1 V
currentpoint stroke M
-2 0 V
-2 1 V
-2 0 V
-2 1 V
-3 0 V
-2 1 V
-2 0 V
-2 1 V
-2 0 V
-2 0 V
-2 1 V
-2 0 V
-3 1 V
-2 0 V
-2 1 V
-2 0 V
-2 1 V
-2 0 V
-2 1 V
-2 0 V
-2 1 V
-2 0 V
-2 1 V
-2 0 V
-2 1 V
-2 0 V
-2 0 V
-2 1 V
-2 0 V
-2 1 V
-2 0 V
-2 1 V
-2 0 V
-2 1 V
-2 0 V
-2 1 V
-2 0 V
-2 1 V
-1 0 V
-2 1 V
-2 0 V
-2 1 V
-2 0 V
-2 1 V
-2 0 V
-2 1 V
-2 0 V
-1 1 V
-2 0 V
-2 1 V
-2 0 V
-2 1 V
-2 0 V
-2 1 V
-1 0 V
-2 1 V
-2 0 V
-2 1 V
-2 0 V
-1 1 V
-2 0 V
-2 1 V
-2 0 V
-2 1 V
-1 0 V
-2 1 V
-2 0 V
-2 1 V
-1 0 V
-2 1 V
-2 0 V
-2 1 V
-1 0 V
-2 1 V
-2 1 V
-1 0 V
-2 1 V
-2 0 V
-2 1 V
-1 0 V
-2 1 V
-2 0 V
-1 1 V
-2 0 V
-2 1 V
-1 0 V
-2 1 V
-2 0 V
-1 1 V
-2 1 V
-2 0 V
-1 1 V
-2 0 V
-1 1 V
-2 0 V
-2 1 V
-1 0 V
-2 1 V
-1 0 V
-2 1 V
-2 1 V
-1 0 V
-2 1 V
-1 0 V
-2 1 V
-2 0 V
-1 1 V
-2 0 V
-1 1 V
-2 1 V
-1 0 V
-2 1 V
-1 0 V
-2 1 V
-2 0 V
-1 1 V
-2 0 V
-1 1 V
-2 1 V
-1 0 V
-2 1 V
-1 0 V
-2 1 V
-1 0 V
-2 1 V
-1 1 V
-2 0 V
-1 1 V
-2 0 V
-1 1 V
-1 1 V
-2 0 V
-1 1 V
-2 0 V
-1 1 V
-2 0 V
-1 1 V
-2 1 V
-1 0 V
-1 1 V
-2 0 V
-1 1 V
-2 1 V
-1 0 V
-2 1 V
-1 0 V
-1 1 V
-2 1 V
-1 0 V
-2 1 V
-1 0 V
-1 1 V
-2 1 V
-1 0 V
-1 1 V
-2 0 V
-1 1 V
-2 1 V
-1 0 V
-1 1 V
-2 0 V
-1 1 V
-1 1 V
-2 0 V
-1 1 V
-1 1 V
-2 0 V
-1 1 V
-1 0 V
-2 1 V
-1 1 V
-1 0 V
-2 1 V
-1 1 V
-1 0 V
-1 1 V
-2 0 V
-1 1 V
-1 1 V
-2 0 V
-1 1 V
-1 1 V
-1 0 V
-2 1 V
-1 1 V
-1 0 V
-1 1 V
-2 1 V
-1 0 V
-1 1 V
-1 0 V
-2 1 V
-1 1 V
-1 0 V
-1 1 V
-2 1 V
-1 0 V
-1 1 V
-1 1 V
-2 0 V
-1 1 V
-1 1 V
-1 0 V
-1 1 V
-2 1 V
-1 0 V
-1 1 V
-1 1 V
-1 0 V
-2 1 V
-1 1 V
-1 0 V
-1 1 V
-1 1 V
-2 0 V
-1 1 V
-1 1 V
-1 0 V
-1 1 V
-1 1 V
-2 0 V
-1 1 V
-1 1 V
-1 0 V
-1 1 V
-1 1 V
-1 1 V
-2 0 V
-1 1 V
-1 1 V
-1 0 V
-1 1 V
-1 1 V
-1 0 V
-2 1 V
-1 1 V
-1 0 V
-1 1 V
-1 1 V
-1 1 V
-1 0 V
-1 1 V
-1 1 V
-1 0 V
-2 1 V
-1 1 V
-1 1 V
-1 0 V
-1 1 V
-1 1 V
-1 0 V
-1 1 V
-1 1 V
-1 1 V
-1 0 V
-1 1 V
-2 1 V
-1 0 V
-1 1 V
-1 1 V
-1 1 V
-1 0 V
-1 1 V
-1 1 V
-1 1 V
-1 0 V
-1 1 V
-1 1 V
-1 0 V
-1 1 V
-1 1 V
-1 1 V
-1 0 V
-1 1 V
-1 1 V
-1 1 V
-1 0 V
-1 1 V
-1 1 V
-1 1 V
-1 0 V
-1 1 V
-1 1 V
-1 1 V
-1 0 V
-1 1 V
-1 1 V
-1 1 V
-1 0 V
-1 1 V
-1 1 V
-1 1 V
-1 0 V
-1 1 V
-1 1 V
-1 1 V
-1 1 V
-1 0 V
-1 1 V
-1 1 V
-1 1 V
-1 0 V
-1 1 V
-1 1 V
-1 1 V
-1 1 V
-1 0 V
-1 1 V
-1 1 V
-1 1 V
-1 1 V
-1 1 V
-1 1 V
-1 1 V
-1 0 V
-1 1 V
-1 1 V
-1 1 V
-1 1 V
-1 0 V
-1 1 V
-1 1 V
-1 1 V
0 1 V
-1 0 V
-1 1 V
-1 1 V
-1 1 V
-1 1 V
-1 0 V
-1 1 V
-1 1 V
-1 1 V
0 1 V
-1 1 V
-1 0 V
-1 1 V
-1 1 V
-1 1 V
-1 1 V
-1 0 V
-1 1 V
0 1 V
-1 1 V
-1 1 V
-1 1 V
-1 0 V
-1 1 V
-1 1 V
0 1 V
-1 1 V
-1 1 V
-1 0 V
-1 1 V
-1 1 V
-1 1 V
0 1 V
-1 1 V
-1 1 V
-1 0 V
-1 1 V
-1 1 V
-1 1 V
0 1 V
-1 1 V
-1 1 V
-1 0 V
-1 1 V
-1 1 V
0 1 V
-1 1 V
-1 1 V
-1 1 V
-1 0 V
0 1 V
-1 1 V
-1 1 V
-1 1 V
-1 1 V
-1 1 V
0 1 V
-1 0 V
-1 1 V
-1 1 V
-1 1 V
0 1 V
-1 1 V
-1 1 V
-1 1 V
-1 0 V
0 1 V
-1 1 V
-1 1 V
-1 1 V
-1 1 V
0 1 V
-1 1 V
-1 1 V
-1 1 V
-1 1 V
currentpoint stroke M
-1 1 V
1.000 UL
LT2
2125 1528 M
263 0 V
3550 593 M
-19 2 V
-19 2 V
-19 2 V
-19 2 V
-19 1 V
-18 2 V
-18 2 V
-18 2 V
-18 2 V
-18 2 V
-17 2 V
-17 2 V
-18 3 V
-17 2 V
-16 2 V
-17 2 V
-17 2 V
-16 2 V
-16 2 V
-16 2 V
-16 2 V
-16 2 V
-16 2 V
-15 3 V
-16 2 V
-15 2 V
-15 2 V
-15 2 V
-15 2 V
-14 3 V
-15 2 V
-14 2 V
-15 2 V
-14 2 V
-14 3 V
-14 2 V
-14 2 V
-14 2 V
-13 3 V
-14 2 V
-13 2 V
-14 3 V
-13 2 V
-13 2 V
-13 3 V
-13 2 V
-12 2 V
-13 3 V
-13 2 V
-12 2 V
-12 3 V
-13 2 V
-12 3 V
-12 2 V
-12 2 V
-12 3 V
-11 2 V
-12 3 V
-12 2 V
-11 3 V
-12 2 V
-11 3 V
-11 2 V
-11 2 V
-11 3 V
-11 3 V
-11 2 V
-11 3 V
-11 2 V
-11 3 V
-10 2 V
-11 3 V
-10 2 V
-11 3 V
-10 3 V
-10 2 V
-10 3 V
-10 2 V
-10 3 V
-10 3 V
-10 2 V
-10 3 V
-9 3 V
-10 2 V
-10 3 V
-9 3 V
-10 2 V
-9 3 V
-9 3 V
-10 2 V
-9 3 V
-9 3 V
-9 2 V
-9 3 V
-9 3 V
-9 3 V
-9 2 V
-8 3 V
-9 3 V
-9 3 V
-8 3 V
-9 2 V
-8 3 V
-9 3 V
-8 3 V
-8 3 V
-9 2 V
-8 3 V
-8 3 V
-8 3 V
-8 3 V
-8 3 V
-8 3 V
-8 2 V
-8 3 V
-7 3 V
-8 3 V
-8 3 V
-8 3 V
-7 3 V
-8 3 V
-7 3 V
-8 3 V
-7 3 V
-7 2 V
-8 3 V
-7 3 V
-7 3 V
-7 3 V
-8 3 V
-7 3 V
-7 3 V
-7 3 V
-7 3 V
-7 3 V
-6 3 V
-7 3 V
-7 3 V
-7 3 V
-7 3 V
-6 3 V
-7 3 V
-7 3 V
-6 3 V
-7 3 V
-6 3 V
-7 3 V
-6 3 V
-6 3 V
-7 3 V
-6 3 V
-6 3 V
-7 3 V
-6 3 V
-6 3 V
-6 3 V
-6 3 V
-6 3 V
-6 3 V
-6 3 V
-6 3 V
-6 3 V
-6 3 V
-6 3 V
-6 3 V
-6 3 V
-5 3 V
-6 3 V
-6 3 V
-5 3 V
-6 3 V
-6 3 V
-5 3 V
-6 3 V
-5 2 V
-6 3 V
-5 3 V
-6 3 V
-5 3 V
-6 3 V
-5 3 V
-5 2 V
-6 3 V
-5 3 V
-5 3 V
-5 2 V
-5 3 V
-6 3 V
-5 2 V
-5 3 V
-5 3 V
-5 2 V
-5 3 V
-5 2 V
-5 3 V
-5 2 V
-5 3 V
-5 2 V
-5 3 V
-4 2 V
-5 3 V
-5 2 V
-5 2 V
-5 2 V
-4 3 V
-5 2 V
-5 2 V
-4 2 V
-5 2 V
-5 2 V
-4 2 V
-5 2 V
-4 2 V
-5 2 V
-5 2 V
-4 2 V
-4 2 V
-5 2 V
-4 1 V
-5 2 V
-4 2 V
-5 1 V
-4 2 V
-4 1 V
-4 2 V
-5 1 V
-4 2 V
-4 1 V
-4 2 V
-5 1 V
-4 2 V
-4 1 V
-4 1 V
-4 1 V
-4 2 V
-4 1 V
-5 1 V
-4 1 V
-4 2 V
-4 1 V
-4 1 V
-4 1 V
-4 1 V
-4 1 V
-3 1 V
-4 1 V
-4 1 V
-4 1 V
-4 1 V
-4 1 V
-4 1 V
-3 1 V
-4 1 V
-4 1 V
-4 1 V
-3 1 V
-4 1 V
-4 1 V
-4 0 V
-3 1 V
-4 1 V
-4 1 V
-3 1 V
-4 1 V
-3 1 V
-4 0 V
-3 1 V
-4 1 V
-4 1 V
-3 1 V
-4 1 V
-3 0 V
-4 1 V
-3 1 V
-3 1 V
-4 1 V
-3 0 V
-4 1 V
-3 1 V
-3 1 V
-4 1 V
-3 0 V
-3 1 V
-4 1 V
-3 1 V
-3 1 V
-4 0 V
-3 1 V
-3 1 V
-3 1 V
-4 0 V
-3 1 V
-3 1 V
-3 1 V
-3 1 V
-3 0 V
-4 1 V
-3 1 V
-3 1 V
-3 1 V
-3 0 V
-3 1 V
-3 1 V
-3 1 V
-3 0 V
-3 1 V
-3 1 V
-3 1 V
-3 1 V
-3 0 V
-3 1 V
-3 1 V
-3 1 V
-3 1 V
-3 0 V
-3 1 V
-3 1 V
-3 1 V
-3 1 V
-3 0 V
-2 1 V
-3 1 V
-3 1 V
-3 1 V
-3 0 V
-3 1 V
-2 1 V
-3 1 V
-3 1 V
-3 0 V
-2 1 V
-3 1 V
-3 1 V
-3 1 V
-2 1 V
-3 0 V
-3 1 V
-2 1 V
-3 1 V
-3 1 V
-2 1 V
-3 0 V
-3 1 V
-2 1 V
-3 1 V
-3 1 V
-2 1 V
-3 0 V
-2 1 V
-3 1 V
-2 1 V
-3 1 V
-3 1 V
-2 1 V
-3 0 V
-2 1 V
-3 1 V
-2 1 V
-3 1 V
-2 1 V
-3 1 V
-2 0 V
-2 1 V
-3 1 V
-2 1 V
-3 1 V
-2 1 V
-3 1 V
-2 1 V
-2 0 V
-3 1 V
-2 1 V
-2 1 V
-3 1 V
-2 1 V
-3 1 V
-2 1 V
-2 1 V
-2 0 V
-3 1 V
-2 1 V
-2 1 V
-3 1 V
-2 1 V
-2 1 V
-2 1 V
-3 1 V
-2 1 V
-2 1 V
-2 0 V
-3 1 V
-2 1 V
-2 1 V
-2 1 V
-2 1 V
-3 1 V
-2 1 V
currentpoint stroke M
-2 1 V
-2 1 V
-2 1 V
-2 1 V
-3 1 V
-2 1 V
-2 1 V
-2 0 V
-2 1 V
-2 1 V
-2 1 V
-2 1 V
-3 1 V
-2 1 V
-2 1 V
-2 1 V
-2 1 V
-2 1 V
-2 1 V
-2 1 V
-2 1 V
-2 1 V
-2 1 V
-2 1 V
-2 1 V
-2 1 V
-2 1 V
-2 1 V
-2 1 V
-2 1 V
-2 1 V
-2 1 V
-2 1 V
-2 1 V
-2 1 V
-2 1 V
-2 1 V
-2 1 V
-1 1 V
-2 1 V
-2 1 V
-2 1 V
-2 1 V
-2 1 V
-2 1 V
-2 1 V
-2 1 V
-1 1 V
-2 1 V
-2 1 V
-2 1 V
-2 1 V
-2 1 V
-2 1 V
-1 1 V
-2 1 V
-2 2 V
-2 1 V
-2 1 V
-1 1 V
-2 1 V
-2 1 V
-2 1 V
-2 1 V
-1 1 V
-2 1 V
-2 1 V
-2 1 V
-1 1 V
-2 1 V
-2 1 V
-2 2 V
-1 1 V
-2 1 V
-2 1 V
-1 1 V
-2 1 V
-2 1 V
-2 1 V
-1 1 V
-2 1 V
-2 2 V
-1 1 V
-2 1 V
-2 1 V
-1 1 V
-2 1 V
-2 1 V
-1 1 V
-2 1 V
-2 2 V
-1 1 V
-2 1 V
-1 1 V
-2 1 V
-2 1 V
-1 1 V
-2 2 V
-1 1 V
-2 1 V
-2 1 V
-1 1 V
-2 1 V
-1 1 V
-2 2 V
-2 1 V
-1 1 V
-2 1 V
-1 1 V
-2 1 V
-1 2 V
-2 1 V
-1 1 V
-2 1 V
-2 1 V
-1 1 V
-2 2 V
-1 1 V
-2 1 V
-1 1 V
-2 1 V
-1 2 V
-2 1 V
-1 1 V
-2 1 V
-1 1 V
-2 2 V
-1 1 V
-2 1 V
-1 1 V
-1 2 V
-2 1 V
-1 1 V
-2 1 V
-1 1 V
-2 2 V
-1 1 V
-2 1 V
-1 1 V
-1 2 V
-2 1 V
-1 1 V
-2 1 V
-1 2 V
-2 1 V
-1 1 V
-1 1 V
-2 2 V
-1 1 V
-2 1 V
-1 1 V
-1 2 V
-2 1 V
-1 1 V
-1 1 V
-2 2 V
-1 1 V
-2 1 V
-1 2 V
-1 1 V
-2 1 V
-1 1 V
-1 2 V
-2 1 V
-1 1 V
-1 2 V
-2 1 V
-1 1 V
-1 1 V
-2 2 V
-1 1 V
-1 1 V
-2 2 V
-1 1 V
-1 1 V
-1 2 V
-2 1 V
-1 1 V
-1 2 V
-2 1 V
-1 1 V
-1 2 V
-1 1 V
-2 1 V
-1 2 V
-1 1 V
-1 1 V
-2 2 V
-1 1 V
-1 1 V
-1 2 V
-2 1 V
-1 1 V
-1 2 V
-1 1 V
-2 2 V
-1 1 V
-1 1 V
-1 2 V
-2 1 V
-1 1 V
-1 2 V
-1 1 V
-1 2 V
-2 1 V
-1 1 V
-1 2 V
-1 1 V
-1 2 V
-2 1 V
-1 1 V
-1 2 V
-1 1 V
-1 2 V
-2 1 V
-1 1 V
-1 2 V
-1 1 V
-1 2 V
-1 1 V
-2 2 V
-1 1 V
-1 1 V
-1 2 V
-1 1 V
-1 2 V
-1 1 V
-2 2 V
-1 1 V
-1 2 V
-1 1 V
-1 1 V
-1 2 V
-1 1 V
-2 2 V
-1 1 V
-1 2 V
-1 1 V
-1 2 V
-1 1 V
-1 2 V
-1 1 V
-1 2 V
-1 1 V
-2 2 V
-1 1 V
-1 2 V
-1 1 V
-1 2 V
-1 1 V
-1 2 V
-1 1 V
-1 2 V
-1 1 V
-1 2 V
-1 1 V
-2 2 V
-1 1 V
-1 2 V
-1 1 V
-1 2 V
-1 1 V
-1 2 V
-1 1 V
-1 2 V
-1 1 V
-1 2 V
-1 2 V
-1 1 V
-1 2 V
-1 1 V
-1 2 V
-1 1 V
-1 2 V
-1 1 V
-1 2 V
-1 2 V
-1 1 V
-1 2 V
-1 1 V
-1 2 V
-1 1 V
-1 2 V
-1 2 V
-1 1 V
-1 2 V
-1 1 V
-1 2 V
-1 2 V
-1 1 V
-1 2 V
-1 1 V
-1 2 V
-1 2 V
-1 1 V
-1 2 V
-1 1 V
-1 2 V
-1 2 V
-1 1 V
-1 2 V
-1 1 V
-1 2 V
-1 2 V
-1 1 V
-1 2 V
-1 2 V
-1 1 V
-1 2 V
-1 2 V
0 1 V
-1 2 V
-1 2 V
-1 1 V
-1 2 V
-1 2 V
-1 1 V
-1 2 V
-1 2 V
-1 1 V
-1 2 V
-1 2 V
-1 1 V
-1 2 V
0 2 V
-1 1 V
-1 2 V
-1 2 V
-1 1 V
-1 2 V
-1 2 V
-1 2 V
-1 1 V
-1 2 V
0 2 V
-1 1 V
-1 2 V
-1 2 V
-1 2 V
-1 1 V
-1 2 V
-1 2 V
-1 1 V
0 2 V
-1 2 V
-1 2 V
-1 1 V
-1 2 V
-1 2 V
-1 2 V
0 1 V
-1 2 V
-1 2 V
-1 2 V
-1 1 V
-1 2 V
-1 2 V
0 2 V
-1 1 V
-1 2 V
-1 2 V
-1 2 V
-1 2 V
-1 1 V
0 2 V
-1 2 V
-1 2 V
-1 2 V
-1 1 V
-1 2 V
0 2 V
-1 2 V
-1 2 V
-1 1 V
-1 2 V
0 2 V
-1 2 V
-1 2 V
-1 1 V
-1 2 V
-1 2 V
0 2 V
-1 2 V
-1 2 V
-1 1 V
-1 2 V
0 2 V
-1 2 V
-1 2 V
-1 2 V
-1 2 V
0 1 V
-1 2 V
-1 2 V
-1 2 V
-1 2 V
0 2 V
-1 2 V
-1 1 V
-1 2 V
currentpoint stroke M
0 2 V
-1 2 V
-1 2 V
stroke
grestore
end
showpage
}
\put(2438,1528){\makebox(0,0)[l]{$\omega\tau = 0.01$}}
\put(2438,1628){\makebox(0,0)[l]{$\omega\tau = 0.005$}}
\put(2438,1728){\makebox(0,0)[l]{$\omega\tau = 0.0001$}}
\put(2000,150){\makebox(0,0){$r/R(t)$}}
\put(100,1230){\makebox(0,0)[b]{\shortstack{$\displaystyle{\tilde s \over \mu}$}}}
\put(3550,300){\makebox(0,0){5}}
\put(2775,300){\makebox(0,0){4}}
\put(2000,300){\makebox(0,0){3}}
\put(1225,300){\makebox(0,0){2}}
\put(450,300){\makebox(0,0){1}}
\put(400,1728){\makebox(0,0)[r]{0.2}}
\put(400,1064){\makebox(0,0)[r]{0.1}}
\put(400,400){\makebox(0,0)[r]{0}}
\end{picture}

%% file: sim_lambda.tex
\setlength{\unitlength}{0.1bp}
\special{!
/gnudict 120 dict def
gnudict begin
/Color false def
/Solid false def
/gnulinewidth 5.000 def
/userlinewidth gnulinewidth def
/vshift -33 def
/dl {10 mul} def
/hpt_ 31.5 def
/vpt_ 31.5 def
/hpt hpt_ def
/vpt vpt_ def
/M {moveto} bind def
/L {lineto} bind def
/R {rmoveto} bind def
/V {rlineto} bind def
/vpt2 vpt 2 mul def
/hpt2 hpt 2 mul def
/Lshow { currentpoint stroke M
  0 vshift R show } def
/Rshow { currentpoint stroke M
  dup stringwidth pop neg vshift R show } def
/Cshow { currentpoint stroke M
  dup stringwidth pop -2 div vshift R show } def
/UP { dup vpt_ mul /vpt exch def hpt_ mul /hpt exch def
  /hpt2 hpt 2 mul def /vpt2 vpt 2 mul def } def
/DL { Color {setrgbcolor Solid {pop []} if 0 setdash }
 {pop pop pop Solid {pop []} if 0 setdash} ifelse } def
/BL { stroke gnulinewidth 2 mul setlinewidth } def
/AL { stroke gnulinewidth 2 div setlinewidth } def
/UL { gnulinewidth mul /userlinewidth exch def } def
/PL { stroke userlinewidth setlinewidth } def
/LTb { BL [] 0 0 0 DL } def
/LTa { AL [1 dl 2 dl] 0 setdash 0 0 0 setrgbcolor } def
/LT0 { PL [] 0 1 0 DL } def
/LT1 { PL [4 dl 2 dl] 0 0 1 DL } def
/LT2 { PL [2 dl 3 dl] 1 0 0 DL } def
/LT3 { PL [1 dl 1.5 dl] 1 0 1 DL } def
/LT4 { PL [5 dl 2 dl 1 dl 2 dl] 0 1 1 DL } def
/LT5 { PL [4 dl 3 dl 1 dl 3 dl] 1 1 0 DL } def
/LT6 { PL [2 dl 2 dl 2 dl 4 dl] 0 0 0 DL } def
/LT7 { PL [2 dl 2 dl 2 dl 2 dl 2 dl 4 dl] 1 0.3 0 DL } def
/LT8 { PL [2 dl 2 dl 2 dl 2 dl 2 dl 2 dl 2 dl 4 dl] 0.5 0.5 0.5 DL } def
/Pnt { stroke [] 0 setdash
   gsave 1 setlinecap M 0 0 V stroke grestore } def
/Dia { stroke [] 0 setdash 2 copy vpt add M
  hpt neg vpt neg V hpt vpt neg V
  hpt vpt V hpt neg vpt V closepath stroke
  Pnt } def
/Pls { stroke [] 0 setdash vpt sub M 0 vpt2 V
  currentpoint stroke M
  hpt neg vpt neg R hpt2 0 V stroke
  } def
/Box { stroke [] 0 setdash 2 copy exch hpt sub exch vpt add M
  0 vpt2 neg V hpt2 0 V 0 vpt2 V
  hpt2 neg 0 V closepath stroke
  Pnt } def
/Crs { stroke [] 0 setdash exch hpt sub exch vpt add M
  hpt2 vpt2 neg V currentpoint stroke M
  hpt2 neg 0 R hpt2 vpt2 V stroke } def
/TriU { stroke [] 0 setdash 2 copy vpt 1.12 mul add M
  hpt neg vpt -1.62 mul V
  hpt 2 mul 0 V
  hpt neg vpt 1.62 mul V closepath stroke
  Pnt  } def
/Star { 2 copy Pls Crs } def
/BoxF { stroke [] 0 setdash exch hpt sub exch vpt add M
  0 vpt2 neg V  hpt2 0 V  0 vpt2 V
  hpt2 neg 0 V  closepath fill } def
/TriUF { stroke [] 0 setdash vpt 1.12 mul add M
  hpt neg vpt -1.62 mul V
  hpt 2 mul 0 V
  hpt neg vpt 1.62 mul V closepath fill } def
/TriD { stroke [] 0 setdash 2 copy vpt 1.12 mul sub M
  hpt neg vpt 1.62 mul V
  hpt 2 mul 0 V
  hpt neg vpt -1.62 mul V closepath stroke
  Pnt  } def
/TriDF { stroke [] 0 setdash vpt 1.12 mul sub M
  hpt neg vpt 1.62 mul V
  hpt 2 mul 0 V
  hpt neg vpt -1.62 mul V closepath fill} def
/DiaF { stroke [] 0 setdash vpt add M
  hpt neg vpt neg V hpt vpt neg V
  hpt vpt V hpt neg vpt V closepath fill } def
/Pent { stroke [] 0 setdash 2 copy gsave
  translate 0 hpt M 4 {72 rotate 0 hpt L} repeat
  closepath stroke grestore Pnt } def
/PentF { stroke [] 0 setdash gsave
  translate 0 hpt M 4 {72 rotate 0 hpt L} repeat
  closepath fill grestore } def
/Circle { stroke [] 0 setdash 2 copy
  hpt 0 360 arc stroke Pnt } def
/CircleF { stroke [] 0 setdash hpt 0 360 arc fill } def
/C0 { BL [] 0 setdash 2 copy moveto vpt 90 450  arc } bind def
/C1 { BL [] 0 setdash 2 copy        moveto
       2 copy  vpt 0 90 arc closepath fill
               vpt 0 360 arc closepath } bind def
/C2 { BL [] 0 setdash 2 copy moveto
       2 copy  vpt 90 180 arc closepath fill
               vpt 0 360 arc closepath } bind def
/C3 { BL [] 0 setdash 2 copy moveto
       2 copy  vpt 0 180 arc closepath fill
               vpt 0 360 arc closepath } bind def
/C4 { BL [] 0 setdash 2 copy moveto
       2 copy  vpt 180 270 arc closepath fill
               vpt 0 360 arc closepath } bind def
/C5 { BL [] 0 setdash 2 copy moveto
       2 copy  vpt 0 90 arc
       2 copy moveto
       2 copy  vpt 180 270 arc closepath fill
               vpt 0 360 arc } bind def
/C6 { BL [] 0 setdash 2 copy moveto
      2 copy  vpt 90 270 arc closepath fill
              vpt 0 360 arc closepath } bind def
/C7 { BL [] 0 setdash 2 copy moveto
      2 copy  vpt 0 270 arc closepath fill
              vpt 0 360 arc closepath } bind def
/C8 { BL [] 0 setdash 2 copy moveto
      2 copy vpt 270 360 arc closepath fill
              vpt 0 360 arc closepath } bind def
/C9 { BL [] 0 setdash 2 copy moveto
      2 copy  vpt 270 450 arc closepath fill
              vpt 0 360 arc closepath } bind def
/C10 { BL [] 0 setdash 2 copy 2 copy moveto vpt 270 360 arc closepath fill
       2 copy moveto
       2 copy vpt 90 180 arc closepath fill
               vpt 0 360 arc closepath } bind def
/C11 { BL [] 0 setdash 2 copy moveto
       2 copy  vpt 0 180 arc closepath fill
       2 copy moveto
       2 copy  vpt 270 360 arc closepath fill
               vpt 0 360 arc closepath } bind def
/C12 { BL [] 0 setdash 2 copy moveto
       2 copy  vpt 180 360 arc closepath fill
               vpt 0 360 arc closepath } bind def
/C13 { BL [] 0 setdash  2 copy moveto
       2 copy  vpt 0 90 arc closepath fill
       2 copy moveto
       2 copy  vpt 180 360 arc closepath fill
               vpt 0 360 arc closepath } bind def
/C14 { BL [] 0 setdash 2 copy moveto
       2 copy  vpt 90 360 arc closepath fill
               vpt 0 360 arc } bind def
/C15 { BL [] 0 setdash 2 copy vpt 0 360 arc closepath fill
               vpt 0 360 arc closepath } bind def
/Rec   { newpath 4 2 roll moveto 1 index 0 rlineto 0 exch rlineto
       neg 0 rlineto closepath } bind def
/Square { dup Rec } bind def
/Bsquare { vpt sub exch vpt sub exch vpt2 Square } bind def
/S0 { BL [] 0 setdash 2 copy moveto 0 vpt rlineto BL Bsquare } bind def
/S1 { BL [] 0 setdash 2 copy vpt Square fill Bsquare } bind def
/S2 { BL [] 0 setdash 2 copy exch vpt sub exch vpt Square fill Bsquare } bind def
/S3 { BL [] 0 setdash 2 copy exch vpt sub exch vpt2 vpt Rec fill Bsquare } bind def
/S4 { BL [] 0 setdash 2 copy exch vpt sub exch vpt sub vpt Square fill Bsquare } bind def
/S5 { BL [] 0 setdash 2 copy 2 copy vpt Square fill
       exch vpt sub exch vpt sub vpt Square fill Bsquare } bind def
/S6 { BL [] 0 setdash 2 copy exch vpt sub exch vpt sub vpt vpt2 Rec fill Bsquare } bind def
/S7 { BL [] 0 setdash 2 copy exch vpt sub exch vpt sub vpt vpt2 Rec fill
       2 copy vpt Square fill
       Bsquare } bind def
/S8 { BL [] 0 setdash 2 copy vpt sub vpt Square fill Bsquare } bind def
/S9 { BL [] 0 setdash 2 copy vpt sub vpt vpt2 Rec fill Bsquare } bind def
/S10 { BL [] 0 setdash 2 copy vpt sub vpt Square fill 2 copy exch vpt sub exch vpt Square fill
       Bsquare } bind def
/S11 { BL [] 0 setdash 2 copy vpt sub vpt Square fill 2 copy exch vpt sub exch vpt2 vpt Rec fill
       Bsquare } bind def
/S12 { BL [] 0 setdash 2 copy exch vpt sub exch vpt sub vpt2 vpt Rec fill Bsquare } bind def
/S13 { BL [] 0 setdash 2 copy exch vpt sub exch vpt sub vpt2 vpt Rec fill
       2 copy vpt Square fill Bsquare } bind def
/S14 { BL [] 0 setdash 2 copy exch vpt sub exch vpt sub vpt2 vpt Rec fill
       2 copy exch vpt sub exch vpt Square fill Bsquare } bind def
/S15 { BL [] 0 setdash 2 copy Bsquare fill Bsquare } bind def
/D0 { gsave translate 45 rotate 0 0 S0 stroke grestore } bind def
/D1 { gsave translate 45 rotate 0 0 S1 stroke grestore } bind def
/D2 { gsave translate 45 rotate 0 0 S2 stroke grestore } bind def
/D3 { gsave translate 45 rotate 0 0 S3 stroke grestore } bind def
/D4 { gsave translate 45 rotate 0 0 S4 stroke grestore } bind def
/D5 { gsave translate 45 rotate 0 0 S5 stroke grestore } bind def
/D6 { gsave translate 45 rotate 0 0 S6 stroke grestore } bind def
/D7 { gsave translate 45 rotate 0 0 S7 stroke grestore } bind def
/D8 { gsave translate 45 rotate 0 0 S8 stroke grestore } bind def
/D9 { gsave translate 45 rotate 0 0 S9 stroke grestore } bind def
/D10 { gsave translate 45 rotate 0 0 S10 stroke grestore } bind def
/D11 { gsave translate 45 rotate 0 0 S11 stroke grestore } bind def
/D12 { gsave translate 45 rotate 0 0 S12 stroke grestore } bind def
/D13 { gsave translate 45 rotate 0 0 S13 stroke grestore } bind def
/D14 { gsave translate 45 rotate 0 0 S14 stroke grestore } bind def
/D15 { gsave translate 45 rotate 0 0 S15 stroke grestore } bind def
/DiaE { stroke [] 0 setdash vpt add M
  hpt neg vpt neg V hpt vpt neg V
  hpt vpt V hpt neg vpt V closepath stroke } def
/BoxE { stroke [] 0 setdash exch hpt sub exch vpt add M
  0 vpt2 neg V hpt2 0 V 0 vpt2 V
  hpt2 neg 0 V closepath stroke } def
/TriUE { stroke [] 0 setdash vpt 1.12 mul add M
  hpt neg vpt -1.62 mul V
  hpt 2 mul 0 V
  hpt neg vpt 1.62 mul V closepath stroke } def
/TriDE { stroke [] 0 setdash vpt 1.12 mul sub M
  hpt neg vpt 1.62 mul V
  hpt 2 mul 0 V
  hpt neg vpt -1.62 mul V closepath stroke } def
/PentE { stroke [] 0 setdash gsave
  translate 0 hpt M 4 {72 rotate 0 hpt L} repeat
  closepath stroke grestore } def
/CircE { stroke [] 0 setdash 
  hpt 0 360 arc stroke } def
/BoxFill { gsave Rec 1 setgray fill grestore } def
end
}
\begin{picture}(3600,2160)(0,0)
\special{"
gnudict begin
gsave
0 0 translate
0.100 0.100 scale
0 setgray
newpath
LTb
450 400 M
63 0 V
3037 0 R
-63 0 V
450 1783 M
63 0 V
3037 0 R
-63 0 V
450 400 M
0 63 V
0 1597 R
0 -63 V
1225 400 M
0 63 V
0 1597 R
0 -63 V
2000 400 M
0 63 V
0 1597 R
0 -63 V
2775 400 M
0 63 V
0 1597 R
0 -63 V
3550 400 M
0 63 V
0 1597 R
0 -63 V
LTb
450 400 M
3100 0 V
0 1660 V
-3100 0 V
450 400 L
1.000 UL
LT0
2512 1783 M
263 0 V
3550 783 M
-19 4 V
-19 4 V
-19 4 V
-19 4 V
-19 3 V
-18 4 V
-18 4 V
-18 4 V
-18 4 V
-18 4 V
-17 4 V
-17 4 V
-18 4 V
-17 4 V
-16 4 V
-17 5 V
-17 4 V
-16 4 V
-16 4 V
-16 4 V
-16 4 V
-16 5 V
-16 4 V
-15 4 V
-16 4 V
-15 5 V
-15 4 V
-15 4 V
-15 5 V
-14 4 V
-15 4 V
-14 5 V
-15 4 V
-14 5 V
-14 4 V
-14 4 V
-14 5 V
-14 4 V
-13 5 V
-14 5 V
-13 4 V
-14 5 V
-13 4 V
-13 5 V
-13 5 V
-13 4 V
-12 5 V
-13 4 V
-13 5 V
-12 5 V
-12 5 V
-13 4 V
-12 5 V
-12 5 V
-12 5 V
-12 5 V
-11 4 V
-12 5 V
-12 5 V
-11 5 V
-12 5 V
-11 5 V
-11 5 V
-11 5 V
-11 5 V
-11 5 V
-11 5 V
-11 5 V
-11 5 V
-11 5 V
-10 5 V
-11 5 V
-10 5 V
-11 5 V
-10 5 V
-10 6 V
-10 5 V
-10 5 V
-10 5 V
-10 5 V
-10 6 V
-10 5 V
-9 5 V
-10 5 V
-10 6 V
-9 5 V
-10 5 V
-9 6 V
-9 5 V
-10 5 V
-9 6 V
-9 5 V
-9 6 V
-9 5 V
-9 6 V
-9 5 V
-9 6 V
-8 5 V
-9 6 V
-9 5 V
-8 6 V
-9 5 V
-8 6 V
-9 5 V
-8 6 V
-8 5 V
-9 6 V
-8 6 V
-8 5 V
-8 6 V
-8 6 V
-8 5 V
-8 6 V
-8 6 V
-8 5 V
-7 6 V
-8 6 V
-8 6 V
-8 5 V
-7 6 V
-8 6 V
-7 6 V
-8 6 V
-7 5 V
-7 6 V
-8 6 V
-7 6 V
-7 6 V
-7 6 V
-8 5 V
-7 6 V
-7 6 V
-7 6 V
-7 6 V
-7 6 V
-6 6 V
-7 6 V
-7 5 V
-7 6 V
-7 6 V
-6 6 V
-7 6 V
-7 6 V
-6 6 V
-7 6 V
-6 6 V
-7 6 V
-6 5 V
-6 6 V
-7 6 V
-6 6 V
-6 6 V
-7 6 V
-6 6 V
-6 5 V
-6 6 V
-6 6 V
-6 6 V
-6 6 V
-6 5 V
-6 6 V
-6 6 V
-6 6 V
-6 5 V
-6 6 V
-6 5 V
-5 6 V
-6 5 V
-6 6 V
-5 5 V
-6 6 V
-6 5 V
-5 5 V
-6 6 V
-5 5 V
-6 5 V
-5 5 V
-6 5 V
-5 5 V
-6 4 V
-5 5 V
-5 5 V
-6 4 V
-5 4 V
-5 5 V
-5 4 V
-5 4 V
-6 3 V
-5 4 V
-5 3 V
-5 4 V
-5 3 V
-5 3 V
-5 2 V
-5 3 V
-5 2 V
-5 3 V
-5 2 V
-5 2 V
-4 1 V
-5 2 V
-5 1 V
-5 1 V
-5 1 V
-4 1 V
-5 1 V
-5 1 V
-4 0 V
-5 1 V
-5 0 V
-4 1 V
-5 0 V
-4 0 V
-5 1 V
-5 0 V
-4 0 V
-4 0 V
-5 0 V
-4 0 V
-5 1 V
-4 0 V
-5 0 V
-4 0 V
-4 0 V
-4 0 V
-5 0 V
-4 0 V
-4 0 V
-4 0 V
-5 0 V
-4 0 V
-4 0 V
-4 0 V
-4 0 V
-4 0 V
-4 0 V
-5 0 V
-4 0 V
-4 0 V
-4 0 V
-4 0 V
-4 0 V
-4 0 V
-4 0 V
-3 0 V
-4 0 V
-4 0 V
-4 0 V
-4 0 V
-4 0 V
-4 0 V
-3 0 V
-4 0 V
-4 0 V
-4 0 V
-3 0 V
-4 0 V
-4 0 V
-4 0 V
-3 0 V
-4 0 V
-4 0 V
-3 0 V
-4 1 V
-3 0 V
-4 0 V
-3 0 V
-4 0 V
-4 0 V
-3 0 V
-4 0 V
-3 0 V
-4 0 V
-3 0 V
-3 0 V
-4 0 V
-3 0 V
-4 0 V
-3 0 V
-3 0 V
-4 0 V
-3 0 V
-3 0 V
-4 0 V
-3 0 V
-3 0 V
-4 0 V
-3 0 V
-3 0 V
-3 0 V
-4 0 V
-3 0 V
-3 0 V
-3 0 V
-3 0 V
-3 0 V
-4 0 V
-3 0 V
-3 0 V
-3 0 V
-3 0 V
-3 0 V
-3 0 V
-3 0 V
-3 0 V
-3 0 V
-3 0 V
-3 0 V
-3 0 V
-3 0 V
-3 0 V
-3 0 V
-3 0 V
-3 0 V
-3 0 V
-3 0 V
-3 0 V
-3 0 V
-3 0 V
-3 0 V
-2 0 V
-3 0 V
-3 0 V
-3 0 V
-3 0 V
-3 0 V
-2 0 V
-3 0 V
-3 0 V
-3 0 V
-2 0 V
-3 0 V
-3 0 V
-3 1 V
-2 0 V
-3 0 V
-3 0 V
-2 0 V
-3 0 V
-3 0 V
-2 0 V
-3 0 V
-3 0 V
-2 0 V
-3 0 V
-3 0 V
-2 0 V
-3 0 V
-2 0 V
-3 0 V
-2 0 V
-3 0 V
-3 0 V
-2 0 V
-3 0 V
-2 0 V
-3 0 V
-2 0 V
-3 0 V
-2 0 V
-3 0 V
-2 0 V
-2 0 V
-3 0 V
-2 0 V
-3 0 V
-2 0 V
-3 0 V
-2 0 V
-2 0 V
-3 0 V
-2 0 V
-2 0 V
-3 0 V
-2 0 V
-3 0 V
-2 0 V
-2 0 V
-2 0 V
-3 0 V
-2 0 V
-2 0 V
-3 0 V
-2 0 V
-2 0 V
-2 0 V
-3 0 V
-2 0 V
-2 0 V
-2 0 V
-3 0 V
-2 0 V
-2 0 V
-2 0 V
-2 0 V
-3 1 V
-2 0 V
currentpoint stroke M
-2 0 V
-2 0 V
-2 0 V
-2 0 V
-3 0 V
-2 0 V
-2 0 V
-2 0 V
-2 0 V
-2 0 V
-2 0 V
-2 0 V
-3 0 V
-2 0 V
-2 0 V
-2 0 V
-2 0 V
-2 0 V
-2 0 V
-2 0 V
-2 0 V
-2 0 V
-2 0 V
-2 0 V
-2 0 V
-2 0 V
-2 0 V
-2 0 V
-2 0 V
-2 0 V
-2 0 V
-2 0 V
-2 0 V
-2 0 V
-2 0 V
-2 0 V
-2 0 V
-2 0 V
-1 0 V
-2 0 V
-2 0 V
-2 0 V
-2 0 V
-2 0 V
-2 0 V
-2 0 V
-2 0 V
-1 0 V
-2 0 V
-2 0 V
-2 0 V
-2 0 V
-2 0 V
-2 1 V
-1 0 V
-2 0 V
-2 0 V
-2 0 V
-2 0 V
-1 0 V
-2 0 V
-2 0 V
-2 0 V
-2 0 V
-1 0 V
-2 0 V
-2 0 V
-2 0 V
-1 0 V
-2 0 V
-2 0 V
-2 0 V
-1 0 V
-2 0 V
-2 0 V
-1 0 V
-2 0 V
-2 0 V
-2 0 V
-1 0 V
-2 0 V
-2 0 V
-1 0 V
-2 0 V
-2 0 V
-1 0 V
-2 0 V
-2 0 V
-1 0 V
-2 0 V
-2 0 V
-1 0 V
-2 0 V
-1 0 V
-2 0 V
-2 0 V
-1 0 V
-2 0 V
-1 0 V
-2 0 V
-2 0 V
-1 0 V
-2 0 V
-1 0 V
-2 1 V
-2 0 V
-1 0 V
-2 0 V
-1 0 V
-2 0 V
-1 0 V
-2 0 V
-1 0 V
-2 0 V
-2 0 V
-1 0 V
-2 0 V
-1 0 V
-2 0 V
-1 0 V
-2 0 V
-1 0 V
-2 0 V
-1 0 V
-2 0 V
-1 0 V
-2 0 V
-1 0 V
-2 0 V
-1 0 V
-1 0 V
-2 0 V
-1 0 V
-2 0 V
-1 0 V
-2 0 V
-1 0 V
-2 0 V
-1 0 V
-1 0 V
-2 0 V
-1 0 V
-2 0 V
-1 0 V
-2 0 V
-1 0 V
-1 0 V
-2 0 V
-1 0 V
-2 0 V
-1 0 V
-1 0 V
-2 1 V
-1 0 V
-1 0 V
-2 0 V
-1 0 V
-2 0 V
-1 0 V
-1 0 V
-2 0 V
-1 0 V
-1 0 V
-2 0 V
-1 0 V
-1 0 V
-2 0 V
-1 0 V
-1 0 V
-2 0 V
-1 0 V
-1 0 V
-2 0 V
-1 0 V
-1 0 V
-1 0 V
-2 0 V
-1 0 V
-1 0 V
-2 0 V
-1 0 V
-1 0 V
-1 0 V
-2 0 V
-1 0 V
-1 0 V
-1 0 V
-2 0 V
-1 0 V
-1 0 V
-1 0 V
-2 0 V
-1 0 V
-1 0 V
-1 0 V
-2 0 V
-1 0 V
-1 1 V
-1 0 V
-2 0 V
-1 0 V
-1 0 V
-1 0 V
-1 0 V
-2 0 V
-1 0 V
-1 0 V
-1 0 V
-1 0 V
-2 0 V
-1 0 V
-1 0 V
-1 0 V
-1 0 V
-2 0 V
-1 0 V
-1 0 V
-1 0 V
-1 0 V
-1 0 V
-2 0 V
-1 0 V
-1 0 V
-1 0 V
-1 0 V
-1 0 V
-1 0 V
-2 0 V
-1 0 V
-1 0 V
-1 0 V
-1 0 V
-1 0 V
-1 0 V
-2 0 V
-1 0 V
-1 0 V
-1 0 V
-1 0 V
-1 1 V
-1 0 V
-1 0 V
-1 0 V
-1 0 V
-2 0 V
-1 0 V
-1 0 V
-1 0 V
-1 0 V
-1 0 V
-1 0 V
-1 0 V
-1 0 V
-1 0 V
-1 0 V
-1 0 V
-2 0 V
-1 0 V
-1 0 V
-1 0 V
-1 0 V
-1 0 V
-1 0 V
-1 0 V
-1 0 V
-1 0 V
-1 0 V
-1 0 V
-1 0 V
-1 0 V
-1 0 V
-1 0 V
-1 0 V
-1 0 V
-1 0 V
-1 0 V
-1 0 V
-1 0 V
-1 0 V
-1 1 V
-1 0 V
-1 0 V
-1 0 V
-1 0 V
-1 0 V
-1 0 V
-1 0 V
-1 0 V
-1 0 V
-1 0 V
-1 0 V
-1 0 V
-1 0 V
-1 0 V
-1 0 V
-1 0 V
-1 0 V
-1 0 V
-1 0 V
-1 0 V
-1 0 V
-1 0 V
-1 0 V
-1 0 V
-1 0 V
-1 0 V
-1 0 V
-1 0 V
-1 0 V
-1 0 V
-1 0 V
-1 0 V
-1 0 V
-1 0 V
-1 0 V
-1 0 V
-1 0 V
-1 1 V
-1 0 V
-1 0 V
-1 0 V
-1 0 V
-1 0 V
-1 0 V
-1 0 V
-1 0 V
-1 0 V
-1 0 V
-1 0 V
-1 0 V
-1 0 V
-1 0 V
-1 0 V
-1 0 V
-1 0 V
-1 0 V
-1 0 V
-1 0 V
-1 0 V
-1 0 V
-1 0 V
-1 0 V
-1 0 V
-1 0 V
-1 0 V
-1 0 V
-1 0 V
-1 0 V
-1 0 V
-1 0 V
-1 1 V
-1 0 V
-1 0 V
-1 0 V
-1 0 V
-1 0 V
-1 0 V
-1 0 V
-1 0 V
-1 0 V
-1 0 V
-1 0 V
-1 0 V
-1 0 V
-1 0 V
-1 0 V
-1 0 V
-1 0 V
-1 0 V
-1 0 V
-1 0 V
-1 0 V
-1 0 V
-1 0 V
-1 0 V
-1 0 V
-1 0 V
-1 0 V
-1 0 V
-1 1 V
-1 0 V
-1 0 V
-1 0 V
-1 0 V
-1 0 V
-1 0 V
-1 0 V
-1 0 V
-1 0 V
1.000 UL
LT1
2512 1683 M
263 0 V
3550 639 M
-19 2 V
-19 3 V
-19 2 V
-19 2 V
-19 3 V
-18 2 V
-18 3 V
-18 2 V
-18 3 V
-18 2 V
-17 3 V
-17 2 V
-18 3 V
-17 2 V
-16 3 V
-17 2 V
-17 3 V
-16 2 V
-16 3 V
-16 2 V
-16 3 V
-16 3 V
-16 2 V
-15 3 V
-16 2 V
-15 3 V
-15 3 V
-15 2 V
-15 3 V
-14 3 V
-15 3 V
-14 2 V
-15 3 V
-14 3 V
-14 2 V
-14 3 V
-14 3 V
-14 3 V
-13 3 V
-14 2 V
-13 3 V
-14 3 V
-13 3 V
-13 3 V
-13 2 V
-13 3 V
-12 3 V
-13 3 V
-13 3 V
-12 3 V
-12 3 V
-13 3 V
-12 3 V
-12 3 V
-12 2 V
-12 3 V
-11 3 V
-12 3 V
-12 3 V
-11 3 V
-12 3 V
-11 3 V
-11 3 V
-11 3 V
-11 4 V
-11 3 V
-11 3 V
-11 3 V
-11 3 V
-11 3 V
-10 3 V
-11 3 V
-10 3 V
-11 3 V
-10 3 V
-10 4 V
-10 3 V
-10 3 V
-10 3 V
-10 3 V
-10 4 V
-10 3 V
-9 3 V
-10 3 V
-10 3 V
-9 4 V
-10 3 V
-9 3 V
-9 3 V
-10 4 V
-9 3 V
-9 3 V
-9 4 V
-9 3 V
-9 3 V
-9 3 V
-9 4 V
-8 3 V
-9 4 V
-9 3 V
-8 3 V
-9 4 V
-8 3 V
-9 3 V
-8 4 V
-8 3 V
-9 4 V
-8 3 V
-8 3 V
-8 4 V
-8 3 V
-8 4 V
-8 3 V
-8 4 V
-8 3 V
-7 4 V
-8 3 V
-8 4 V
-8 3 V
-7 4 V
-8 3 V
-7 4 V
-8 3 V
-7 4 V
-7 3 V
-8 4 V
-7 3 V
-7 4 V
-7 3 V
-8 4 V
-7 3 V
-7 4 V
-7 4 V
-7 3 V
-7 4 V
-6 3 V
-7 4 V
-7 4 V
-7 3 V
-7 4 V
-6 3 V
-7 4 V
-7 4 V
-6 3 V
-7 4 V
-6 4 V
-7 3 V
-6 4 V
-6 4 V
-7 3 V
-6 4 V
-6 4 V
-7 3 V
-6 4 V
-6 4 V
-6 3 V
-6 4 V
-6 4 V
-6 3 V
-6 4 V
-6 4 V
-6 3 V
-6 4 V
-6 4 V
-6 4 V
-6 3 V
-5 4 V
-6 4 V
-6 3 V
-5 4 V
-6 4 V
-6 4 V
-5 3 V
-6 4 V
-5 4 V
-6 3 V
-5 4 V
-6 4 V
-5 4 V
-6 3 V
-5 4 V
-5 4 V
-6 4 V
-5 3 V
-5 4 V
-5 4 V
-5 4 V
-6 3 V
-5 4 V
-5 4 V
-5 3 V
-5 4 V
-5 4 V
-5 4 V
-5 3 V
-5 4 V
-5 4 V
-5 3 V
-5 4 V
-4 4 V
-5 4 V
-5 3 V
-5 4 V
-5 4 V
-4 3 V
-5 4 V
-5 4 V
-4 3 V
-5 4 V
-5 4 V
-4 3 V
-5 4 V
-4 4 V
-5 3 V
-5 4 V
-4 4 V
-4 3 V
-5 4 V
-4 4 V
-5 3 V
-4 4 V
-5 3 V
-4 4 V
-4 4 V
-4 3 V
-5 4 V
-4 3 V
-4 4 V
-4 4 V
-5 3 V
-4 4 V
-4 3 V
-4 4 V
-4 3 V
-4 4 V
-4 3 V
-5 4 V
-4 3 V
-4 4 V
-4 3 V
-4 4 V
-4 3 V
-4 3 V
-4 4 V
-3 3 V
-4 4 V
-4 3 V
-4 3 V
-4 4 V
-4 3 V
-4 3 V
-3 4 V
-4 3 V
-4 3 V
-4 4 V
-3 3 V
-4 3 V
-4 3 V
-4 4 V
-3 3 V
-4 3 V
-4 3 V
-3 3 V
-4 4 V
-3 3 V
-4 3 V
-3 3 V
-4 3 V
-4 3 V
-3 3 V
-4 3 V
-3 3 V
-4 3 V
-3 3 V
-3 3 V
-4 3 V
-3 3 V
-4 3 V
-3 3 V
-3 3 V
-4 3 V
-3 3 V
-3 3 V
-4 2 V
-3 3 V
-3 3 V
-4 3 V
-3 2 V
-3 3 V
-3 3 V
-4 3 V
-3 2 V
-3 3 V
-3 3 V
-3 2 V
-3 3 V
-4 2 V
-3 3 V
-3 2 V
-3 3 V
-3 2 V
-3 3 V
-3 2 V
-3 3 V
-3 2 V
-3 2 V
-3 3 V
-3 2 V
-3 2 V
-3 3 V
-3 2 V
-3 2 V
-3 2 V
-3 3 V
-3 2 V
-3 2 V
-3 2 V
-3 2 V
-3 2 V
-3 2 V
-2 2 V
-3 2 V
-3 2 V
-3 2 V
-3 2 V
-3 2 V
-2 2 V
-3 2 V
-3 2 V
-3 2 V
-2 1 V
-3 2 V
-3 2 V
-3 2 V
-2 1 V
-3 2 V
-3 2 V
-2 2 V
-3 1 V
-3 2 V
-2 1 V
-3 2 V
-3 1 V
-2 2 V
-3 2 V
-3 1 V
-2 1 V
-3 2 V
-2 1 V
-3 2 V
-2 1 V
-3 1 V
-3 2 V
-2 1 V
-3 1 V
-2 2 V
-3 1 V
-2 1 V
-3 1 V
-2 2 V
-3 1 V
-2 1 V
-2 1 V
-3 1 V
-2 1 V
-3 1 V
-2 2 V
-3 1 V
-2 1 V
-2 1 V
-3 1 V
-2 1 V
-2 1 V
-3 1 V
-2 0 V
-3 1 V
-2 1 V
-2 1 V
-2 1 V
-3 1 V
-2 1 V
-2 1 V
-3 0 V
-2 1 V
-2 1 V
-2 1 V
-3 0 V
-2 1 V
-2 1 V
-2 1 V
-3 0 V
-2 1 V
-2 1 V
-2 0 V
-2 1 V
-3 0 V
-2 1 V
currentpoint stroke M
-2 1 V
-2 0 V
-2 1 V
-2 0 V
-3 1 V
-2 1 V
-2 0 V
-2 1 V
-2 0 V
-2 1 V
-2 0 V
-2 1 V
-3 0 V
-2 0 V
-2 1 V
-2 0 V
-2 1 V
-2 0 V
-2 1 V
-2 0 V
-2 0 V
-2 1 V
-2 0 V
-2 1 V
-2 0 V
-2 0 V
-2 1 V
-2 0 V
-2 0 V
-2 1 V
-2 0 V
-2 0 V
-2 0 V
-2 1 V
-2 0 V
-2 0 V
-2 1 V
-2 0 V
-1 0 V
-2 0 V
-2 1 V
-2 0 V
-2 0 V
-2 0 V
-2 1 V
-2 0 V
-2 0 V
-1 0 V
-2 0 V
-2 1 V
-2 0 V
-2 0 V
-2 0 V
-2 0 V
-1 1 V
-2 0 V
-2 0 V
-2 0 V
-2 0 V
-1 0 V
-2 1 V
-2 0 V
-2 0 V
-2 0 V
-1 0 V
-2 0 V
-2 0 V
-2 0 V
-1 1 V
-2 0 V
-2 0 V
-2 0 V
-1 0 V
-2 0 V
-2 0 V
-1 0 V
-2 1 V
-2 0 V
-2 0 V
-1 0 V
-2 0 V
-2 0 V
-1 0 V
-2 0 V
-2 0 V
-1 0 V
-2 0 V
-2 0 V
-1 1 V
-2 0 V
-2 0 V
-1 0 V
-2 0 V
-1 0 V
-2 0 V
-2 0 V
-1 0 V
-2 0 V
-1 0 V
-2 0 V
-2 0 V
-1 0 V
-2 0 V
-1 0 V
-2 1 V
-2 0 V
-1 0 V
-2 0 V
-1 0 V
-2 0 V
-1 0 V
-2 0 V
-1 0 V
-2 0 V
-2 0 V
-1 0 V
-2 0 V
-1 0 V
-2 0 V
-1 0 V
-2 0 V
-1 0 V
-2 0 V
-1 0 V
-2 0 V
-1 0 V
-2 0 V
-1 0 V
-2 0 V
-1 0 V
-1 0 V
-2 0 V
-1 0 V
-2 0 V
-1 1 V
-2 0 V
-1 0 V
-2 0 V
-1 0 V
-1 0 V
-2 0 V
-1 0 V
-2 0 V
-1 0 V
-2 0 V
-1 0 V
-1 0 V
-2 0 V
-1 0 V
-2 0 V
-1 0 V
-1 0 V
-2 0 V
-1 0 V
-1 0 V
-2 0 V
-1 0 V
-2 0 V
-1 0 V
-1 0 V
-2 0 V
-1 0 V
-1 0 V
-2 0 V
-1 0 V
-1 0 V
-2 0 V
-1 0 V
-1 0 V
-2 0 V
-1 0 V
-1 0 V
-2 0 V
-1 0 V
-1 0 V
-1 0 V
-2 0 V
-1 0 V
-1 0 V
-2 0 V
-1 0 V
-1 0 V
-1 0 V
-2 0 V
-1 0 V
-1 0 V
-1 0 V
-2 0 V
-1 0 V
-1 0 V
-1 0 V
-2 0 V
-1 0 V
-1 0 V
-1 0 V
-2 0 V
-1 0 V
-1 0 V
-1 0 V
-2 0 V
-1 0 V
-1 0 V
-1 0 V
-1 0 V
-2 0 V
-1 0 V
-1 0 V
-1 0 V
-1 0 V
-2 0 V
-1 0 V
-1 0 V
-1 0 V
-1 0 V
-2 0 V
-1 0 V
-1 0 V
-1 0 V
-1 0 V
-1 0 V
-2 0 V
-1 0 V
-1 0 V
-1 0 V
-1 0 V
-1 0 V
-1 0 V
-2 0 V
-1 0 V
-1 0 V
-1 0 V
-1 0 V
-1 0 V
-1 0 V
-2 0 V
-1 0 V
-1 0 V
-1 0 V
-1 0 V
-1 0 V
-1 0 V
-1 0 V
-1 0 V
-1 0 V
-2 0 V
-1 0 V
-1 0 V
-1 0 V
-1 0 V
-1 0 V
-1 0 V
-1 0 V
-1 0 V
-1 0 V
-1 0 V
-1 0 V
-2 0 V
-1 0 V
-1 0 V
-1 0 V
-1 0 V
-1 0 V
-1 0 V
-1 0 V
-1 0 V
-1 0 V
-1 0 V
-1 0 V
-1 0 V
-1 0 V
-1 0 V
-1 0 V
-1 0 V
-1 0 V
-1 0 V
-1 0 V
-1 0 V
-1 0 V
-1 1 V
-1 0 V
-1 0 V
-1 0 V
-1 0 V
-1 0 V
-1 0 V
-1 0 V
-1 0 V
-1 0 V
-1 0 V
-1 0 V
-1 0 V
-1 0 V
-1 0 V
-1 0 V
-1 0 V
-1 0 V
-1 0 V
-1 0 V
-1 0 V
-1 0 V
-1 0 V
-1 0 V
-1 0 V
-1 0 V
-1 0 V
-1 0 V
-1 0 V
-1 0 V
-1 0 V
-1 0 V
-1 0 V
-1 0 V
-1 0 V
-1 0 V
-1 0 V
-1 0 V
-1 0 V
-1 0 V
-1 0 V
-1 0 V
-1 0 V
-1 0 V
-1 0 V
-1 0 V
-1 0 V
-1 0 V
-1 0 V
-1 0 V
-1 0 V
-1 0 V
-1 0 V
-1 0 V
-1 0 V
-1 0 V
-1 0 V
-1 0 V
-1 0 V
-1 0 V
-1 0 V
-1 0 V
-1 0 V
-1 0 V
-1 0 V
-1 0 V
-1 0 V
-1 0 V
-1 0 V
-1 0 V
-1 0 V
-1 0 V
-1 0 V
-1 0 V
-1 0 V
-1 0 V
-1 0 V
-1 0 V
-1 0 V
-1 0 V
-1 0 V
-1 0 V
-1 0 V
-1 0 V
-1 0 V
-1 0 V
-1 0 V
-1 0 V
-1 0 V
-1 0 V
-1 0 V
-1 0 V
-1 0 V
-1 0 V
-1 0 V
-1 0 V
-1 0 V
-1 0 V
-1 0 V
-1 0 V
-1 0 V
-1 0 V
-1 0 V
-1 0 V
-1 0 V
-1 0 V
-1 0 V
-1 0 V
-1 0 V
-1 0 V
-1 0 V
1.000 UL
LT2
2512 1583 M
263 0 V
3550 450 M
-19 0 V
-19 1 V
-19 0 V
-19 1 V
-19 0 V
-18 1 V
-18 0 V
-18 1 V
-18 1 V
-18 0 V
-17 1 V
-17 0 V
-18 1 V
-17 0 V
-16 1 V
-17 0 V
-17 1 V
-16 0 V
-16 1 V
-16 0 V
-16 1 V
-16 1 V
-16 0 V
-15 1 V
-16 0 V
-15 1 V
-15 0 V
-15 1 V
-15 0 V
-14 1 V
-15 1 V
-14 0 V
-15 1 V
-14 0 V
-14 1 V
-14 1 V
-14 0 V
-14 1 V
-13 0 V
-14 1 V
-13 0 V
-14 1 V
-13 1 V
-13 0 V
-13 1 V
-13 1 V
-12 0 V
-13 1 V
-13 0 V
-12 1 V
-12 1 V
-13 0 V
-12 1 V
-12 1 V
-12 0 V
-12 1 V
-11 0 V
-12 1 V
-12 1 V
-11 0 V
-12 1 V
-11 1 V
-11 0 V
-11 1 V
-11 1 V
-11 0 V
-11 1 V
-11 1 V
-11 0 V
-11 1 V
-10 1 V
-11 0 V
-10 1 V
-11 1 V
-10 0 V
-10 1 V
-10 1 V
-10 0 V
-10 1 V
-10 1 V
-10 0 V
-10 1 V
-9 1 V
-10 1 V
-10 0 V
-9 1 V
-10 1 V
-9 0 V
-9 1 V
-10 1 V
-9 1 V
-9 0 V
-9 1 V
-9 1 V
-9 1 V
-9 0 V
-9 1 V
-8 1 V
-9 0 V
-9 1 V
-8 1 V
-9 1 V
-8 0 V
-9 1 V
-8 1 V
-8 1 V
-9 0 V
-8 1 V
-8 1 V
-8 1 V
-8 1 V
-8 0 V
-8 1 V
-8 1 V
-8 1 V
-7 0 V
-8 1 V
-8 1 V
-8 1 V
-7 1 V
-8 0 V
-7 1 V
-8 1 V
-7 1 V
-7 1 V
-8 0 V
-7 1 V
-7 1 V
-7 1 V
-8 1 V
-7 0 V
-7 1 V
-7 1 V
-7 1 V
-7 1 V
-6 1 V
-7 0 V
-7 1 V
-7 1 V
-7 1 V
-6 1 V
-7 1 V
-7 0 V
-6 1 V
-7 1 V
-6 1 V
-7 1 V
-6 1 V
-6 1 V
-7 0 V
-6 1 V
-6 1 V
-7 1 V
-6 1 V
-6 1 V
-6 1 V
-6 1 V
-6 0 V
-6 1 V
-6 1 V
-6 1 V
-6 1 V
-6 1 V
-6 1 V
-6 1 V
-6 1 V
-5 1 V
-6 0 V
-6 1 V
-5 1 V
-6 1 V
-6 1 V
-5 1 V
-6 1 V
-5 1 V
-6 1 V
-5 1 V
-6 1 V
-5 1 V
-6 0 V
-5 1 V
-5 1 V
-6 1 V
-5 1 V
-5 1 V
-5 1 V
-5 1 V
-6 1 V
-5 1 V
-5 1 V
-5 1 V
-5 1 V
-5 1 V
-5 1 V
-5 1 V
-5 1 V
-5 1 V
-5 1 V
-5 1 V
-4 1 V
-5 1 V
-5 1 V
-5 1 V
-5 1 V
-4 1 V
-5 1 V
-5 1 V
-4 1 V
-5 1 V
-5 1 V
-4 1 V
-5 1 V
-4 1 V
-5 1 V
-5 1 V
-4 1 V
-4 1 V
-5 1 V
-4 1 V
-5 1 V
-4 1 V
-5 1 V
-4 1 V
-4 1 V
-4 1 V
-5 1 V
-4 1 V
-4 1 V
-4 1 V
-5 1 V
-4 1 V
-4 1 V
-4 1 V
-4 1 V
-4 1 V
-4 1 V
-5 1 V
-4 2 V
-4 1 V
-4 1 V
-4 1 V
-4 1 V
-4 1 V
-4 1 V
-3 1 V
-4 1 V
-4 1 V
-4 1 V
-4 1 V
-4 1 V
-4 2 V
-3 1 V
-4 1 V
-4 1 V
-4 1 V
-3 1 V
-4 1 V
-4 1 V
-4 1 V
-3 2 V
-4 1 V
-4 1 V
-3 1 V
-4 1 V
-3 1 V
-4 1 V
-3 1 V
-4 1 V
-4 2 V
-3 1 V
-4 1 V
-3 1 V
-4 1 V
-3 1 V
-3 1 V
-4 2 V
-3 1 V
-4 1 V
-3 1 V
-3 1 V
-4 1 V
-3 2 V
-3 1 V
-4 1 V
-3 1 V
-3 1 V
-4 1 V
-3 1 V
-3 2 V
-3 1 V
-4 1 V
-3 1 V
-3 1 V
-3 2 V
-3 1 V
-3 1 V
-4 1 V
-3 1 V
-3 1 V
-3 2 V
-3 1 V
-3 1 V
-3 1 V
-3 1 V
-3 2 V
-3 1 V
-3 1 V
-3 1 V
-3 2 V
-3 1 V
-3 1 V
-3 1 V
-3 1 V
-3 2 V
-3 1 V
-3 1 V
-3 1 V
-3 1 V
-3 2 V
-3 1 V
-2 1 V
-3 1 V
-3 2 V
-3 1 V
-3 1 V
-3 1 V
-2 2 V
-3 1 V
-3 1 V
-3 1 V
-2 2 V
-3 1 V
-3 1 V
-3 1 V
-2 2 V
-3 1 V
-3 1 V
-2 1 V
-3 2 V
-3 1 V
-2 1 V
-3 2 V
-3 1 V
-2 1 V
-3 1 V
-3 2 V
-2 1 V
-3 1 V
-2 1 V
-3 2 V
-2 1 V
-3 1 V
-3 2 V
-2 1 V
-3 1 V
-2 2 V
-3 1 V
-2 1 V
-3 1 V
-2 2 V
-3 1 V
-2 1 V
-2 2 V
-3 1 V
-2 1 V
-3 2 V
-2 1 V
-3 1 V
-2 2 V
-2 1 V
-3 1 V
-2 1 V
-2 2 V
-3 1 V
-2 1 V
-3 2 V
-2 1 V
-2 1 V
-2 2 V
-3 1 V
-2 1 V
-2 2 V
-3 1 V
-2 2 V
-2 1 V
-2 1 V
-3 2 V
-2 1 V
-2 1 V
-2 2 V
-3 1 V
-2 1 V
-2 2 V
-2 1 V
-2 1 V
-3 2 V
-2 1 V
currentpoint stroke M
-2 1 V
-2 2 V
-2 1 V
-2 2 V
-3 1 V
-2 1 V
-2 2 V
-2 1 V
-2 1 V
-2 2 V
-2 1 V
-2 2 V
-3 1 V
-2 1 V
-2 2 V
-2 1 V
-2 1 V
-2 2 V
-2 1 V
-2 2 V
-2 1 V
-2 1 V
-2 2 V
-2 1 V
-2 2 V
-2 1 V
-2 1 V
-2 2 V
-2 1 V
-2 2 V
-2 1 V
-2 1 V
-2 2 V
-2 1 V
-2 2 V
-2 1 V
-2 1 V
-2 2 V
-1 1 V
-2 2 V
-2 1 V
-2 2 V
-2 1 V
-2 1 V
-2 2 V
-2 1 V
-2 2 V
-1 1 V
-2 1 V
-2 2 V
-2 1 V
-2 2 V
-2 1 V
-2 2 V
-1 1 V
-2 1 V
-2 2 V
-2 1 V
-2 2 V
-1 1 V
-2 2 V
-2 1 V
-2 2 V
-2 1 V
-1 1 V
-2 2 V
-2 1 V
-2 2 V
-1 1 V
-2 2 V
-2 1 V
-2 2 V
-1 1 V
-2 1 V
-2 2 V
-1 1 V
-2 2 V
-2 1 V
-2 2 V
-1 1 V
-2 2 V
-2 1 V
-1 2 V
-2 1 V
-2 1 V
-1 2 V
-2 1 V
-2 2 V
-1 1 V
-2 2 V
-2 1 V
-1 2 V
-2 1 V
-1 2 V
-2 1 V
-2 2 V
-1 1 V
-2 1 V
-1 2 V
-2 1 V
-2 2 V
-1 1 V
-2 2 V
-1 1 V
-2 2 V
-2 1 V
-1 2 V
-2 1 V
-1 2 V
-2 1 V
-1 2 V
-2 1 V
-1 2 V
-2 1 V
-2 2 V
-1 1 V
-2 2 V
-1 1 V
-2 2 V
-1 1 V
-2 1 V
-1 2 V
-2 1 V
-1 2 V
-2 1 V
-1 2 V
-2 1 V
-1 2 V
-2 1 V
-1 2 V
-1 1 V
-2 2 V
-1 1 V
-2 2 V
-1 1 V
-2 2 V
-1 1 V
-2 2 V
-1 1 V
-1 2 V
-2 1 V
-1 2 V
-2 1 V
-1 2 V
-2 1 V
-1 2 V
-1 1 V
-2 2 V
-1 1 V
-2 2 V
-1 1 V
-1 2 V
-2 1 V
-1 2 V
-1 1 V
-2 2 V
-1 1 V
-2 2 V
-1 1 V
-1 2 V
-2 1 V
-1 2 V
-1 1 V
-2 2 V
-1 1 V
-1 2 V
-2 1 V
-1 2 V
-1 1 V
-2 2 V
-1 1 V
-1 2 V
-2 1 V
-1 2 V
-1 1 V
-1 2 V
-2 1 V
-1 2 V
-1 1 V
-2 2 V
-1 1 V
-1 2 V
-1 1 V
-2 2 V
-1 1 V
-1 2 V
-1 1 V
-2 2 V
-1 1 V
-1 2 V
-1 1 V
-2 2 V
-1 1 V
-1 2 V
-1 1 V
-2 2 V
-1 1 V
-1 2 V
-1 1 V
-2 2 V
-1 1 V
-1 2 V
-1 1 V
-1 2 V
-2 1 V
-1 2 V
-1 1 V
-1 2 V
-1 1 V
-2 2 V
-1 1 V
-1 2 V
-1 1 V
-1 2 V
-2 1 V
-1 2 V
-1 1 V
-1 2 V
-1 1 V
-1 2 V
-2 1 V
-1 2 V
-1 1 V
-1 2 V
-1 1 V
-1 1 V
-1 2 V
-2 1 V
-1 2 V
-1 1 V
-1 2 V
-1 1 V
-1 2 V
-1 1 V
-2 2 V
-1 1 V
-1 2 V
-1 1 V
-1 2 V
-1 1 V
-1 2 V
-1 1 V
-1 2 V
-1 1 V
-2 2 V
-1 1 V
-1 2 V
-1 1 V
-1 2 V
-1 1 V
-1 1 V
-1 2 V
-1 1 V
-1 2 V
-1 1 V
-1 2 V
-2 1 V
-1 2 V
-1 1 V
-1 2 V
-1 1 V
-1 2 V
-1 1 V
-1 2 V
-1 1 V
-1 1 V
-1 2 V
-1 1 V
-1 2 V
-1 1 V
-1 2 V
-1 1 V
-1 2 V
-1 1 V
-1 2 V
-1 1 V
-1 1 V
-1 2 V
-1 1 V
-1 2 V
-1 1 V
-1 2 V
-1 1 V
-1 2 V
-1 1 V
-1 1 V
-1 2 V
-1 1 V
-1 2 V
-1 1 V
-1 2 V
-1 1 V
-1 1 V
-1 2 V
-1 1 V
-1 2 V
-1 1 V
-1 2 V
-1 1 V
-1 1 V
-1 2 V
-1 1 V
-1 2 V
-1 1 V
-1 2 V
-1 1 V
-1 1 V
-1 2 V
-1 1 V
-1 2 V
0 1 V
-1 1 V
-1 2 V
-1 1 V
-1 2 V
-1 1 V
-1 1 V
-1 2 V
-1 1 V
-1 2 V
-1 1 V
-1 1 V
-1 2 V
-1 1 V
0 2 V
-1 1 V
-1 1 V
-1 2 V
-1 1 V
-1 2 V
-1 1 V
-1 1 V
-1 2 V
-1 1 V
0 1 V
-1 2 V
-1 1 V
-1 2 V
-1 1 V
-1 1 V
-1 2 V
-1 1 V
-1 1 V
0 2 V
-1 1 V
-1 2 V
-1 1 V
-1 1 V
-1 2 V
-1 1 V
0 1 V
-1 2 V
-1 1 V
-1 1 V
-1 2 V
-1 1 V
-1 1 V
0 2 V
-1 1 V
-1 1 V
-1 2 V
-1 1 V
-1 1 V
-1 2 V
0 1 V
-1 1 V
-1 2 V
-1 1 V
-1 1 V
-1 2 V
0 1 V
-1 1 V
-1 2 V
-1 1 V
-1 1 V
0 1 V
-1 2 V
-1 1 V
-1 1 V
-1 2 V
-1 1 V
0 1 V
-1 2 V
-1 1 V
-1 1 V
-1 1 V
0 2 V
-1 1 V
-1 1 V
-1 2 V
-1 1 V
0 1 V
-1 1 V
-1 2 V
-1 1 V
-1 1 V
0 1 V
-1 2 V
-1 1 V
-1 1 V
currentpoint stroke M
0 1 V
-1 2 V
-1 1 V
stroke
grestore
end
showpage
}
\put(2825,1583){\makebox(0,0)[l]{$\lambda s_y= 0.5$}}
\put(2825,1683){\makebox(0,0)[l]{$\lambda s_y= 0.05$}}
\put(2825,1783){\makebox(0,0)[l]{$\lambda s_y = 0.005$}}
\put(2000,150){\makebox(0,0){$r/R(t)$}}
\put(100,1230){\makebox(0,0)[b]{\shortstack{$\displaystyle{\tilde s \over \mu}$}}}
\put(3550,300){\makebox(0,0){5}}
\put(2775,300){\makebox(0,0){4}}
\put(2000,300){\makebox(0,0){3}}
\put(1225,300){\makebox(0,0){2}}
\put(450,300){\makebox(0,0){1}}
\put(400,1783){\makebox(0,0)[r]{0.1}}
\put(400,400){\makebox(0,0)[r]{0}}
\end{picture}